\documentclass[pra,twocolumn,superscriptaddress,preprintnumbers,nofootinbib]{revtex4-1}

\usepackage[utf8]{inputenc}
\usepackage{graphicx}
\usepackage{amsmath}
\usepackage{amsthm}
\usepackage{bm}
\usepackage{float}
\usepackage{amsfonts}
\usepackage{enumerate}
\usepackage{amssymb}%
\usepackage{array}%
\usepackage[margin=0.75in]{geometry}
\usepackage{color}
\usepackage{soul}
\usepackage{mathtools}
\theoremstyle{definition}

\usepackage[colorlinks=true,urlcolor=blue,citecolor=blue]{hyperref} 

\usepackage{physics}

\usepackage{ulem}
\usepackage{hyperref}

\renewcommand{\emph}[1]{\textit{#1}}

\newcounter{para}

\makeatletter
\newcommand*\bigcdot{\mathpalette\bigcdot@{.5}}
\newcommand*\bigcdot@[2]{\mathbin{\vcenter{\hbox{\scalebox{#2}{$\m@th#1\bullet$}}}}}

\newcommand{\llangle}[1][]{\savebox{\@brx}{\(\m@th{#1\langle}\)}%
	\mathopen{\copy\@brx\kern-0.5\wd\@brx\usebox{\@brx}}}
\newcommand{\rrangle}[1][]{\savebox{\@brx}{\(\m@th{#1\rangle}\)}%
	\mathclose{\copy\@brx\kern-0.5\wd\@brx\usebox{\@brx}}}

\makeatother

\usepackage{array}
\newcolumntype{L}{>{$}l<{$}} 
\newcolumntype{C}{>{$}c<{$}} 
\newcolumntype{R}{>{$}r<{$}} 

\usepackage{xcolor}

\usepackage{mathtools}

\usepackage{physics}
\usepackage{booktabs}
\usepackage{multirow}

\newtheorem*{theorem*}{Theorem}
\newtheorem*{lemma*}{Lemma}

\usepackage{comment}
\usepackage{ragged2e}
\usepackage{atbegshi,picture}
\usepackage{array}
\usepackage{placeins}
\newcolumntype{M}[1]{>{\centering\arraybackslash}m{#1}}
\usepackage{soul}
 \makeatletter
 \usepackage{booktabs}

\begin{document}
	\newcommand{\Caltech}{California Institute of Technology, Pasadena, CA, USA}
	\newcommand{\MIT}{
		Massachusetts Institute of Technology, Cambridge, MA, USA}
	\newcommand{\WalterBurke}{Walter Burke Institute for Theoretical Physics, California Institute of Technology, Pasadena, CA, USA}
	\newcommand{\GoogleQuantum}{Google Quantum AI, Venice, CA, USA}
	\newcommand{\Stanford}{
		Stanford University, Stanford, CA, USA}
	\newcommand{\TCD}{School of Physics, Trinity College Dublin, Dublin 2, Ireland}
	\newcommand{\TKcite}{{\color{red}[CITE]}}
	\newcommand{\nocontentsline}[3]{}
	\let\origcontentsline\addcontentsline
	\newcommand\stoptoc{\let\addcontentsline\nocontentsline}
	\newcommand\resumetoc{\let\addcontentsline\origcontentsline}
	
	\preprint{MIT-CTP/5931}
	\title{
		Observation of ballistic plasma and memory in high-energy gauge theory dynamics
	}
	\author{Daniel~K.~Mark}
	\affiliation{\MIT}
	
	\author{Federica~M.~Surace}
	\affiliation{\Caltech}
	\affiliation{\TCD}
	
	\author{Thomas~Schuster}
	\affiliation{\Caltech}

	\author{Adam~L.~Shaw}
	\affiliation{\Caltech}
	\affiliation{\Stanford}
	
	\author{Wenjie~Gong}
	\affiliation{\MIT}
	
	\author{Soonwon~Choi}
	\altaffiliation{Corresponding author: \href{mailto:soonwon@mit.edu}{soonwon@mit.edu}}
	\affiliation{\MIT}
	
	\author{Manuel~Endres}
	\altaffiliation{Corresponding author: \href{mailto:mendres@caltech.edu}{mendres@caltech.edu}}
	\affiliation{\Caltech}
	
\maketitle

\begin{figure*}[t!]
	\includegraphics[width=\textwidth]{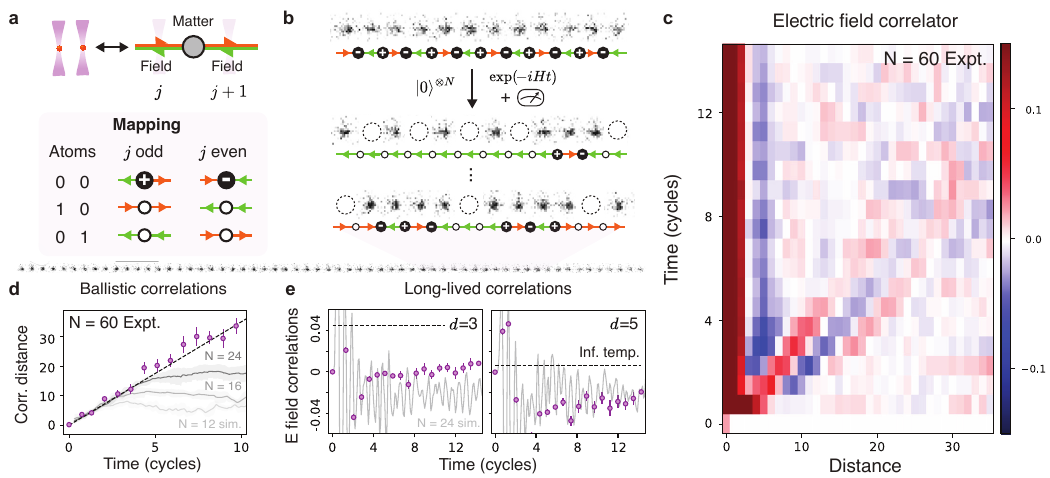}
	\caption{\textbf{Experimental observations.} \textbf{a.} Correspondence between the Rydberg atom array and a lattice gauge theory.  
		Atomic states map onto configurations of electric fields and charges. The Rydberg blockaded spin configurations map onto the matter-gauge configurations that satisfy Gauss' Law and the Kogut-Susskind fermion staggering.
		\textbf{b.} We perform an out-of-equilibrium quench experiment on a Rydberg atom array, equivalent to a one-dimensional quantum link model. The initial state $|0\rangle^{\otimes N}$ corresponds to  state fully filled with particles and anti-particles.
		Experimental measurements yield atomic fluorescence images, with bright and dark (dashed circles) indicating $|0\rangle/|1\rangle$, which translate to field and charge configurations. 
		\textbf{c.} Connected electric-field correlations  $\langle E_j E_{{j+d}}\rangle_c = (-1)^d(\langle Z_j Z_{{j+d}}\rangle-\langle Z_j \rangle \langle Z_{{j+d}}\rangle)$ reveal correlations propagating ballistically to large distances
		and short-range correlations which persist to long times,
		both defying conventional expectations of thermalization.
		\textbf{d.} Consistent with ballistic transport, the average distance of the correlations (Methods) grows with a velocity that quantitatively agrees with analytical predictions (dashed line) as well as exact simulations of small system sizes (gray).
		\textbf{e.} Late-time correlations deviate from the infinite-temperature value (black dashed) at short distances, in agreement with numerical simulations (grey), which reveal further oscillations over time.
	}
	\label{fig:experimental_data} 
\end{figure*}

\noindent
\textbf{Gauge theories describe the fundamental forces of nature~\cite{kogut1979introduction, kogut1983lattice, oraifeartaigh2000gauge}.
	However, high-energy dynamics, such as the formation of quark--gluon plasmas~\cite{shuryak2017strongly}, is notoriously difficult to model with classical methods. Quantum simulation offers a promising alternative in this regime, yet experiments have mainly probed low energies~\cite{banuls2020simulating, aidelsburger2022cold, bauer2023quantum,halimeh2025quantum,martinez2016real,yang2020observation,zhang2024observation,farrell2024quantum,de2024observation,ciavarella2024quantum,cochran2025visualizing,gonzalez2025observation,schuhmacher2025observation,datla2025statistical}. Here, we observe the formation of a ballistic plasma and long-time memory effects in high-energy gauge theory dynamics on a high-precision quantum simulator~\cite{shaw2024benchmarking}. Both observations are unexpected, as the initial state -- fully filled with particle-antiparticle pairs -- was thought to rapidly thermalize~\cite{zhou2022thermalization,bernien2017probing,turner2018weak}. Instead, we find correlations spreading ballistically to long distances and a memory of charge clusters. 
	Our observations cannot be explained by many-body scars~\cite{bernien2017probing,turner2018weak,chandran2023quantum}, but are captured by a new theory of plasma oscillations between electric field and current operators~\cite{kluger1992fermion,hebenstreit2013simulating,buyens2014matrix,buyens2017real}, persisting all the way to the continuum limit of the (1+1)D Schwinger model, of which we simulate a lattice version ~\cite{surace2020lattice}. Adapting techniques from quantum optics, we visualize plasma oscillations as rotations of Wigner distributions, leading to a novel set of predictions which we test in experiment and numerics. The new framework encompasses both our scenario and scars, which show up as coherent states of the plasma. 
	The experimental surprises we observe in the high-energy dynamics of a simple gauge theory point to the potential of high-precision quantum simulations of gauge theories for general scientific discovery.
}

Gauge theories encapsulate the fundamental forces of nature such as in quantum electrodynamics (QED) or quantum chromodynamics (QCD)~\cite{kogut1979introduction,kogut1983lattice,oraifeartaigh2000gauge}. Studies of gauge theories underpin much of modern research in high-energy physics, spanning lattice QCD simulations~\cite{beane2011nuclear,kronfeld2012twenty,nagata2022finite} to effective field theories that model complex phenomenology in high-energy experiment~\cite{burgess2007introduction,hammer2020nuclear}.
However, many questions remain unanswered due to long-standing challenges facing conventional analytic and numerical approaches. 

Quantum technologies provide new ways to study gauge theories.
Quantum simulations allow for controlled explorations of lattice gauge theories~\cite{banuls2020simulating,aidelsburger2022cold,bauer2023quantum}. They may answer questions difficult to address with traditional approaches, such as about real-time dynamics that are particularly challenging to simulate with classical computing approaches~\cite{halimeh2025quantum}. Recent experiments have observed non-trivial phenomena in several lattice gauge theories (LGTs), including string breaking, confinement, and the effects of topological terms arising from a background field~\cite{martinez2016real,yang2020observation,zhang2024observation,farrell2024quantum,de2024observation,ciavarella2024quantum,cochran2025visualizing,gonzalez2025observation,schuhmacher2025observation,datla2025statistical}. However, most experiments probe the dynamics of low-energy excitations atop a ground state. Dynamics at high energies, relevant for physical processes such as quark-gluon plasmas~\cite{shuryak2017strongly}, remain largely unexplored.

Here we report two unexpected experimental observations in a gauge theory at high energies realized by a Rydberg quantum simulator. We perform a quantum quench at infinite effective temperature under the paradigmatic PXP model~\cite{bernien2017probing,turner2018weak,fendley2004competing} which is equivalent to the \textit{quantum link model}~\cite{surace2020lattice} describing (1+1)D QED~\cite{kogut1983lattice,banuls2020simulating}. From an initial state filled with particle-antiparticle charges~\cite{yang2020observation,zhou2022thermalization}, we observe the ballistic propagation of plasma quasiparticles on an infinite temperature background, as well as persistent athermal short-range correlations.

The plasma quasiparticles arise from a collective mode that has a sharply-defined band structure even at infinite effective temperature. 
Underlying this is the fact that at the operator level, the electric field and current form approximate canonical pairs with a momentum-dependent coupling frequency. 
Adapting the Wigner distribution from quantum optics~\cite{scully1997quantum} provides a common, quantitative framework for both our observations and previous ones typically associated with many-body scars in the PXP model. 
They illustrate that our experimental state is analogous to a squeezed vacuum state,  while scar initial states are analogous to coherent states in quantum optics. Finally, we find that the observed athermal correlations arise from a long memory of three-body charge clusters: the Liouvillian graph from quantum information theory~\cite{white2023effective,yithomas2024comparing,parker2019universal,cao2021statistical} reveals the mechanism as destructive interference in operator space.

\vspace{0.25cm}
\noindent\textbf{Experimental lattice gauge theory dynamics}\newline
Our experimental system consists of a Rydberg quantum simulator in which up to $N=60$ strontium-88 atoms are arranged in a linear chain~\cite{madjarov2020high,shaw2024benchmarking}. Each atom is well-described by a two-level system comprising a low-lying electronic state $|0\rangle$ and a highly-excited Rydberg state $|1\rangle$. 
The experiment performs a high-fidelity quantum quench~\cite{shaw2024benchmarking} on an initial state $|\psi_0\rangle \equiv|0\rangle^{\otimes N}$, with a time-independent Hamiltonian that is close to the well-studied PXP model~\cite{bernien2017probing,turner2018weak,fendley2004competing}, before projective measurement in the computational basis. 

The model was expected to show typical chaotic behavior~\cite{srednicki1999approach,bertini2021finite} except for rare initial states that lead to long-lasting oscillations (so-called many-body scars~\cite{bernien2017probing,turner2018weak,chandran2023quantum}).
In this context, our initial state, $|\psi_0\rangle \equiv|0\rangle^{\otimes N}$, was regarded as a counterpoint to a prototypical scarred initial state, the N{\'e}el state $|\mathbb{Z}_2\rangle\equiv|1010\dots\rangle$~\cite{bernien2017probing,turner2018quantum}, and its local expectation values were expected to quickly thermalize to their infinite temperature values according to chaotic dynamics~\cite{bernien2017probing,turner2018weak,bertini2021finite,zhou2022thermalization} that governs a generic initial state.

Among various approaches developed~\cite{turner2018weak,turner2018quantum,choi2019emergent,khemani2019signatures,chandran2023quantum,giudici2024unraveling}, the PXP model can be described as a lattice gauge theory~\cite{surace2020lattice}, a perspective extremely useful for understanding our new and previous observations. 
The Rydberg-blockade constraint exactly corresponds to a Gauss law gauge constraint~\cite{surace2020lattice} that couples charged particles with gauge fields on a lattice. Positively and negatively charged fermions occupy alternating sites, in accordance with the \textit{Kogut-Susskind fermion staggering}~\cite{kogut1983lattice}, while electric fields occupy the links between matter sites, taking values in accordance to Gauss' law. In a full $U(1)$ gauge theory, the electric field takes unbounded, integer-spaced values. Our setting corresponds to the \textit{quantum link model}, which truncates this to a spin-1/2 variable $E_j = \pm 1/2$: such truncations are necessary to simulate lattice gauge theories on systems with finite local Hilbert spaces~\cite{bauer2023quantum}.

Electric fields $E_j$ in the link model correspond to spins $Z_j$ in the Rydberg array as $E_j = (-1)^j Z_j$ (Fig.~\ref{fig:experimental_data}a). The allowed configurations under the Rydberg blockade~\cite{urban2009observation} correspond exactly to all allowed field configurations under Gauss' law and the charged fermion staggering. 
The initial state $|0\rangle^{\otimes N}$ maps onto a state fully filled with alternating particle and anti-particle pairs, and 
subsequent experimental measurements can be translated into configurations of fields and charges (Fig.~\ref{fig:experimental_data}b). The $\mathbb{Z}_2$ N{\'e}el state corresponds to an initial state with a homogeneous electric field and no charges~\cite{surace2020lattice}.

The dynamics of the PXP model and the quantum link model are also equivalent. The link model is specified by the more general \textit{massless lattice Schwinger model}, 
\begin{equation}
	H_\text{Schw.} = \sum_{j = 2}^{N} \left(c^\dagger_{j-1,j} U_{j}^{}c^{}_{j,j+1} + \text{h.c.} \right) + \mathcal{J} \sum_{j=1}^{N} E_{j}^2.
	\label{eq:lattice_Schwinger_Hamiltonian} \end{equation}
in the limit of strong matter-field coupling $\mathcal{J}$~\cite{surace2020lattice}. The (lattice) Schwinger model provides a controlled setting to study nonperturbative phenomena in gauge theories, such as confinement and string-breaking~\cite{coleman1976more}.

This model describes fermions (with creation operators $c^\dagger_{j-1,j}$) coupled to a vector potential $U_{j}$, along with a Maxwell field energy term for the electric field $E_j$.
Its coefficient $\mathcal{J}$ is a tuning parameter which allows us to study the generality of our observations: the limit $\mathcal{J}\rightarrow \infty$
, along with the boundary condition $E_0 = 1/2$, 
restricts $E_j = \pm 1/2$, corresponds to the quantum link model~\cite{surace2020lattice}, while $\mathcal{J}\rightarrow 0$ corresponds to the exactly-solvable massless Schwinger field theory limit~\cite{coleman1976more,banuls2013mass}. 

\begin{figure*}[t!]
	\includegraphics[width=\textwidth]{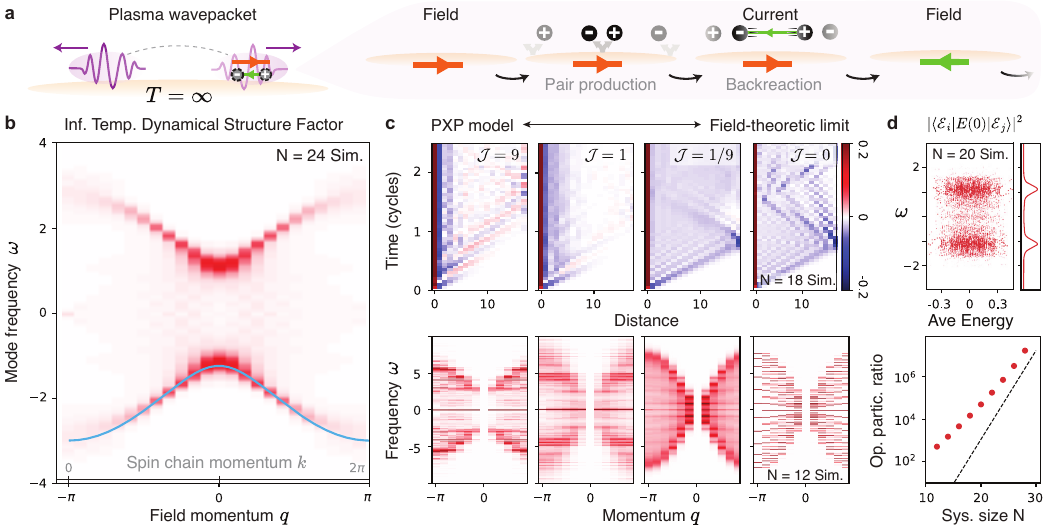}
	\caption{\textbf{Plasma band structure.} \textbf{a.} Ballistic correlations are due to plasma oscillations on an infinite temperature background. Plasma oscillations are induced by pair production and backreaction processes which couple electric field and charge current.
		\textbf{b.} Plasma oscillations are evident in the narrow band $\omega_0(q)$ of the infinite-temperature dynamical structure factor (iDSF) of the electric field operator (computed at $N=24$, but whose features are independent of system size), well approximated by a mean-field theory (blue line, Methods).
		\textbf{c.} The lattice Schwinger model [Eq.~\eqref{eq:lattice_Schwinger_Hamiltonian}] provides a class of models containing plasma oscillations, spanning the large $\mathcal{J}$ PXP limit to the $\mathcal{J}\rightarrow 0$ field-theoretic limit. This is evident in the ballistic propagation of correlators (top row) and the band structure of the iDSF (bottom) of the fermion occupation operator.
		\textbf{d.} Our phenomenology is not due to exceptional eigenstates (so-called scars) and requires all eigenstates to participate. Top: The iDSF is the sum of a large number of terms $|\langle \mathcal{E}_i |E(q)|\mathcal{E}_j\rangle|^2$ for eigenstate pairs with energy difference $\omega = \mathcal{E}_i - \mathcal{E}_j$. Summing over all average energies $(\mathcal{E}_i+\mathcal{E}_j)/2$ gives the iDSF (data for $N=20$ system plotted for momentum $q=0$). Bottom: the iDSF is delocalized among generic eigenstate pairs, as confirmed by the operator participation ratio (Methods) which asymptotically saturates the maximum expected growth rate of $\varphi^{2N}$ (dashed), where $\varphi \approx 1.618$ is the golden ratio. 
	}
	\label{fig:Lattice_Schwinger} 
\end{figure*}

\vspace{0.25cm}
\noindent\textbf{Experimental observations}\newline
We examine the connected two-point correlators $\langle E_j(t),E_{j+d}(t)\rangle_c$ at different times $t$. Since the system is approximately homogeneous, we average over positions $j$ away from the boundaries (Methods) and study the correlators as a function of the distance $d$ and time $t$ (Fig.~\ref{fig:experimental_data}c). 
The spacetime correlations reveal two surprising findings. First, correlations propagate to large distances with constant velocity (Fig.~\ref{fig:experimental_data}d) while also undergoing internal oscillations, as seen in the red-and-blue stripes.
Second, short-distance correlations in our experimental data do not reach thermal equilibrium (Fig.~\ref{fig:experimental_data}e) within experimental timescales.
We verify through numerical simulation that this behavior is seen in the PXP model, and deviations arising from the long-range $1/R^6$ Ising interactions do not qualitatively affect our observations.

Both phenomena are unexpected in a generic many-body system for several reasons. 
Non-conserved quantities are expected to relax exponentially to their thermal values~\cite{srednicki1999approach,rigol2008thermalization}, irrespective of its initial conditions.
Furthermore, although the growth of correlations is bounded within a Lieb-Robinson light-cone~\cite{lieb1972finite}, it is unusual in a nonintegrable, chaotic system to exhibit ballistic transport \textit{on} a light cone (in fact, these correlations also carry a considerable amount of non-local entanglement [Ext.~Dat.~Fig.~\ref{fig:entanglement_dynamics_ED}]). Instead featureless, diffusive propagation of conserved quantities~\cite{bertini2021finite} is expected. This expectation is exemplified by the mixed field Ising model (MFIM), a canonical model of many-body chaos~\cite{kim2014testing,rodriguez2024quantifying} (Ext.~Dat.~Fig.~\ref{fig:MFIM_TFIM_XXZ_structure_factors}). 

\begin{figure*}[t!]
	\includegraphics[width=\textwidth]{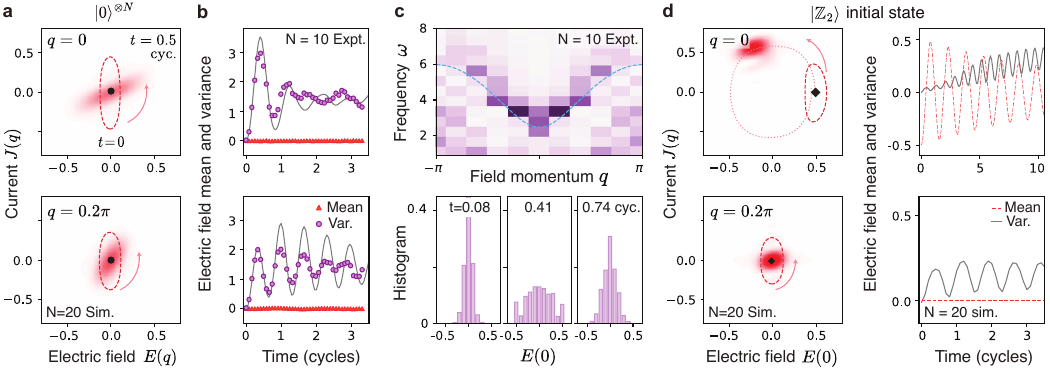}
	\caption{\textbf{Wigner distributions.} 
		\textbf{a.} The many-body Wigner distribution reveals semi-classical structure in a phase space of electric field and current. The evolved $|0\rangle^{\otimes N}$ state is akin to a squeezed state: its Wigner distribution is always centered about the origin, with a stretched axis (initially along the current direction [dashed ovals]) that rotates over time. This holds in all momentum sectors, illustrated for $q=0$ (top) and $q=0.2\pi$ (bottom).
		\textbf{b.} 
		Experimental data of the field mean and variance for the evolved $|0\rangle^{\otimes N}$ state confirms the phase-space oscillations. While $\langle E(q)\rangle$ (orange triangles) is always zero, $\text{var}[E(q)]$ oscillates over time for all momenta $q$.
		\textbf{c.} Top: The spacetime Fourier transform of the variance oscillations (normalized for each $q$, Methods) reflect their momentum-dependent frequencies and experimentally reconstructs the iDSF band-structure $2\omega_0(q)$ (blue dashed line).
		Bottom: The experimental full-counting-statistics of electric field broadens before sharpening again, in accordance with the Wigner distribution dynamics.
		\textbf{d.} In contrast, the $|\mathbb{Z}_2\rangle$ state is analogous to a coherent state in the $q=0$ mode. Top: Its Wigner distribution is displaced in phase space and moves along the circular trajectory traced by its expectation values (dotted line). Accordingly, both the mean and variance of the uniform electric field $E(0)$ oscillate.
		The $|\mathbb{Z}_2\rangle$ state is not displaced for $q\neq 0$ modes and like $|0\rangle^{\otimes N}$, behaves as a squeezed state. Accordingly, numerical simulations reveal variance oscillations for $E(q\neq0)$.
	}
	\label{fig:wigner} 
\end{figure*}

\vspace{0.25cm}
\noindent\textbf{Plasma oscillations}\newline
We attribute the ballistic correlations to the propagation of wavepackets of plasma oscillations (Fig.~\ref{fig:Lattice_Schwinger}a).
These are coherent, collective oscillations of the electric field coupled to particle–antiparticle pairs: they arise from the hybridization between the electric field and pair-creation (i.e.,~current) operators (in spin language, the $Z$ and $PYP$ operators~\cite{iadecola2019quantum}, Methods).
Plasma oscillations are a generic feature of electrodynamics. However, they have thus far only been studied in certain regimes of the lattice Schwinger model, such as at low energies, where the effects of backreaction on the electric field due to pair creation can be suitably  approximated~\cite{kluger1992fermion,hebenstreit2013simulating,buyens2014matrix,buyens2017real}. 
We find these collective modes at all effective temperatures (Ext.~Dat.~Fig.~\ref{fig:PXP_model_operator_structure_factors}), including infinite temperature.

The plasma modes in our system are evident in the \textit{infinite-temperature dynamical structure factor} (iDSF) of the electric field operator $E(q)\equiv \sum_j e^{iqj}E_j$ in the PXP model (Fig.~\ref{fig:Lattice_Schwinger}b, Methods). 
While the DSF is most commonly used to probe the physics of low-energy excitations~\cite{baez2020dynamical},
here we consider it at infinite temperature.
For all momenta $q$, the iDSF displays sharp peaks, unlike standard interacting chaotic quantum systems, whose iDSFs are typically featureless (Ext. Dat. Fig.~\ref{fig:MFIM_TFIM_XXZ_structure_factors}). The peak frequencies and linewidths are independent of system size, indicating velocities and lifetimes that we expect to persist in the thermodynamic limit.The iDSF of the current operator $J(q)$ shows an identical band structure (Ext.~Dat.~Fig.~\ref{fig:PXP_model_operator_structure_factors}).
The peak frequencies $\omega_0(q)$ grow with field momentum $|q|$ and are well approximated by a mean-field theory which derives equations of motion that couple field and current, analogous to Maxwell's equations (Methods).

The momentum dependence of the plasma frequency is crucial to our observed ballistic correlations. The group velocity $\partial_q\omega_0(q)$ (taken as a simple uniform average over $q\in[0,\pi]$) agrees with the observed correlation velocity (Fig.~\ref{fig:experimental_data}d, Ext.~Dat.~Fig.~\ref{fig:data_processing_correlators}). Moreover, this structure --- the ballistic spreading of correlations and band structure in the iDSF --- persists across all values of $\mathcal{J}$ in the lattice Schwinger model [Eq.~\eqref{eq:lattice_Schwinger_Hamiltonian}, Fig.~\ref{fig:Lattice_Schwinger}c], with certain differences: the presence of a gap $\omega_0(0) \neq 0$ in the iDSF for large $\mathcal{J}$ results in a distinct phase velocity, manifesting as the red/blue stripes in the correlation function, absent in the continuum limit (Methods). 

Finally, we numerically confirm that these features of the iDSF are generic: they originate from all eigenstates of the PXP model rather than from a small number of special ones such as many-body scars. To illustrate this, in Fig.~\ref{fig:Lattice_Schwinger}d we plot the eigenstate overlaps $|\langle \mathcal{E}_i|E(q)|\mathcal{E}_j\rangle|^2$ which sum to the iDSF: the size of each point  $(\mathcal{E}_i, \mathcal{E}_j)$ is proportional to the (logarithm of the) above overlap. The electric field operator maps each eigenstate $|\mathcal{E}_j\rangle$ onto exponentially many other eigenstates $|\mathcal{E}_i\rangle$.
The participation ratio, which quantifies the degree of delocalization of the iDSF (Methods), confirms this exponential scaling.

\vspace{0.25cm}
\noindent\textbf{Many-body Wigner distributions}\newline
We now show how the dynamics of various initial states can be visualized using tools adapted from quantum optics. This is inspired by our mean-field theory (Methods), where the electric field and current operators form approximate canonical operator pairs with a momentum-dependent coupling frequency --- similar to the typical quantum optics setting, where the electric and magnetic fields form canonical pairs.
Plasma oscillations become exact in the exactly-solvable continuum limit of the Schwinger model~\cite{coleman1976more,iso1988all}. 

Specifically, we adapt \textit{Wigner distributions} from quantum optics to the strongly interacting many-body problem.
These allow picturing plasma oscillations as rotations in a phase-space spanned by electric field and current operators $(E(q),J(q))$ for each field mode $q$. 
Beginning from a first-principles definition of the Wigner distribution
--- in terms of the 
(inverse) \textit{Wigner-Weyl transform}~\cite{scully1997quantum}
---we introduce the Wigner distribution $W(E, J)$ for our system (Methods).  Although the electric field and current are only approximately canonically conjugate, we show that $W(E, J)$ satisfies many desired properties of a quasiprobability distribution. In particular, the two-dimensional phase-space distributions accurately reproduce the marginal probability distributions on $E(q)$ or $J(q)$. 
At each momentum $q$ (corresponding to a field mode in quantum optics), the Wigner distribution exists, and the Hamiltonian dynamics approximately rotates the distribution in phase space with frequency $\omega(q)$.

The Wigner distributions of the initial state $|0\rangle^{\otimes N}$ are akin to \textit{squeezed vacuum states}: centered on the origin and squeezed initially along the current axis. For all $q$, the electric field variance $\text{var}[E(q)]$ of $|0\rangle^{\otimes N}$ is initially zero. The distribution and the squeezed axis subsequently rotate with frequency $\omega_0(q)$ during the quench dynamics (Fig.~\ref{fig:wigner}a). 

This picture has measurable consequences. For each $q$, we can evaluate the variance and mean of $E(q)$, as shown in Fig.~\ref{fig:wigner}b for $q=0$ and $q=\pi/5$. While the means $\langle E(q)\rangle$ are always zero, the \textit{variances} $\text{var}[E(q)]$ oscillate with frequency $2\omega_0(q)$. These oscillations are evident in experimental data of a $N=10$ system, for which we have high temporal resolution (Fig.~\ref{fig:wigner}b). The frequency of the variance oscillations increases with $q$, and their Fourier transform allows us to experimentally reconstruct the iDSF band structure (Fig.~\ref{fig:wigner}c). Full-counting statistics~\cite{wei2022quantum} also show an initially sharp electric field distribution which broadens before partially refocusing. 

\vspace{0.25cm}
\noindent\textbf{Many-body scarring from plasma oscillations} \newline
Our plasma oscillation framework also yields new insights into quantum many-body scars in the PXP model.
In contrast to our initial state $\ket{0}^{\otimes N}$, the N{\'e}el state $|\mathbb{Z}_2\rangle$ is analogous to a squeezed and displaced coherent state at $q=0$. Thus, its Wigner distribution is maximally displaced from the origin and moves along an approximately circular trajectory (Fig.~\ref{fig:wigner}d). Consequently, in contrast to $|0\rangle^{\otimes N}$, there is a clear oscillation of the \emph{mean} electric field, $\langle E(q) \rangle$, a signal typically associated with scar behavior~\cite{bernien2017probing, surace2020lattice}. 
This picture also predicts oscillations of the variance $\text{var}[E(q)]$,  to our knowledge previously unobserved. Note that at $q\neq0$ the distribution is not displaced and behaves like a squeezed vacuum state, with oscillations only in the variance, similar to $|0\rangle^{\otimes N}$ (Fig.~\ref{fig:wigner}d). 

The iDSF and these observations lead to an intriguing hypothesis: There is a more general class of $q$-dependent many-body scarred initial states, corresponding to displaced Wigner functions for different momentum $q$. The N{\'e}el state corresponds to the special case $q=0$ ($k=\pi$ in the original PXP language). While the N{\'e}el state has a maximum and fully homogeneous electric field, the $q$-dependent scar-states are characterized by a spatially modulated electric field, but retain high overlap with product states in the computational basis. Indeed, numerical simulations reveal long-lived oscillations of $\langle E(q\neq0) \rangle$ for such computational basis states at various momenta $q$ (Ext.~Dat.~Fig.~\ref{fig:ED_Wigner}).
These states define families of many-body scars, wholly distinct from $\mathbb{Z}_2$ scars, with energy spacings that agree with the iDSF band structure $\omega_0(q)$ (Ext.~Dat.~Fig.~\ref{fig:scars}). Similar families had been numerically observed in Refs.~\cite{surace2020lattice,szoldra2022unsupervised,ljubotina2023superdiffusive}, but have not been analyzed in-depth.

Further results from conventional quantum optics carry over: we show that for a given value of $q$, scar behavior also appears for states with a weak electric field that are hence only weakly displaced from the Wigner function origin (Ext.~Dat.~Fig.~\ref{fig:ED_Wigner}), in contrast to the $\mathbb{Z}_2$ state which has a maximum electric field and displacement. We find a family of oscillatory initial states, organized within a Bloch sphere, whose energy and field displacements can be continually tuned, consistent with the approximate SU(2) symmetry in the system~\cite{iadecola2019quantum,choi2019emergent,iadecola2019quantum,maskara2021discrete,bluvstein2021controlling,ljubotina2023superdiffusive,kerschbaumer2024quantum}.

Our approach based on the Wigner function provides a new theoretical framework for the PXP model. Rotations of the Wigner function occur at the purely operator level and serve as an organizing principle for known phenomena. For example, the exact many-body scar states identified in Ref.~\cite{ivanov2025exact} have Wigner distributions similar to Fock states (Ext.~Dat.~Fig.~\ref{fig:ED_Wigner}) which may be an intuitive, physically-motivated picture for such exceptional states. This may also provide insights to other recent observations such as of ballistic, soliton-like objects~\cite{giudici2024unraveling,kerschbaumer2025discrete} or anomalous energy transport~\cite{ljubotina2023superdiffusive,bhakuni2025anomalously} in the PXP model. In Ext.~Dat.~Fig.~\ref{fig:energy_transport}, we observe signatures of a ballistic-to-diffusive crossover in energy transport in the PXP model which may explain observations in Ref.~\cite{ljubotina2023superdiffusive}.

\begin{figure}[t!]
	\includegraphics[width=0.48\textwidth]{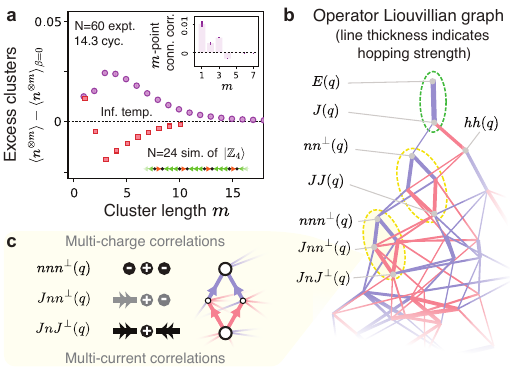}
	\caption{\textbf{Long-lived correlations and charge clusters.} 
		\textbf{a.} The long-lived correlations correspond to a long memory of charge clusters: experimental data (purple circles) reveal clustering in excess of the thermal value (dashed line). This is a memory of initial conditions and not simply a tendency to cluster: there is \textit{negative} excess clustering in the dynamics of the  $|\mathbb{Z}_4\rangle$ initial state, which has no clusters of three charges (red squares). Inset: Connected $m$-point correlators of matter density (Methods) reveal that this is primarily a three-body effect, with comparatively small $m$-point correlators for $m>3$.
		\textbf{b.} The memory mechanism is revealed by the \textit{Liouvillian graph} which represents operator dynamics as a hopping process on a graph of local operators: red and blue colors indicate positive and negative hopping, while thickness represents hopping strength.
		Graph vertices denote operators such as field, charge, current, or energy ($h$) densities, with `$\perp$' indicating operators orthogonalized to remove their overlap with smaller ones 
		(Methods).
		There are clusters of operators that are strongly coupled.  Electric field and current (green oval) are strongly coupled, giving rise to plasma oscillations,while larger diamond-shaped clusters (yellow ovals) give rise to the long-lived correlations.
		\textbf{c.} The memory of charge clustering arises from destructive interference between the multi-charge and multi-current operators, allowing for ``dark" operator superpositions that retain memory of a state's initial conditions.
	}
	\label{fig:clustering} 
\end{figure}

\vspace{0.25cm}
\noindent\textbf{Long-lived correlations: charge clustering}\newline
We now analyze the long-lived correlations observed in Fig.~\ref{fig:experimental_data}e. 
Even at late times, experimental correlations deviate from their expected infinte-temperature values for distances $d=3$ to $d=5$. Numerical simulations indicate that this deviation slowly disappears at very late times for large system sizes (Ext.~Dat.~Fig.~\ref{fig:ED_clustering}).

In both real and momentum space, slow thermalization is visible at the level of two-body observables. 
In momentum space, field variance oscillations persist to long times, with characteristic peaks in momentum that agree with a toy model of fully-packed electric field strings (Ext.~Dat.~Fig.~\ref{fig:ED_clustering}). 

However, a real-space analysis is more revealing: we find a tendency for charged particles to remain in small clusters. In the experimental data, we see larger-than-thermal probabilities to observe clusters of $m$ continuous charges, i.e. of $\langle n_{j,j+1} \dots n_{j+m-1,j+m}\rangle$, peaked at cluster lengths of $m=3$ (Fig.~\ref{fig:clustering}a), where $n_{j,j+1}$ is the matter density. 
Connected $m$-point correlations~\cite{chalopin2024probing} reveal that this clustering is largely limited to genuine three-point correlations.
More accurately, this clustering is a \textit{memory} of the initial state --- the excess or lack of charge clusters in the initial state persists for long times. For example, the anti-clustered $|\mathbb{Z}_4\rangle = |10001000\dots\rangle$ initial state instead shows smaller-than-thermal clustering at late times (Fig.~\ref{fig:clustering}a).

The clustering memory is intrinsically quantum-mechanical, arising from destructive interference in operator dynamics. We understand this mechanism through the \textit{Liouvillian graph}, which formulates the dynamics of many-body operators as a single-particle hopping problem (Methods)~\cite{white2023effective,yithomas2024comparing,cao2021statistical,parker2019universal}.
In a typical many-body system, this graph is densely interconnected without clear features (Ext.~Dat.~Fig.~\ref{fig:MFIM_TFIM_XXZ_structure_factors}). In our system, however, the constrained Hilbert space limits the number of small operators, leading to clear substructures in the Liouvillian graph (Fig.~\ref{fig:clustering}b). 

For example, the field operator $E(q)$ is strongly coupled to the current operator $J(q)$, which is only weakly connected to the rest of the graph that acts as a dissipative ``bath"~\cite{white2023effective}. This quantitatively justifies our mean-field theory for the plasma band structure (Methods). Meanwhile, there are diamond-shaped sub-structures which represent multi-step hybridization pathways for two- and three-charge clusters (Fig.~\ref{fig:clustering}c). 
These steps can destructively interfere, resulting in a ``dark" superposition of operators which experiences slow dynamics, leading to a long memory of initial conditions (Ext.~Dat.~Fig.~\ref{fig:ED_clustering}). Superpositions that remain `dark' to dynamics appear in other contexts~\cite{medenjak2020isolated,buvca2022algebraic}, see e.g.~recent work on `Fock space cages,'~\cite{jonay2025localized}. However, to our knowledge, this interference in operator space is a novel mechanism for slow thermalization.

These features appear to be absent in the $\mathcal{J}\rightarrow 0$ continuum limit (Ext.~Dat.~Fig.~\ref{fig:ED_clustering}), and may be due to the electric field truncation in the quantum link model. The truncation means that in the link model, two $+$ charges cannot be two sites apart without a third $-$ charge between them, i.e.~only the $+,-,+$ charge configuration is allowed, which may give rise to unique three-body effects. Such observations shed light on the effects of field truncation in lattice gauge theories~\cite{zache2022toward,ciavarella2025efficient}. 

\vspace{0.25cm}
\noindent\textbf{Outlook}\newline
This work stands at the intersection of several areas of research: in a neutral atom experiment, we perform a quantum simulation of a lattice gauge theory. We discover phenomena that defy the expectations of many-body thermalization and hydrodynamic theory. Using tools from quantum optics and quantum information, we find simple, semi-classical degrees of freedom which describe our observations as well as previous observations in the literature.

Our theoretical framework opens several avenues of investigation. Within this model, the effects of particle mass (controlled in experiment with staggered optical detuning), dimensionality~\cite{gonzalez2025observation}, and generalizations to other gauge theories~\cite{celi2020emerging,desaules2023weak,desaules2023prominent,shah2025quantum} may be interesting avenues of further study.
More generally, the gauge-theoretic perspective proved to be a fruitful source of intuition: while quantitative calculations are most convenient in the spin language, the gauge theory provided a clear physical picture of the nature of the infinite-temperature quasiparticle, which inspired the tools we introduce, including the Wigner function. Indeed, this may point to deeper connections: our plasma dynamics arises from a collective mode which is robust up to high temperatures. Collective modes have deep field-theoretic origins, ranging from spontaneous symmetry breaking, quantum anomalies, to long-range forces in gauge theories. Meanwhile, the Liouvillian graph provides a lattice-level description of the plasma oscillations, suggesting the possibility of interfacing between field-theoretic and lattice descriptions of many-body dynamics. 

Such connections were enabled by an unexpected experimental observation, and we expect experiments of the dynamics of large-scale, highly controlled quantum systems~\cite{darbha2025emergent} to yield further connections across branches of physics.

\vspace{0.25cm}
\noindent\textbf{Acknowledgements}\newline
We thank the following for insightful discussions: 
Gil Refael, Olexei I.~Motrunich, Andrew N.~Ivanov, Dmitry Abanin, Sarang Gopalakrishnan, David Huse,
Stuart Yi-Thomas, Christopher White, Julian Bender, Maksym Serbyn, Vedika Khemani, Anthony N. Ciavarella, Christian W. Bauer,
Zohreh Davoudi, Zhengyan (Darius) Shi, Wen Wei Ho and Zhuo Chen.
We thank Joonhee Choi, Ran Finkelstein and Pascal Scholl for experimental assistance in previous work.
We acknowledge support by the NSF QLCI Award OMA-2016245, the DOE (DE-SC0021951), the Institute for Quantum Information and Matter, an NSF Physics Frontiers Center (NSF Grants PHY-1733907, PHY-2317110),
the Center for Ultracold Atoms, an NSF Physics Frontiers Center (NSF Grants PHY-1734011 and PHY-2317134), Army Research Office MURI program (W911NF2010136), NSF CAREER award 2237244. Support is also acknowledged from the U.S. Department of Energy, Office of Science, National Quantum Information Science Research Centers, Quantum Systems Accelerator. WG is supported by the Hertz Foundation Fellowship.  
TS acknowledges support from the Walter Burke Institute for Theoretical Physics, Caltech. ALS is supported by the Stanford Science Fellowship, and additionally by the Felix Bloch Fellowship and the Urbanek-Chodorow Fellowship.

\vspace{0.25cm}
\noindent\textbf{Author Contributions}\newline
DKM, FMS, ME, and TS developed the theory.
DKM and ALS performed the data analysis. DKM, FMS, TS and WG performed numerical studies. ALS performed the experiments. DKM, SC and ME wrote the
manuscript with contributions and input from all authors. SC and ME supervised this project.

\vspace{0.25cm}
\noindent\textbf{Data and code availability}\newline
The data and code supporting this study are available from the corresponding authors upon reasonable request.

\newpage
\stoptoc

\section*{Methods}
\noindent\textbf{Details of experiment}
\noindent The experiment consists of a Rydberg atom array quantum simulator
in which strontium-88 atoms are trapped with optical tweezers. The trapped atoms are rearrangeable into defect-free arrays in one dimension~\cite{madjarov2020high,shaw2024benchmarking}. 

The data studied in this work were originally used in two other works: the large $N=60$ experimental data in Ref.~\cite{shaw2024benchmarking}, and the small $N=10$ data in Ref.~\cite{shaw2025experimental}. In both cases, atoms are initially in the $5s^2~^{1}S_0$ state, are cooled on the narrow-line $5s^2~{}^1S_0\leftrightarrow 5s5p~{}^3P_1$ transition close to their motional ground state, and then prepared into the long-lived $5s5p~{}^3P_0$ clock state which we take to be the metastable ground state $|0\rangle$.
The quantum simulation of the PXP model consists of driving the atoms globally to the Rydberg state $|1\rangle$, the $5s61s~{}^3S_1,m_J=0$, while the traps are briefly blinked off. Following Hamiltonian evolution, measurements are performed by autoionizing the atoms in the Rydberg state~\cite{madjarov2020high}. See Ref.~\cite{shaw2024benchmarking} for more details.

Each experimental run results in a single bitstring measurement.  We perform the same data processing scheme as Ref.~\cite{shaw2024benchmarking,shaw2025experimental}: we discard any experimental bitstrings for which initial rearrangement failed, and we post-select on bitstrings which satisfy the Rydberg blockade constraint. The post-selection rate monotonically decreases with system size and evolution time, and for $N=60$ at the final time point, $45\%$ of such bitstrings satisfy the blockade constraint~\cite{shaw2024benchmarking}.

The Hamiltonian of our system is described by the long-range Ising model
\begin{equation}
	H/h =\Omega\sum_jS^x_j-\Delta\sum_jn_j+\frac{C_6}{a^6}\sum_{i>j}\frac{n_in_j}{|i-j|^6}
	\label{eq:rydberg_Hamiltonian}
\end{equation}
describing interacting two-level systems driven by a laser with Rabi frequency $\Omega$ and detuning $\Delta$, with long-range interactions determined by the $C_6$ coefficient and lattice spacing $a$. $S^x_j=(|1\rangle \langle 0|_j + |0\rangle\langle 1|_j)/2$ is the spin matrix and $n_j=|1\rangle\langle1|_j$ is the Rybderg occupation number. The Hamiltonian parameters are: $\Omega=6.9\text{ MHz},\Delta=0.5\text{ MHz}$, $C_6=254\text{ GHz}\times\mu\text{m}^6$, lattice spacing $a=3.77\mu\text{m}$, see Ref.~\cite{shaw2024benchmarking} for details. This Hamiltonian evolution was benchmarked in Ref.~\cite{shaw2024benchmarking} to be of very high fidelity: with a global quantum fidelity of 0.095(11) at $N=60$ at the final time point of 14.3 cycles, corresponding to a per-atom evolution fidelity of 0.9973 (Ext.~Dat.~Fig.~3 of Ref.~\cite{shaw2024benchmarking}). 

Our time unit of a ``cycle" refers to the Rabi cycle of an $1~\text{cycle}=2\pi/\Omega$ ($1/\Omega$ when $\Omega$ has units of $\text{Hz}$). For the $N=60$ dataset, an average of 2187 data points were taken (before post-selction) per time point; at two time points --- 5.9 and 14.3 cycles (the final time point) --- 7305 and 8264 measurements were taken respectively. Data was taken at time intervals of 0.75 cycles up to 14.3 cycles. In this work, we also utilize $N=10$ quench evolution data with fine time resolution, taken at intervals of 0.08 cycles, up to 3.3 cycles, with an average of 2391 data points at each time point. In this dataset, we have parameters $\Omega = 5.3 \text{ MHz}$, $\Delta= 0.6 \text{ MHz}$, $C_6=254\text{ GHz}\times\mu\text{m}^6$, and $a = 3.77 \mu\text{m}$.
Note that the standard $PXP$ model $H=\sum_jP_{j-1}X_jP_{j+1}$ (where $P_j\equiv|0\rangle\langle0|_j$ and $X_j = |0\rangle\langle1|_j+\text{h.c.}$) is obtained by setting the parameter $\Omega=2$, which has one cycle of duration $\pi$.

\noindent\textbf{Analysis of experimental data} 

\noindent\textit{Estimating correlation distance ---}
In Ext.~Dat.~Fig.~\ref{fig:data_processing_correlators}, we extract effective distances and amplitudes from the experimental correlations $C^{EE}(d,t)\equiv\langle E_jE_{j+d}(t)\rangle_c=(-1)^d\langle Z_jZ_{j+d}(t)\rangle_c$. This distance grows linearly, while the amplitude decays exponentially over time. We first approximately isolate the time-dependent, ballistically propagating correlations from the static ``background" $C^{EE}(d,t\rightarrow\infty)$ (Ext.~Dat.~Fig.~\ref{fig:data_processing_correlators}). We estimate the background by averaging the experimental correlators over the last five points in time.

The distance and amplitude are related to the squared correlators $\Delta C^{EE}(d,t)^2\equiv|C^{EE}(d,t)-C^{EE}(d,\infty)|^2$. The amplitude of the ballistic part is the sum $\sum_d\Delta C^{EE}(d,t)^2$. To find the distance, we normalize the squared correlator $p_{EE}(d)\equiv\Delta C^{EE}(d,t)^2/\sum_d\Delta C^{EE}(d,t)^2$ and treat it as a probability distribution. The average distance is $\bar{d}(t)\equiv\sum_dp_{EE}(d)d$.
$\bar{d}(t)$ grows linearly in time (Fig.~\ref{fig:experimental_data}d), in agreement with numerical simulations, as well as a simple theoretical estimate $\bar{d}(t)\approx2\bar{v}t$, where $\bar{v}\equiv(\pi)^{-1}\int_0^\pi\partial_q\omega_0(q)dq$ is the average group velocity of the iDSF. The root-mean-square distance $\sqrt{\sum_dp_{EE}(d)d^2}$, as used in Ref.~\cite{gopalakrishnan2023anomalous} (not plotted), agrees quantitatively with $\bar{d}(t)$.

The normalization also decays exponentially in time, agreeing with the simple theoretical prediction $\sum_d\Delta C^{EE}(d,t)^2\propto\exp(-2\bar{\gamma}t)$, where $\bar{\gamma}$ is the average decay constant estimated from linewidths $\gamma(q)$ of Lorentzian fits of the iDSF (Ext.~Dat.~Fig.~\ref{fig:data_processing_correlators}b). 

\noindent\textit{Fourier-transformed correlators for band structure---} In Fig.~\ref{fig:wigner}, we experimentally observe the plasmon band structure from the Fourier transform of the two point correlators:
\begin{align}
	\mathcal{F}_{d,t}[C^{EE}(d,t)]=\mathcal{F}_t[\langle E(q)E(-q)\rangle(t)-|\langle E(q)\rangle(t)|^2]\nonumber
\end{align}
In practice, we directly evaluate $\langle E(q)E(-q)\rangle$ on the data, and then take its temporal Fourier transform. In Fig.~\ref{fig:wigner}c, we normalize each column as $\hat{C}^{EE}(q,\omega)/\int d\omega \hat{C}^{EE}(q,\omega)$, since the weights $\int d\omega \hat{C}^{EE}(q,\omega)$ are large for $q\sim \pi$ and would otherwise overwhelm the signal at $q\sim 0$.

\noindent\textit{Multipoint correlations---} In Fig.~\ref{fig:clustering}, we compute multi-point correlation functions~\cite{chalopin2024probing} of matter density. Matter density $n_{j,j+1}$ is equal to $P_jP_{j+1}$, and we compute the $m$-point connected correlators of the variables $\langle n_{j,j+1},n_{j+1,j+2},\dots n_{j+m-1,j+m}\rangle$. These correlators, also known as Ursell functions, are given by~\cite{camia2023monotonicity}
\begin{equation}
	C^{(m)}(o_1,\dots,o_m)=\sum_{\lambda }(-1)^{|\lambda|-1}(\lambda-1)!C^{(m)}_\lambda(o_1,\dots,o_m)
	\nonumber
\end{equation}
where $\lambda$ is a partition of the set $[m]\equiv\{1,\dots m\}$ into $|\lambda|$ subsets and $C^{(m)}_\lambda(o_1,\dots,o_m)$ is the correlation function associated with this partition. It is given by $C^{(m)}_\lambda(o_1,\dots,o_m)=\prod_{j=1}^{|\lambda|}\langle\prod_{i\in S_j}o_{i}\rangle$, where $\lambda=\{S_1,...,S_{|\lambda|}\}$ is a partition of $[m]$ into subsets $S_j$. These partitions can be grouped by equivalence classes under permutation, indexed by the \textit{partitions of the integer $m$}.
The simplest correlators are $C^{(2)}(o_1,o_2)=\langle o_1 o_2\rangle-\langle o_1\rangle\langle o_2\rangle$ (the covariance) and
$C^{(3)}(o_1,o_2,o_3)=\langle o_1o_2o_3\rangle-\langle o_1\rangle\langle o_2o_3\rangle-\langle o_2\rangle\langle o_1o_3\rangle-\langle o_3\rangle\langle o_1o_2\rangle+2\langle o_1\rangle\langle o_2\rangle\langle o_3\rangle$, associated with the partitions $\lambda=(1,2,3),(1)(2,3)$, and $(1),(2),(3)$.

\noindent\textit{Error bars---} Error bars in this work are estimated by bootstrapping. The variance of the estimated quantities from repeated bootstrap resamples provides an estimate of its statistical uncertainty~\cite{efron1994introduction}.

\noindent\textbf{Exact time-evolution numerical simulations}

\noindent We compare the experimental results with exact numerical simulation of the Hamiltonian dynamics of Eq.~\eqref{eq:rydberg_Hamiltonian} in the blockaded subspace, obtained by the Lanczos algorithm. We also verify that the power-law-decaying Ising interactions and detuning terms in Eq.~\eqref{eq:rydberg_Hamiltonian}, as well as periodic boundary conditions (PBC) vs.~open boundary conditions (OBC) are not essential to the phenomena we discover, and we restrict our theoretical analysis to the PXP model under PBC.

\noindent\textbf{Infinite temperature expectation values}

\noindent The expectation value of an observable $O$ at infinite temperature is proportional to its trace: $\langle O\rangle_{\beta=0}\equiv\text{tr}(O~\mathbb{I}/D)$, where $D$ is the Hilbert space dimension. In an unconstrained spin-1/2 system, the trace of any Pauli operator is zero. However, in the PXP model, even the smallest operator $Z_j$ has non-zero trace: $\langle Z_j\rangle_{\beta=0}\overset{N\rightarrow\infty}{=}\frac{\varphi^2-1}{\varphi^2+1}\approx 0.447$. The infinite temperature values of any operator can be exactly computed in the thermodynamic limit [Supplementary Information (SI)~\cite{SI}]. For example, the infinite-temperature thermal values of the connected two-point correlator, plotted in Fig.~\ref{fig:experimental_data}, are $\text{tr}(E_0E_{d})/D-(\text{tr}(E_0)/D)^2=(4/5)\varphi^{-2d}$.

\noindent\textbf{Lattice Schwinger model}

\noindent The lattice Schwinger model describes QED in 1+1D with the Kogut-Susskind staggered fermion formulation~\cite{kogut1983lattice,surace2020lattice,martinez2016real}. It has Hamiltonian
\begin{align}
	H_\text{Schw.}=&-w\sum_{j=1}^{N}\left(c^\dagger_{j-1,j}U_{j}^{}c^{}_{j,j+1}+\text{h.c.}\right) \label{eq:KG_Hamiltonian}\\
	&+m\sum_{j=1}^{N}(-1)^jc^\dagger_{j-1,j}c^{}_{j-1,j}+\mathcal{J}\sum_{j=1}^{N}E_{j}^2~,
	\nonumber
\end{align}
where $c^\dagger_{j,j+1}$ is the fermion creation operator on site $(j,j+1)$, $E_{j}$ is the electric field on the link $j$, and $U_{j}\equiv ie^{i\vartheta_{j}}$, with $\vartheta_{j}$ the vector potential satisfying $[\vartheta_{j},E_{k}]=i\delta_{j,k}$. $w$ and $\mathcal{J}$ are related to the field-theoretic gauge-matter coupling strength $g$ and lattice spacing $a$ by the relations $w=1/(2a)$ and $\mathcal{J}=g^2a/2$~\cite{banuls2013mass}. We denote the matter sites as $(j,j+1)$ and the gauge links as $j$ to emphasize the mapping onto the PXP model with atomic spin variables $Z_j$. 

The alternating signs $(-1)^j$ on the mass term are due to the staggered fermion formulation, and the electric fields and charges satisfy the Gauss' Law gauge constraint  $E_{j+1}-E_{j}=(-1)^jn_{j,j+1}$, with $n_{j,j+1}$ the matter density (Table~\ref{tab:LGT_to_spin})  of the alternating charges.
With OBC, the electric field $E_{j}$ is determined by the charge configuration (and vice versa) after the boundary condition $E_0$ is specified. This background electric field corresponds to the \textit{$\Theta-$term} which tunes between confined and deconfined phases when the mass $m$ is non-zero~\cite{coleman1976more}.

In Fig.~\ref{fig:Lattice_Schwinger}, we simulate the dynamics as well as the iDSF of the lattice Schwinger model by following Ref.~\cite{martinez2016real} which exactly transforms Eq.~\eqref{eq:KG_Hamiltonian} (with OBC) into a spin-1/2 Hamiltonian with long-range interactions.

\noindent \textit{Quantum link model---}  We obtain the quantum link model from Eq.~\eqref{eq:KG_Hamiltonian} by taking the limit $\mathcal{J}\rightarrow\infty$ along with the boundary condition $E_0=-1/2$, equivalent to a $\Theta$-term with $\theta=\pi$~\cite{coleman1976more}, thereby restricting $E_{j}=\pm 1/2$. The quantum link model is directly equivalent to a family of Rydberg blockaded models, with the massless case $m=0$ corresponding to the PXP model (non-zero mass can be introduced with a staggered atom detuning)~\cite{surace2020lattice}. This equivalence is given by identifying electric field with atomic spin as $E_{j}=(-1)^jZ_j$: the allowed states satisfying Gauss' Law and the staggered fermions prescription map onto spin configurations satisfying the Rydberg blockade. The $(-1)^j$ staggering factor shifts momenta by $\pi$ between the spin and LGT pictures, i.e.~$E(q)=Z(k=q+\pi)$. In Table~\ref{tab:LGT_to_spin}, we state corresponding lattice gauge theory (LGT) and spin expressions for several quantities of interest.

\begin{table}
	\renewcommand{\arraystretch}{1.5}
	\centering
	\begin{tabular}{M{0.07\textwidth} M{0.27\textwidth} M{0.11\textwidth}}
		\hline
		Quantity & Quantum link model &PXP model \\
		\hline
		Field & $E_j$ & $(-1)^jZ_j$  \\
		Current & $J_{j}^{}= ic^{\dagger}_{j-1,j} U_{j}^{}c_{j,j+1}^{}+\text{h.c.}$ & $(-1)^jP_{j-1}Y_jP_{j+1}$\\
		Matter 
		density & $n_{j,j+1} =  \begin{cases}
			c^{\dagger}_{j,j+1} c^{}_{j,j+1} \text{ if }j \text{ even}\\
			c^{}_{j,j+1} c^{\dagger}_{j,j+1}\text{ if }j \text{ odd}
		\end{cases}$ & $P_jP_{j+1}$\\
		Energy density & $h_{j,j+1} = c^{\dagger}_{j-1,j}U_{j}^{}c_{j,j+1}^{}+\text{h.c.}$ & $P_{j-1}X_jP_{j+1}$\\
		\hline
	\end{tabular}
	\caption{Spin and lattice gauge theory expressions for several quantities of interest.}
	\label{tab:LGT_to_spin}
\end{table}

\noindent \textbf{Infinite temperature dynamical structure factor}

\noindent The infinite-temperature dynamical structure factor (iDSF)~\cite{baez2020dynamical}, or spectral function, is a central object of study in this work. 

The iDSF is defined as the spacetime Fourier transform of the infinite-temperature autocorrelator: $S^O(q,\omega)\equiv\mathcal{F}[\text{tr}(O^\dagger_j(t)O_0^{}(0))]$. 
The iDSF $S^O(q,\omega)$ can be expressed in terms of the energy eigenstates as:
\begin{align}
	S^O(q,\omega) =  \sum_{\mathcal{E},\mathcal{E}',p} \delta(\omega -\mathcal{E}' + \mathcal{E})|\langle\mathcal{E}',p+q|O_j|\mathcal{E},p\rangle|^2~,\nonumber
\end{align}
where we have explicitly labeled the momenta $p$ of the energy eigenstates $|\mathcal{E},p\rangle$. In numerical simulations on finite systems, we replace the delta function with a Gaussian of small width to obtain a smooth spectral function. The iDSF can be interpreted as the distribution of energy and momentum transferred by the local operator $O_j$: even at infinite temperature, local operators such as $E_j$ transfer relatively well-defined amounts of energy and momentum.

\noindent \textbf{Participation ratio of iDSF}

\noindent Our phenomenon is consistent with the nonintegrability of the model: local operators do not simply map individual eigenstates to other individual (or a small number of) eigenstates, as may be the case for spectrum-generating algebras \cite{moudgalya2022quantum} in integrable systems or for special eigenstates in many-body scarred systems. We demonstrated this in Fig.~\ref{fig:Lattice_Schwinger} with the participation ratio of the iDSF, which grows exponentially with system size. This is defined as $\text{PR}(O,N)\equiv(\sum_{i,j}|\langle \mathcal{E}_i|O|\mathcal{E}_j\rangle|^2)^2/\sum_{i,j}|\langle\mathcal{E}_i|O|\mathcal{E}_j\rangle|^4$. 

We interpret this as the degree of delocalization of the autocorrelator $\text{tr}(O^\dagger(t) O)$ in frequency space. To see this, note the autocorrelator has Fourier components $|\langle\mathcal{E}_i|O|\mathcal{E}_j\rangle|^2$. These coefficients are always positive and by normalizing by the Hilbert-Schmidt norm $\Vert O\Vert_\text{HS}^2$, we can treat them as a probability distribution over pairs of eigenstates $(\mathcal{E}_i,\mathcal{E}_j)$. $\text{PR}(O,N)$ is the participation ratio of this distribution.

In ideal situations such as with Haar-random eigenstates, the PR scales as $D^2$. Even in our system, which features a sharp peak in the structure factor, every eigenstate $|E_i\rangle$ in the bulk of the spectrum is mapped onto exponentially many eigenstates with the appropriate energy difference. Therefore we expect the PR to also scale as $D^2$, possibly with a small prefactor reflecting the sharpness of the iDSF, consistent with the behavior in Fig.~\ref{fig:Lattice_Schwinger}.

\textit{Energy transport --- } The scaling form of the iDSF of a conserved quantity is related to the universality class of its transport. Hydrodynamic theory predicts that the long-wavelength modes $O(k)$ decay as $k^z$, where $z$ is the dynamical exponent. In turn, the iDSF is predicted to follow the scaling relation~\cite{scheie2021detection} $\lim_{k\rightarrow 0}k^z S^O(k,\omega) = f(\omega/k^z)$. In Ext.~Dat.~Fig.~\ref{fig:energy_transport}, we plot the iDSFs of the energy density $PXP$. Through most of the Brillouin zone, the iDSF has a two-peak structure, with peak frequencies scaling linearly with $k$: $S^{PXP}(k,\omega)$ follows the above scaling relation with $z = 1$. However, finite-size simulations do not provide the iDSFs for arbitrarily small $k$. To this end, we adapt the operator size truncation (OST) method from Ref.~\cite{yithomas2024comparing}, which generates the iDSF in the thermodynamic limit from the spectral function of the Liouvililan graph (below), truncated at operators of size 9. Graph edges to larger operators are removed and replaced with a dissipation of fixed strength, here $\gamma = 3$ (the dissipation strength at which the diffusion constant $D$ is stationary, i.e. $d D /d\gamma  =0$~\cite{yithomas2024comparing}). The iDSF method generated by this method shows diffusive scaling ($z=2$), albeit only at very small momenta $k\leq 0.03$.

\noindent \textbf{Mean-field theory}

\noindent We develop a mean-field theory to model plasma oscillations between the electric field and current operators.
Our mean-field theory is a simple approximation of the Heisenberg equations of motion of these operators, which yields predictions that closely match the observed iDSF band structure (Fig.~\ref{fig:Lattice_Schwinger}). The theory is most conveniently stated in spin language: $Z(k)=E(q=k+\pi)$ and $PYP(k)=J(q=k+\pi)$. These operators obey the equations of motion:
\begin{align}
	\frac{d}{dt}Z(k)=&i[H_\text{PXP},Z(k)]=2PYP(k)~,\label{eq:Zk_EOM}\\
	\frac{d}{dt}PYP(k)=&-2 PZP(k)\label{eq:PYP_eom}\\
	&+2\cos(k/2) [P(\sigma^+\sigma^-+\sigma^-\sigma^+)P](k)\nonumber
\end{align}
where $PYP(k)\equiv\sum_je^{ikj}P_{j-1}Y_jP_{j+1}$, $PZP(k)\equiv\sum_je^{ikj}P_{j-1}Z_jP_{j+1}$, and $P(\sigma^+\sigma^-+\sigma^-\sigma^+)P(k)\equiv\sum_je^{ik(j+1/2)}(P_{j-1}^{}\sigma_j^+\sigma_{j+1}^-P_{j+2}^{}+\text{h.c.})$. 

Our mean-field theory is most transparently presented by treating the operators $PZP$ and $PYP$, instead of $Z$ and $PYP$. The two choices are largely equivalent, due to the high overlap between the $PZP$ and $Z$ operators (see the Liouvillian graph formalism below).
The $PZP$ operator obeys the equation of motion,
\begin{equation}
	\frac{d}{dt}PZP(k)=2PYP(k)+e^{-ik}PPYP(k)+e^{ik}PYPP(k)~, \label{eq:PZP_eom}
\end{equation}
where $PPYP(k)\equiv\sum_je^{ikj}P_{j-2}P_{j-1}Y_jP_{j+1}$ and $PYPP(k)\equiv\sum_je^{ikj}P_{j-1}Y_jP_{j+1}P_{j+2}$. 
In its simplest form, our mean-field theory truncates the $P\sigma^+\sigma^-P$ term in Eq.~(\ref{eq:PYP_eom}) and approximates  $PPY\!P(k)\approx\langle P\rangle_\beta'PY\!P(k)$ in Eq.~(\ref{eq:PZP_eom}). Here, $\langle P \rangle'_\beta=1/\varphi$ is the expectation value of $P$ at infinite temperature, conditioned on a neighboring site being in the `0' state, where $\varphi \equiv (1+\sqrt{5})/2\approx1.618$ is the golden ratio. This yields a closed system of two coupled linear differential equations in terms of the momentum-resolved operators, ${PY\!P}(k)$ and $PZP(k)$, with eigenmodes $\omega_0(k)=\pm 2\sqrt{1+\cos k/\varphi}$.

We find that this approximation yields good agreement with the iDSF of the $Z$ operator (as well as the $PY\!P$ and ${PZP}$ operators, see Ext.~Dat.~Fig.~\ref{fig:PXP_model_operator_structure_factors}), and, moreover, can be improved upon systematically. 
The mean-field theory plotted in Fig.~\ref{fig:Lattice_Schwinger} is the prediction $\omega_0(k)=\pm\sqrt{2}\sqrt{1+\varphi+3\cos k/\varphi}$ which is derived by including the $P(\sigma^+\sigma^-+\sigma^-\sigma^+)P(k)$ operator and its equation of motion, and solving the resulting $3\times3$ linear system. This yields an even better approximation of the iDSF.

The Liouvillian graph formalism (below) provides a systematic way of taking beyond-mean-field fluctuations into account. It also quantitatively justifies our mean-field assumptions, as its solution yields extremely similar eigenmodes as above but with small non-zero linewidths. 

\noindent \textit{Modified Maxwell equations ---} The mean-field equations can be interpreted as Maxwell-like equations. Eq.~\eqref{eq:Zk_EOM} relates the rate of change of the field to the current, analogous to the Maxwell equation $\frac{d}{dt}E=-\mu_0j+\nabla\cross B$, with no magnetic field (absent in one-dimension). Meanwhile, in Eq.~\eqref{eq:PYP_eom}, $PZP$ corresponds to the density of particle, anti-particle pairs which can be annihilated to induce a charge current. The $P\sigma^+\sigma^-P+\text{h.c.}$ term represents isolated charges that can be moved by two lattice sites and also contribute to the current: these terms together represent a kind of current polarizability. 
Eq.~\eqref{eq:PZP_eom} refers to the rate of change of particle pairs: the $PPYP$ terms in Eq.~\eqref{eq:PZP_eom} correspond to processes such as $(q^+,\phi,\phi)\leftrightarrow(q^+,q^-,q^+)$, where $q^\pm$ represent charges and $\phi$ represents the matter vacuum. These create/destroy \textit{two} local pairs and hence are additionally weighted.

\noindent \textbf{Many-body Wigner distribution}

\noindent The Wigner distribution is a common tool in quantum optics to visualize quantum states of light~\cite{scully1997quantum}. Motivated by the harmonic-oscillator-like dynamics of the electric field modes $E(q)$, for each momentum $q$, we generalize the Wigner distribution to the many-body quantum setting.
We consider the Hermitian observables $E_\mathbb{R}(q)\equiv[E(q)+E(-q)]/2$ and $J_\mathbb{R}(q)\equiv[J(q)+J(-q)]/2$ as phase space variables. Unfortunately, standard expressions for the Wigner distribution cannot be directly generalized to the many-body setting, in part because $E_\mathbb{R}(q),J_\mathbb{R}(q)$ are not exactly canonically conjugate. 

Instead, we use the relationship between the Wigner distribution and the (inverse) \textit{Wigner-Weyl transform}~\cite{scully1997quantum} to develop the appropriate many-body Wigner distribution. This yields
\begin{equation}
	W(E,J)\equiv\int d\alpha d\beta~e^{-i(\alpha E + \beta J)}\text{tr}(\rho e^{i[\alpha E_\mathbb{R}(q)+\beta J_\mathbb{R}(q)]}),\label{eq:many_body_Wigner}
\end{equation}
which features many of the desired properties of a quasiprobability distribution: it is real-valued, normalized such that $\int dEdJ~W(E,J)=1$, and has the property that the appropriate marginal distributions, $\int dJ~W(E,J)$ and $\int dE~W(E,J)$, coincide with the spectral distributions of $E_\mathbb{R}(q)$ and $J_\mathbb{R}(q)$, respectively. We derive Eq.~\eqref{eq:many_body_Wigner} in the SI~\cite{SI} and provide proofs of the above properties.

\noindent \textit{Numerical method: Fej{\'e}r kernel}--- We numerically evaluate the integral in Eq.~\eqref{eq:many_body_Wigner}, on an evenly-spaced grid of $(\alpha,\beta)$ values. This suffers from the Gibbs' phenomenon~\cite{korner2022fourier}, in which the Fourier series of a function truncated to a certain order $n$, $f_n(x)=\sum_{|k|\leq n}c_k\exp(ikx)$, reproduces the original function $f$, but with oscillations which are present even as $n\rightarrow\infty$ when $f$ is discontinuous.
Such discontinuities are present in our initial states of interest, $|0\rangle^{\otimes N}$ and $|\mathbb{Z}_2\rangle$, which have zero variance in electric field and hence behave like delta functions along one axis. Na{\"i}ve reconstructions of their Wigner distributions from a finite grid have Gibbs oscillations that may be mistaken as signatures of non-classicality (Ext.~Dat.~Fig.~\ref{fig:ED_Wigner}).

We alleviate this issue with the \textit{Fej{\'e}r kernel}, i.e.~the sum of the partial sums $g_n(x)\equiv\frac{1}{n+1}\sum_{j=0}^nf_j(x)$~\cite{korner2022fourier}. Fej{\'e}r's theorem states that, unlike the Fourier sums $f_n$, $g_n$ uniformly converges to $f$, $\text{lim}_{n\rightarrow \infty}g_n=f$. This allows a more faithful representation of the Wigner distribution. 

\noindent \textbf{Liouvillian graph}

\noindent The \textit{Liouvillian graph}~\cite{white2023effective,yithomas2024comparing} describes operator dynamics under the Heisenberg equations of motion, $\frac{d}{dt}O(t)=i[H,O(t)]\equiv\mathcal{L}[O]$, in terms of a hopping model on a graph. The vertices of the graph consist of an orthonormal basis of operators: we denote orthonormalized operators as $O^\wedge$. The edges in the graph (between vertices $O^\wedge_a$ and $O^\wedge_b$) have weights $\text{tr}(O_b^{\wedge\dagger}\mathcal{L}[O^\wedge_a])$.

For spin-1/2 systems, a convenient orthonormal operator basis is the set of Pauli strings. However, Rydberg blockaded systems do not have such a simple basis: the non-factorizable Hilbert space leads to unconventional properties. 
To this end, we construct an orthonormal Rydberg operator basis by (i) constructing a basis of operators from strings of the symbols $\{P,Z,\sigma^+,\sigma^-\}$ that do not violate the Rydberg blockade, then (ii) orthonormalizing these operators using a Gram-Schmidt procedure, with respect to the Hilbert-Schmidt inner product $\langle A,B\rangle\equiv\text{tr}(A^\dagger B)/D$, where the trace is taken over the blockaded subspace (see SI~\cite{SI} for details).

We can illustrate features of the Rydberg blockaded operator space with a few relevant examples. 
First, in contrast to an infinite-temperature operator space, the operator $Z(k)$ is not traceless, $\langle Z(k), \mathbb{I} \rangle = \text{tr}[Z(k)]/D \neq 0$. This traceful component signifies an overlap with the identity operator, which must be subtracted during orthonormalization. Second, over the blockaded subspace, the one- and three-body operators $Z(k)$ and $PZP(k)$ have a large inner product $\alpha(k)$.
This large inner product justifies our use of the $PZP$ operator instead of $Z$ in our mean-field theory. The vertex $nn^\perp(q)$ in Fig.~\ref{fig:clustering} refers to the orthogonal component $ PZP(k)-\alpha(k)Z(k)$. Similarly, $PY\!P(k)$ and $PY\!PP(k)$ have large inner product, which, appropriately normalized, is precisely the mean-field factor $\langle P'\rangle=1/\varphi$.

\noindent \textit{Symmetries--- }The Liouvillian graph of the PXP model has two connected components, distinguished by their quantum numbers under the particle-hole symmetry (PHS) and time-reversal symmetry (TRS) of the PXP model. The Liouvillian $\mathcal{L}[O]$ flips the sign of the PHS and TRS symmetry quantum numbers, $\mathcal{C}_\text{PHS}, \mathcal{C}_\text{TRS}$, carried by an operator $O$.
Hence, the two connected components are labeled by the \textit{product} of the symmetry charges. For example, $\mathcal{C}_\text{PHS}\times\mathcal{C}_\text{TRS}=+1$ for the $Z$ and $PY\!P$ operators, while $\mathcal{C}_\text{PHS}\times\mathcal{C}_\text{TRS}=-1$ for the $PXP$ operator; hence, they belong to different connected components (Ext.~Dat.~Fig.~\ref{fig:energy_transport}), each of which are bipartite graphs (SI~\cite{SI}).

In translation-invariant systems, the Liouvillian also preserves the momentum of an operator. This further decomposes each connected component into sectors containing operators of each fixed $q$. Each such graph has the same connectivity, but with edge weights that depend on $q$, which lead to the observed iDSF band structure. In our studies, we generate Liouvillian graphs for relevant $q$ separately allowing us to directly study the thermodynamic limit, for example with the OST method~\cite{parker2019universal,white2023effective,yithomas2024comparing}.

\noindent\textit{Representing states on the Liouvillian graph---} The Liouvillian graph also provides a representation of quantum states, through their expectation values of operators. We simply expand any density matrix $\rho$ in the orthonormal operator basis, $\{O^\wedge_a\}$: $\rho = \sum_a \text{tr}(O^{\wedge\dagger}_a\rho)O^\wedge_a$. Choosing $O^\wedge_a$ to be Hermitian, the coefficients $\text{tr}(O^{\wedge}_a \rho)$ are real and represent the operator expectation values. For a time-dependent state $\rho=|\psi(t)\rangle\langle\psi(t)|$, these expectation values evolve over time precisely according to hopping dynamics on the Liouvillian graph. 

In Ext.~Dat.~Fig.~\ref{fig:ED_clustering}, we plot the steady state of this time evolution, given by the \textit{diagonal ensemble}~\cite{pietracaprina2017total,cfiakan2021approximating} $\rho_d \equiv \mathbb{E}_t[|\psi(t)\rangle\langle \psi(t)|]$, specified by the initial state $|\psi_0\rangle$ and energy eigenstates $|\mathcal{E}\rangle$, $\rho_d = \sum_{\mathcal{E}} |\langle \psi_0|\mathcal{E}\rangle|^2 |\mathcal{E}\rangle\langle\mathcal{E}|$. $\rho_d$ is a stationary, equilibrium state and its expectation values also represent a stationary state on the graph. These show the operator dark states (Ext.~Dat.~Fig.~\ref{fig:ED_clustering}): these states not only decay slowly, but for finite system size, have a component with eigenvalue precisely zero, representing deviations from infinite temperature expectation values that persist for all times. These deviations decrease with system size and hence we expect all expectation values to agree with the infinite temperature thermal value in the thermodynamic limit. 


%

\setcounter{figure}{0}
\renewcommand{\figurename}{Ext. Dat. Fig.}
\renewcommand{\thefigure}{\arabic{figure}}
\renewcommand{\theHfigure}{A\arabic{figure}}

\begin{figure*}
	\centering
	\includegraphics[width=\linewidth]{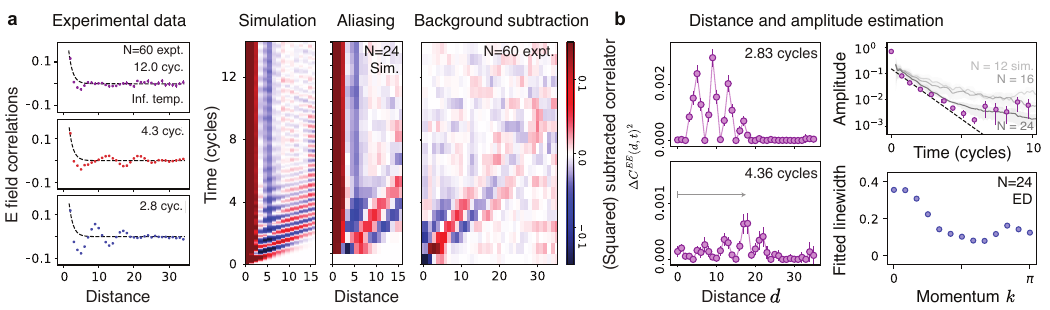}
	\caption{\textbf{Ballistic correlations.} \textbf{a.} Cross sections of experimental  electric field correlators at three points in time: the propagating correlations are visible, along with remaining athermal correlations at long times.
		Numerical simulations corroborate both observations, but also reveal high-frequency oscillations which are not seen in  experimental data due to the interval of time samples. Plotting the numerical simulations only at the experimental times aliases this high-frequency oscillations, in agreement with the experimental data. \textbf{b.} To estimate the effective distance and amplitude of the ballistic propagation, we first subtract a  background $C^{EE}(d,\infty)$ of approximately static correlations at small distances ($d\leq 5$), estimated from the average of the experimental correlators over the last five available time-points (from 11.2---14.3 cycles). This background subtraction removes the long-lived non-thermal correlations, and only the ballistic correlations remain in $\Delta C^{EE}(d,t)$ (Methods). We plot the squared correlations $\Delta C^{EE}(d,t)^2$ at two points in time to illustrate the propagation and decay of the correlations: we use this quantity to define the correlation distance in Fig.~\ref{fig:experimental_data} (Methods). We also see that the experimental amplitudes decay exponentially over time as $\exp(-2\bar{\gamma}t)$ (dashed line), where $\bar{\gamma}\equiv \mathbb{E}_q~\gamma(q)$ is an average decay rate, as well as numerical simulation of small system sizes ($N=16,20,24$, gray filled lines). $\gamma(q)$ is estimated by fitting the numerically-obtained iDSF to a Lorentzian function (bottom).}
	\label{fig:data_processing_correlators}
\end{figure*}

\begin{figure*}
	\centering
	\includegraphics[width=\linewidth]{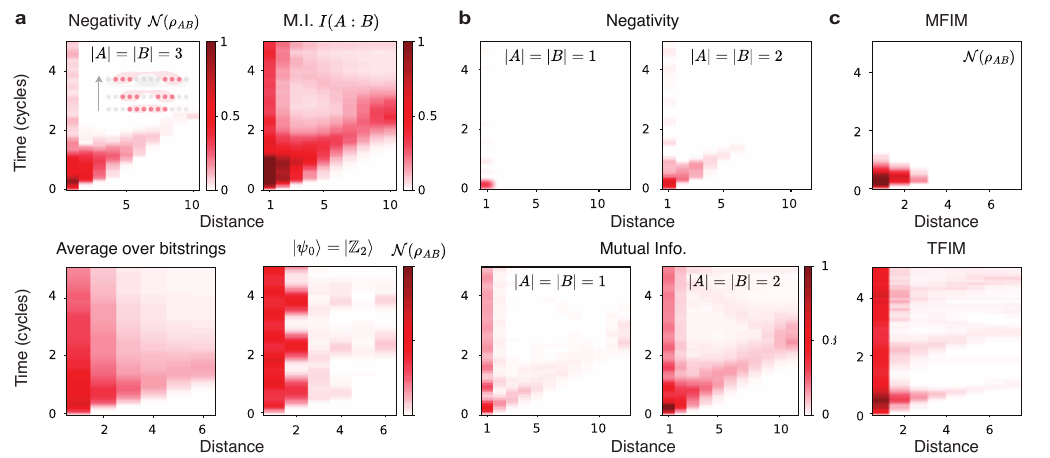}
	\caption{\textbf{Entanglement dynamics.} 
		\textbf{a.} A spacetime plot of the negativity $\mathcal{N}(\rho_{AB})$ for the reduced state $\rho_{AB}$ associated with disjoint subsystems $A$ and $B$ (details in SI~\cite{SI}) reveals the ballistic propagation of entanglement, schematically illustrated in the inset. $A$ and $B$ are contiguous three-site subsystems separated by $d_{AB}$ sites. At a given time, the negativity takes substantial positive values only at a certain $d_{AB}$ which grows linearly with time (in addition to local entanglement that remains at $d_{AB} = 1$). 
		A spacetime plot of the quantum mutual information $I(A:B)$ of $\rho_{AB}$ also shows ballistic spreading of correlations. Unlike the negativity, these correlations may be classical in nature and are more pronounced.
		Bottom row: Averaging the negativity over all bitstring initial states reveals that entanglement transport is a generic feature, independent of initial state. In contrast, the $|\mathbb{Z}_2\rangle$ initial state shows limited entanglement transport. All numerical simulations are for $N=24$ systems with periodic boundary conditions.
		\textbf{b.} Plots of the negativity and mutual information for subsystems of size $|A|=|B|=1,2$ show similar but weaker behavior.
		\textbf{c.} The MFIM (top) does not show such entanglement growth, while the TFIM (bottom) shows repeated entanglement propagation, possibly due to its underlying free-fermion description. Numerical simulations for  $N=20$ systems with parameters described in the caption of Ext.~Dat.~Fig.~\ref{fig:MFIM_TFIM_XXZ_structure_factors}, and with $|A|=|B|=3$.
	}
	\label{fig:entanglement_dynamics_ED}
\end{figure*}

\begin{figure*}
	\centering
	\includegraphics[width=\linewidth]{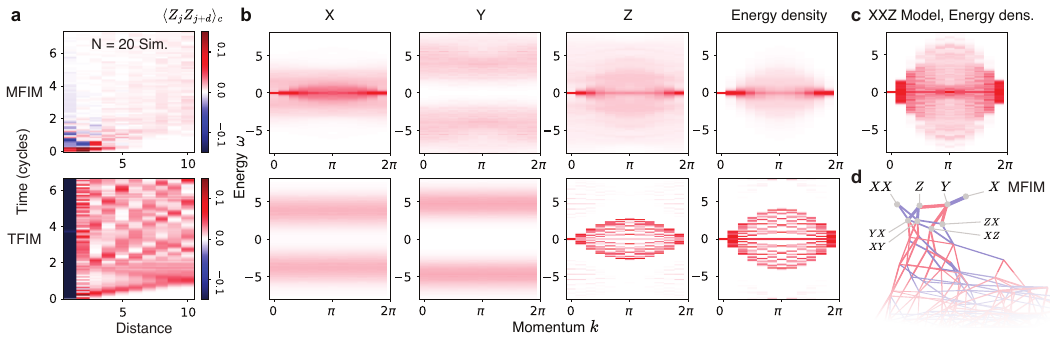}
	\caption{\textbf{Correlators and iDSFs in other models.}  \textbf{a.} The connected $Z$ correlations of the the mixed field Ising model (MFIM) illustrate the expected dynamics of correlations in a nonintegrable system: correlations propagate diffusively and quickly decay, since $Z$ is not a conserved quantity. Meanwhile, the correlations in the transverse field Ising model (TFIM) show more features, with ballistic propagation over all time due to its integrability. The models have Hamiltonians given by $H =\sum_j h_j$, where $h_j \equiv h_x X_j + h_z Z_j + J Z_j Z_{j+1}$ with parameters $(h_x,h_z,J) = (0.8090, 0.9045, 1)$ for the MFIM~\cite{kim2013ballistic,kim2014testing} and  $(0, 2, 1)$ for the TFIM and we consider the $|0\rangle^{\otimes N}$ initial state.
		\textbf{b.}
		iDSFs of the $X,Y,Z$ and energy density operators in the MFIM and TFIM. The MFIM is non-integrable and hence its structure factors are featureless (however, see the SI~\cite{SI} for a fine-tuned regime where the energy density iDSF has non-trivial structure).
		In the TFIM, the $X$ and $Y$ operators are mapped to many-fermion operators under this transformation, and hence show a featureless iDSF, while the $Z$ and energy operators are related to simple fermionic operators (SI~\cite{SI}) and hence have sharp peaks in their iDSF.
		\textbf{c.} We also plot the energy density iDSF of the XXZ model at the isotropic Heisenberg point $\Delta=1$~\cite{bulchandani2021superdiffusion}, which is integrable via Bethe ansatz but has a diffuse energy structure factor.
		All iDSFs obtained for $N=12$ periodic spin chains. \textbf{d.} We plot the Liouvillian graph of the MFIM: even small operators are highly interconnected and show no distinct substructures. Thickness of graph edges are proportional to the Liouvillian matrix element (at generic momentum $k=\pi/6$), with red and blue edges indicating positive and negative elements respectively.}
	\label{fig:MFIM_TFIM_XXZ_structure_factors}
\end{figure*}

\begin{figure*}
	\centering
	\includegraphics[width=\linewidth]{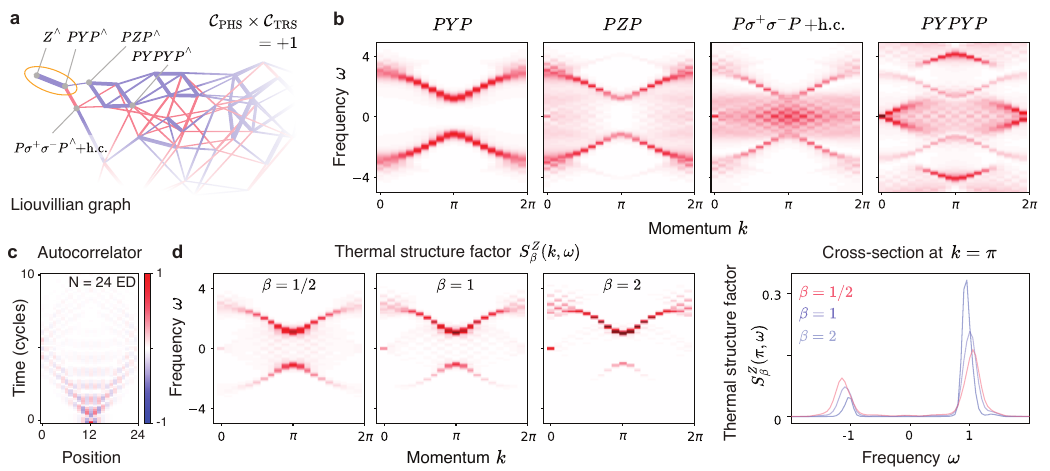}
	\caption{\textbf{iDSFs in the PXP model.} 
		\textbf{a.} The Liouvillian graph (in the $\mathcal{C}_\text{PHS}\times \mathcal{C}_\text{TRS} = 1$ symmetry sector) contains substructures that lead to well-defined band structures in the iDSFs of several operators. Thickness of graph edges are proportional to the Liouvillian matrix element (at generic momentum $k=\pi/6$), with red and blue edges indicating positive and negative elements respectively.
		\textbf{b.} The $PYP(k) = J(k+\pi)$ and $PZP(k)$ operators share the same band structure as the $Z(k) = E(k+\pi)$ operator, since $PYP(k)$ is strongly coupled to $Z(k)$, and $PZP(k)$ has high overlap with $Z(k)$ [the graph vertex labeled by $PZP^{\wedge}(k)$ in \textbf{a}, equal to the matter-matter correlator $nn^\perp(k+\pi)$, is the component of $PZP(k)$ orthogonal to $Z(k)$].
		The $(P\sigma^+\sigma^-P+\text{h.c.})(k)$ operator also shows this band structure, due to its coupling with $PYP(k)$, but additionally contains a broad ``dissipative" band near $\omega=0$ due to its coupling to the rest of the graph. The $PYPYP(k)\equiv JJ(k+\pi)$ operator displays multiple stark features, including a sharp peak at $\omega=0,k=0$, reflecting the slowly decaying athermal correlations in Fig.~\ref{fig:clustering}, along with new band structures near $\omega=0,k=0$ and $\omega=4,k=\pi$.
		\textbf{c.} The autocorrelator $\text{tr}[Z_{j}(0)Z_{N/2}(t)]/D$ is the Fourier transform of the iDSF $S^Z(k,\omega)$. It captures the fact that a $Z$ operator, initially localized at $j=N/2$ (for a $N=24$ periodic chain) has an anomalously high weight remaining on local $Z$ operators (at different positions $j$ over time).
		\textbf{d.} Left: The \textit{finite temperature} DSF (SI~\cite{SI}), $S^Z_\beta(k,\omega)$, equal to the Fourier transform of $\text{tr}[\rho_\beta O^\dagger_j(t)O_0(0)]$, for the Boltzmann states $\rho_\beta \propto \exp(-\beta H)$ at inverse temperatures $\beta = 1/2,1,2$.
		The band structure remains stable as the temperature is lowered, despite the increasing asymmetry between the upper and lower bands due to their imbalance in thermal population. Right: Cross-section of the finite-temperature DSFs at $k=\pi$. The peak frequency changes slightly and the linewidth narrows as the temperature decreases.
	}
	\label{fig:PXP_model_operator_structure_factors}
\end{figure*}

\begin{figure*}
	\centering
	\includegraphics[width=\linewidth]{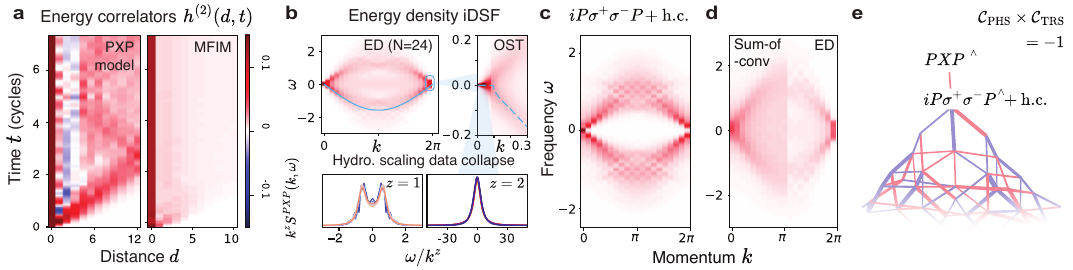}
	\caption{\textbf{Energy transport in the PXP model.} 
		Long-lived plasma oscillations lead to unusual energy transport in the PXP model.
		\textbf{a.} The equal-time energy correlation function, $h^{(2)}(d,t) \equiv \langle \psi(t)|(PX_jP)(PX_{j+d}P)|\psi(t)\rangle$, shows quasi-ballistic energy propagation (exact simulation; $N=24$, periodic boundary conditions, initial state $|0\rangle^{\otimes N}$), which is absent in the MFIM (right, exact simulation with $N=20$. We use an initial state that is at infinite temperature but has maximal energy variance~\cite{maceira2024thermalization}, see SI~\cite{SI}).  
		\textbf{b.} The iDSF of the energy operator, $PX\!P(k)$, displays a clear band structure. Top left: data from exact diagonalization (ED) at $N=24$, mean-field theory (SI) in blue, which has a limited momentum resolution. Top right: operator-size truncation (OST, Methods)~\cite{yithomas2024comparing} approximation for smaller values of $k$ near 0 (or $2\pi$), showing a single-to-double-peak ``hot band second sound"~\cite{yithomas2024comparing} crossover (peak frequencies $\omega_0(k)$ marked in blue).
		Bottom: The iDSF displays a ballistic scaling at larger $k$, manifested by the linearly growing ($z=1$) peak frequencies $\omega_0(k)$ (left, $ \pi/6 \leq k \leq 2\pi/3$) but shows a diffusive scaling collapse for smaller $k$ (right, $k\leq 0.03$, data obtained from OST). This suggests a ballistic-to-diffusive crossover in energy transport which may account for observations in Ref.~\cite{ljubotina2023superdiffusive}. 
		\textbf{c.} The energy current operator $(iP\sigma^+\sigma^-P+\text{h.c.})(k)$ is coupled to the $PXP(k)$ operator and hence shares a similar band structure. 
		\textbf{d.} The energy iDSF $S^{PXP}$ is approximately related to the iDSFs $S^{PYP}$ and $S^Z$ by convolution (SI). This relationship qualitatively captures many features of $S^{PXP}$.
		\textbf{e.} The connected component of the Liouvillian graph with symmetry charges $\mathcal{C}_\text{PHS}\times \mathcal{C}_\text{TRS} = -1$, which includes the energy density and energy current operators at the head of the graph. Thickness of graph edges are proportional to the Liouvillian matrix element (at generic momentum $k=\pi/6$), with red and blue edges indicating positive and negative elements respectively.
	}
	\label{fig:energy_transport}
\end{figure*}

\begin{figure*}
	\centering
	\includegraphics[width=\linewidth]{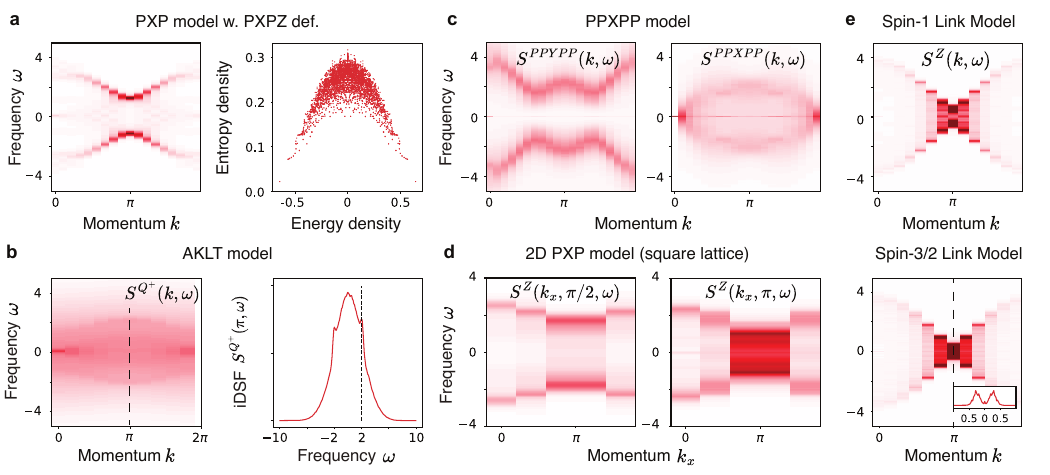}
	\caption{\textbf{iDSFs of constrained and many-body scarred systems.} \textbf{a.} We consider the PXP model perturbed by a $PXPZ$ term (with coefficient $0.071$), noted in Ref.~\cite{khemani2019signatures} to reduce many-body scarring. The $Z$ operator shows a band structure in its iDSF without many-body scarring, as seen in the lack of eigenstates with exceptionally low bipartite entanglement entropy. Simulations for $N=18$ shown. \textbf{b.} We also consider the spin-1 AKLT model~\cite{moudgalya2022quantum} (numerical results for a periodic $N=10$ system), which hosts a tower of exact many-body scarred states generated by the operator $Q^+$. This operator does not show sharp peaks in its iDSF. A cross-section at $k=\pi$ shows a broad diffuse background with small peaks at the scar frequencies $\omega = \pm 2$ (the scar tower energy). \textbf{c.} We plot the iDSFs of the $PPYPP$ and $PPXPP$ operators in the PPXPP model~\cite{giudici2019diagnosing}, obtained from exact diagonalization at $N=20$. Sharp quasiparticle bands are clearly visible. 
		\textbf{d.} We also consider the $Z$ operator iDSF in the 2D PXP model~\cite{bluvstein2021controlling,maskara2021discrete,lin2020quantum,samajdar2020complex}, obtained from ED on a $5\times 4$ periodic square lattice. We plot cross-sections of the structure factor along slices of $k_y = \frac{\pi}{2}$ and $\pi$. A two-dimensional band structure is visible. 
		\textbf{e.} Finally, we plot the $Z$ structure factor in spin-1 and spin-3/2 quantum link models~\cite{zache2022toward}, which generalize the quantum link model. An operator band structure is visible in both models (simulations on $N=14$ and $N=12$ periodic chains, respectively).}
	\label{fig:not_scars}
\end{figure*}

\begin{figure*}
	\centering
	\includegraphics[width=\linewidth]{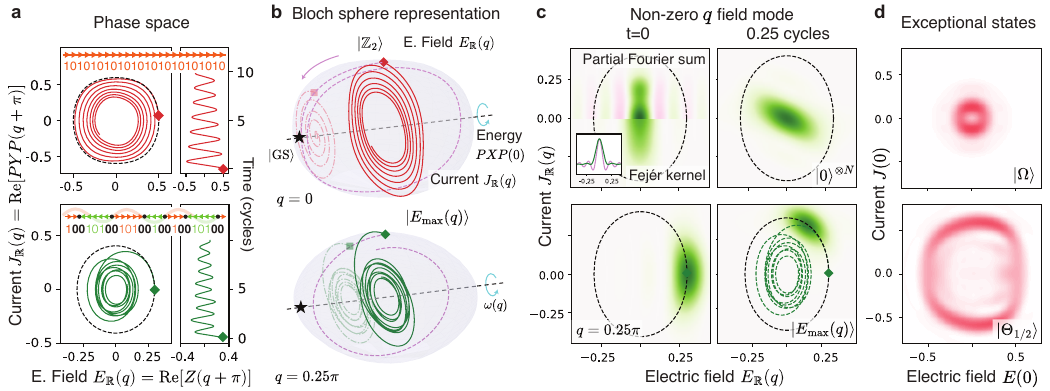}
	\caption{\textbf{Phase space, Bloch spheres and Wigner distributions for plasma oscillations.} 
		\textbf{a.} Plasma oscillations can be represented on a phase space of field $E(q)$ and current $J(q)$, exemplified by the dynamics of the $|\mathbb{Z}_2\rangle$ state for the uniform field mode $q=0$.
		The phase space trajectory (solid line) lies close to the boundary of possible expectation values $\langle E(q)\rangle,\langle J(q)\rangle$ (dashed, SI).
		Oscillations of non-zero momentum field modes also occur. We illustrate this with the maximally field-polarized initial state of $E(q)$ for $q=0.25\pi$, which is a bitstring product state consisting of domains of alternating electric field (inset). Right: time-traces of electric field $\langle E(q)\rangle$, which shows momentum-dependent oscillation frequencies $\omega_0(q)$.
		\textbf{b.} The phase space of the electric field $E(q)$ and current $J(q)$ can be extended to a Bloch sphere which represents the approximate $SU(2)$ relations between $E(q), J(q)$ and the Hamiltonian $H \equiv PXP(k=0)$~ \cite{choi2019emergent,iadecola2019quantum, maskara2021discrete,kerschbaumer2024quantum}. Different positions on the Bloch sphere, including along the energy axis, can be systematically explored, e.g.~by evolving under the current operator $J_\mathbb{R}(q)\equiv [J(q)+J(-q)]/2$ to tune between the maximally field-polarized states and the ground state. We see scar-like, persistent oscillations for all energies. These Bloch spheres are ellipsoidal and are defined as the boundary of allowed expectation values $(\langle E_\mathbb{R}(q) \rangle, \langle J_\mathbb{R}(q)\rangle, \langle H\rangle)$ (SI). 
		\textbf{c.} We plot the Wigner distributions in higher field modes $E(q), J(q)$ for $q = 0.25 \pi$, for the $|0\rangle^{\otimes N}$ state and for $|E_\text{max}(q)\rangle$, the maximal eigenstate of $E_\mathbb{R}(q)$. As in Fig.~\ref{fig:wigner}, these behave like squeezed and coherent states respectively. In the top left panel, we also illustrate the effect of the Fej{\'e}r kernel over the conventional partial Fourier sum. The Fourier sum suffers from the Gibbs phenomenon for discontinuous distributions, such as for $|0\rangle^{\otimes N}$, which might otherwise be mistaken as signs of nonclassicality (top half of panel). The Fej{\'e}r kernel (Methods) mitigates this effect (bottom half of panel). The dashed ovals indicate the maximum allowed values of $\langle E_\mathbb{R}(q)\rangle, \langle J_\mathbb{R}(q)\rangle$.
		\textbf{d.} The Wigner distributions of the exact scar states $|\Omega\rangle$ and $|\Theta_{1/2}\rangle$ from Ref.~\cite{ivanov2025exact} have energy $E=0$  and are analogous to Fock states in quantum optics, with rotationally symmetric Wigner distributions which are hence stationary under time evolution. Numerical simulations in \textbf{a,b} conducted in $N=24$ periodic systems, while simulations in \textbf{c,d} conducted in $N=16$ periodic systems.}
	\label{fig:ED_Wigner}
\end{figure*}

\begin{figure*}
	\centering
	\includegraphics[width=\linewidth]{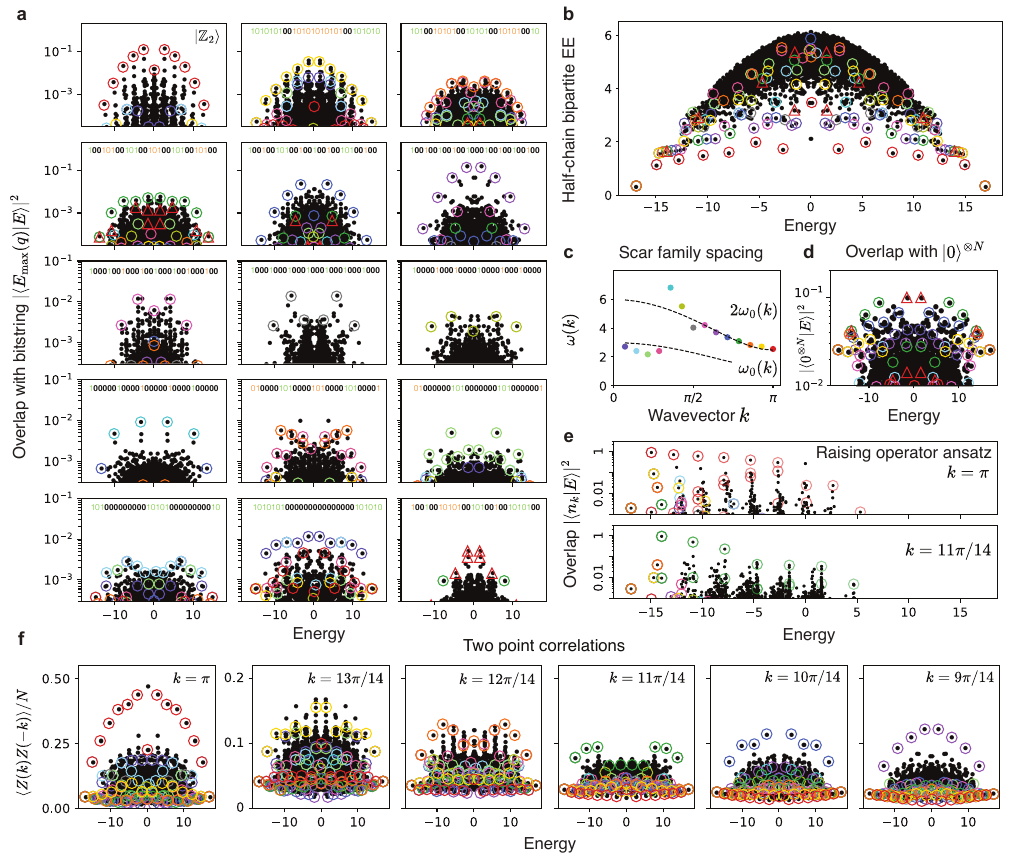}
	\caption{\textbf{Families of many-body scars.} 
		Plasma oscillations in various electric field modes translate to new families of many-body scars, some of which had been identified in Ref.~\cite{szoldra2022unsupervised}.
		\textbf{a.} The overlap of each eigenstate (here, in the $k=0$ inversion symmetric sector for $N=28$) with different bitstrings reveal states with anomalously high overlaps. We choose these bitstrings as the largest eigenstates $|E_\text{max}(q)\rangle$ of each electric field mode $E(q)$ for each momentum $q = 2\pi j/N$. 
		We detect approximately equally spaced peaks (colored circles), which select a different set of eigenstates for each $q$. The $q=0$ case reproduces the well known $|\mathbb{Z}_2\rangle$ case (red circles), $q = 2\pi \times 5/N$ is close to the $|\mathbb{Z}_3\rangle$ scar family (purple circles). In the lower-right panel, we also plot another bitstring consisting of `100' and `10' substrings. This selects a new set of eigenstates (red triangles) which have high overlap with $|0\rangle^{\otimes N}$ (panel d), but do not belong to any of the above scar families. We interpret this bitstring as a typical random \textit{fully-packed} bitstring (SI).
		\textbf{b.} These new scar families include many eigenstates with anomalously low half-chain entanglement entropy (EE). 
		\textbf{c.} The average spacing between scarred eigenstates (within the $k=0$ sector) agrees with twice the iDSF bandstructure $\omega_0(q)$, for $k \in [\pi/2,\pi]$ ($q \in [0,\pi/2]$). Outside this range, the scar spacing is closer to $\omega_0(k)$ (plotted as guide to the eye). 
		\textbf{d.} The $|0\rangle^{\otimes N}$ has modest overlap with a number of exceptional eigenstates, most notably with the families corresponding to $|E_\text{max}(q)\rangle$ for $q = 2\pi j/N$ for $j = 4,5$ (green, blue circles) and for the random fully-packed bitstring (red triangles). This momentum range agrees with our estimate $q = 0.186\pi$ (Ext. Dat. Fig.~\ref{fig:ED_clustering}). 
		\textbf{e.} We use a raising operator ansatz $|n_k\rangle = [Z^+(k)Z^-(-k)]^n|\text{GS}\rangle $ (where $Z^{\pm}(k)\equiv Z(k) \mp i \alpha(k) PYP(k)$ and $\alpha(k)$ is obtained from the mean-field theory, SI) to approximate the families of states. The $q=0$ familly has been studied in Ref.~\cite{iadecola2019quantum}, and the ansatz is moderately successful for other families as well, such as the $k = 11\pi/14$ case plotted.
		\textbf{f.} These identified states also display non-thermal expectation values of the observable $\langle Z(k)Z(-k) \rangle$ (six smallest values of $q$, i.e. $k=\pi$ to $k=9\pi/14$ plotted).}
	\label{fig:scars}
\end{figure*}

\begin{figure*}
	\centering
	\includegraphics[width=\linewidth]{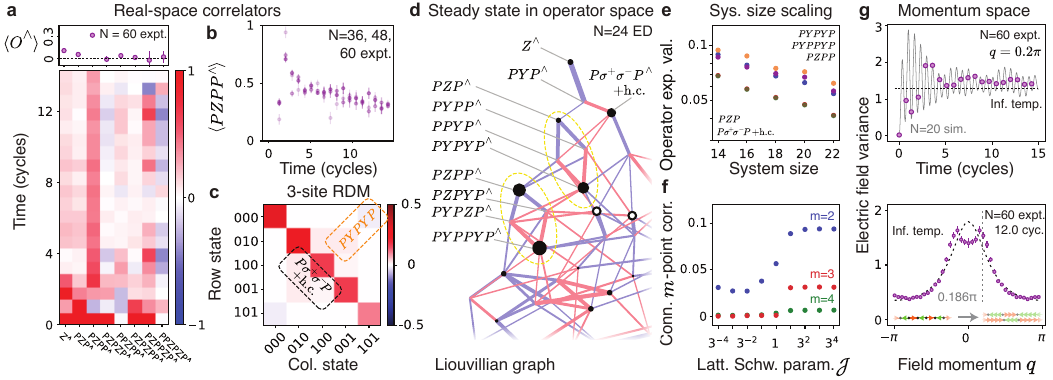}
	\caption{\textbf{Signatures of clustering.} 
		\textbf{a.} Experimentally-observed expectation values of each operator in an orthonormal operator basis up to distance 5. This basis is obtained from the same Gram-Schmidt procedure used to compute the Liouvillian graph; we denote an initial operator $O$ as $O^\wedge$ after orthonormalization.
		We restrict to operators with momentum $k=0$ because the initial state is translation-invariant. These expectation values contain full information about all two-point and multi-point correlation functions up to distance 5.
		This basis largely isolates the athermal signal onto a single operator, $PZPP^\wedge(k = 0)$, i.e.~the component of $PZPP(0)$ orthonormal to $Z(0)$ and $PZP(0)$. \textbf{b.} The athermal signal of $PZPP^\wedge(k = 0)$ gradually decreases over time, in a manner independent of system size (data for $N=36, 48,$ and $60$ plotted from light to dark). 
		\textbf{c.} Numerical simulations reveal additional athermal correlations in non-computational basis observables. The reduced density matrix (RDM) over three continuous sites, computed for the diagonal ensemble at $N=24$ (Methods), shows large off-diagonal coherences not present in infinite-temperature equilibrium. These non-zero matrix elements (boxed)  correspond to the $PYPYP$ and $P\sigma^+\sigma^-P + \text{h.c.}$ operators (relevant matrix elements boxed, e.g.~$PYPYP = -|01010\rangle\langle 00000|+|01000\rangle\langle 00010| + \text{h.c.}$), which large weights on the Liouvillian graph.
		\textbf{d.} We visualize these athermal correlations on the \textit{Liouvillian graph}. On each vertex we plot the late-time expectation values, $\text{tr}(\rho_d O^\wedge)$, of the state $|0\rangle^{\otimes N}$ after time-evolution (Methods).
		The size of the circle denotes the magnitude of the expectation value, with a filled circle being positive and an unfilled circle being negative.
		The operator dark states appear as non-trivial weights on the diamond-shaped substructures, and account for all athermal expectation values.
		\textbf{e.} Athermal expectation values $\text{tr}(\rho_dO)$ (plotted for the raw operators $O$ listed without orthonormalization) decrease with system size, suggesting that these memory effects do not survive in the strict thermodynamic limit, where $\rho_d \equiv \mathbb{E}_t[|\psi(t)\rangle\langle\psi(t)|]$ is the diagonal ensemble, which contains information of expectation values averaged over all times. 
		\textbf{f.} The memory effect also does not persist to the $\mathcal{J}\rightarrow 0$ field-theoretic limit of the lattice Schwinger model. The connected $m$-point correlators, in particular the large signal at $m=3$, vanishes when tuning from  large to small $\mathcal{J}$.     
		Correlators for the diagonal ensemble are computed for an $N=14$ system with open boundary conditions.
		In this simulation of the lattice Schwinger model, the $m=2$ correlators are larger than simulations on the $PXP$ model. We speculate that these may be due to small system size or boundary conditions.
		\textbf{g.} Field variance oscillations persist to long times. The $N=60$ data (a spatial Fourier transform of Fig.~\ref{fig:experimental_data}c) reveals deviations from the infinite-temperature value (dashed lines) both as a function of time (top) and momentum (bottom). Peaks in momentum are consistent with a toy model of randomly packed electric field strings (vertical dashed line and inset, SI).
	}
	\label{fig:ED_clustering}
\end{figure*}

\setcounter{figure}{0}
\renewcommand{\figurename}{Fig.}
\renewcommand{\thefigure}{S\arabic{figure}}
\renewcommand{\theHfigure}{S\arabic{figure}}
	
	\FloatBarrier
	\begin{widetext}
		\appendix
		\begin{center}
			{\large \textbf{Supplementary Information for:~Observation of ballistic plasma and memory in high-energy gauge theory dynamics}} 
		\end{center}
			\resumetoc
		\tableofcontents

		\newpage
		
		\section{Quantities and symbols}
		For reference, here we summarize the quantities that commonly appear in this work:
		\setlength{\tabcolsep}{15pt}
		\renewcommand{\arraystretch}{1.2}
		
		\begin{center}
			\begin{tabular}{c c}
				\hline
				Quantity & Symbol\\
				\hline
				Electric field & $E_j$\\
				Current & $J_j$\\
				Energy & $\mathcal{E}$\\
				Momentum & $p,k$ (spin-chain), $q$ (lattice gauge theory)\\
				Initial state & $|\psi_0\rangle \equiv |0\rangle^{\otimes N}$\\
				Local operator at site $j$ & $O_j$\\
				\multicolumn{2}{c}{for brevity we label the position of an operator string with a single index, e.g.~$PX_jP\equiv P_{j-1}X_jP_{j+1}$}\\
				\\
				Fourier-transformed operator   &  $O(k) \equiv \frac{1}{\sqrt{N}}\sum_j e^{ikj} O_j$\\
				Time-evolved operator & $O(t) \equiv e^{iHt} O e^{-iHt}$\\
				Hilbert-Schmidt inner product & $\langle A, B\rangle \equiv \text{tr}(A^\dagger B)/D$\\
				Hilbert-Schmidt (or Frobenius) norm & $\Vert O \Vert \equiv \langle O, O\rangle^{1/2}$ \\
				Orthonormalized operator & $O^\wedge$\\
				Eigenstates of translationally-invariant Hamiltonian $H$ & $|\mathcal{E},p\rangle$ ($p$ omitted when unnecessary)\\
				Golden ratio & $\varphi \equiv (1+\sqrt{5})/2\approx 1.618$\\
				Hilbert space dimension & $D,D_N^\text{(OBC)},D_N^\text{(PBC)}$ \\
				\multicolumn{2}{c}{ (System size $N$ and boundary conditions occasionally omitted)}\\
				\hline
			\end{tabular}
			\label{tab:symbols}
		\end{center}

		\section{Infinite-temperature dynamical structure factor (iDSF)}
		\noindent The infinite-temperature dynamical structure factor (iDSF) is a central object that we study in this work. In this section, we give a detailed treatment of its properties. For an operator $O$, the iDSF $S^O(k,\omega)$ can be expressed as:
		\begin{align}
			S^O(k,\omega) & \equiv  \sum_{\overset{\mathcal{E},\mathcal{E}'}{p}} \delta(\omega - \mathcal{E} + \mathcal{E}')|\langle \mathcal{E}', p+k|O_j|\mathcal{E},p\rangle|^2~, \label{eq:struct_fac_two}\\
			&={\frac{1}{\sqrt{N}}}\sum_{\mathcal{E},\mathcal{E}'} \delta(\omega - \mathcal{E} + \mathcal{E}') |\langle \mathcal{E}'|O(k)|\mathcal{E}\rangle|^2~,\label{eq:struct_fac_one}
		\end{align}
		where in Eq.~\eqref{eq:struct_fac_two} we have explicitly labelled the momenta $p$ of the eigenstates $|\mathcal{E},p\rangle$ of the Hamiltonian. Eq.~\eqref{eq:struct_fac_one} provides an expression that is also applicable to systems without translational invariance, such as the lattice Schwinger model with open boundary conditions in Fig.~\ref{fig:Lattice_Schwinger}.
		
		\subsection{Properties: Fourier transform of autocorrelation function, sum rule, and symmetries}
		\noindent The iDSF can also be interpreted as the Fourier transform of the \textit{autocorrelator} $\text{tr}[O^\dagger_{j+d}(t) O_j(0)]$, where $O_j(t)$ denotes the operator $O$ on site $j$, Heisenberg-evolved to time $t$. This can be seen from by the following resolutions of identity
		\begin{align}
			\text{tr}(O^\dagger_{j+d} (t) O_j(t)) &= \sum_{\mathcal{E},\mathcal{E}'} \langle \mathcal{E} | O_{j+d}^\dagger |\mathcal{E}'\rangle \langle \mathcal{E}' | O_j | \mathcal{E} \rangle \exp[i(\mathcal{E}-\mathcal{E}')t]\\
			&= \sum_{\overset{\mathcal{E},\mathcal{E}'}{p,p'}} \langle \mathcal{E}, p | O_j^\dagger |\mathcal{E}', p'\rangle \langle \mathcal{E}', p' | O_j | \mathcal{E}, p \rangle \exp[i(\mathcal{E}-\mathcal{E}')t] \exp(i(p-p')d)\\
			&= \int d \omega \sum_{k} S(k,\omega) \exp[-i(\omega t + kd)]
		\end{align}
		The iDSF obeys the following sum rules:
		\begin{equation}
			\int d \omega S(\omega, k) \propto\text{tr}(O^\dagger(-k) O(k)) \equiv D \Vert O(k) \Vert^2
			\label{eq:sum_rule_1}
		\end{equation}
		where $O(k)$ is defined above. When summed over $k$, this gives the overall norm $\sum_k \Vert O(k)\Vert^2 = \Vert O\Vert^2$.
		
		Finally, when $O$ is Hermitian, the iDSF is symmetric under spacetime-inversion, i.e.~$S(k,\omega) = S(-k,-\omega)$. When the Hamiltonian has spatial inversion symmetry, it is also fourfold symmetric: $S(k,\omega) = S(-k,\omega) = S(k,-\omega) = S(-k,-\omega)$.
		
		\subsection{Relationship between iDSF and experimental equal-time correlators}
		\label{app:relation_to_experimental_correlator}
		\begin{figure}
			\centering
			\includegraphics[width=\linewidth]{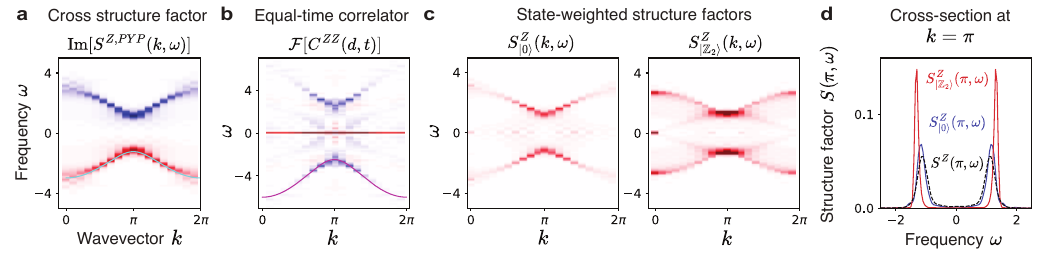}
			\caption{Modified structure factors considered in this work. \textbf{a.} We consider a \textit{cross structure factor} $S^{Z,PYP}(k,\omega) \equiv \mathcal{F}[\text{tr}(Z_{j+d}(t)PYP_j)]$. This quantity is purely imaginary and we plot its imaginary component, which has negative value for the positive frequency branch, and vice-versa. This displays the same band structure as $S^{Z}(k,\omega)$ (mean-field theory plotted in blue). \textbf{b.} We next consider the Fourier transform of the connected correlator $C^{ZZ}(d,t)$ which is measured in experiment, revealing a similar band structure with twice the frequency and linewidth (mean-field theory plotted in magenta). \textbf{c.} We also consider state-weighted structure factors $S^{Z}_{\psi}(k,\omega)$, which minimally incorporate the effects of the initial state $|\psi\rangle$ by weighting the eigenstates $|\mathcal{E}\rangle$ by their overlaps $|\langle \mathcal{E}| \psi\rangle|^2$. While $S^{Z}_{|0\rangle}(k,\omega)$ (left) is similar to $S^{Z}(k,\omega)$, $S^{Z}_{|\mathbb{Z}_2\rangle}(k,\omega)$ is strongly modified, with a different dispersion $\omega_0(k)$ and a smaller linewidth, due to its high overlap with scarred many-body eigenstates. \textbf{d.} Cross-sections of these three structure factors at $k=\pi$. All numerical simulations obtained from exact diagonalization at $N=20$.}
			\label{fig:modified_structure_factors}
		\end{figure}
		
		\noindent While the iDSF is a theoretically simple object and corresponds to the infinte-temperature autocorrelation function $\text{tr}[O_{j+d}(t)O_{j}]$ , in this section we draw connections to the experimentally measured equal-time correlators $C^{OO}(d,t)\equiv \langle\psi(t)|O_{j}O_{j+d}|\psi(t)\rangle - \langle \psi(t)|O_{j}|\psi(t)\rangle\langle \psi(t)|O_{j+d}|\psi(t)\rangle$ of the time-evolved state $|\psi(t)\rangle \equiv \exp(-iHt)|0\rangle^{\otimes N}$.
		
		The connection between $\text{tr}[O_{j+d}(t)O_{j}]$ and 
		$\langle O_{j}O_{j+d} \rangle_c$ has been previously noted~\cite{schonle2021eigenstate}, under ETH assumptions. Here, we discuss their relationship specifically in our system for $O = Z$ and initial state $|0\rangle^{\otimes N}$. The more relevant quantity will be the \textit{cross-correlator} $\text{tr}(Z_{j+d}(t)PYP_{j})$ which is qualitatively similar to the autocorrelator $\text{tr}(Z_{j+d}(t)Z_{j})$. 
		
		We can understand the equal-time correlators in the Heisenberg picture: $\langle\psi(t)|Z_{j}Z_{j+d}|\psi(t)\rangle = \langle\psi_0|Z_{j}(t)Z_{j+d}(t)|\psi_0\rangle$. Unique to our system is that the $Z_j$ operator scrambles anomalously slowly, with anomalously high weight on small operators, oscillating between $Z_{j'}$ and $PYP_{j'}$ operators (see Mean-Field Theory):
		\begin{equation}
			Z_j(t) = \sum_{j'} c_{j'-j}^Z(t) Z_{j'} + c_{j'-j}^{PYP}(t) PY_{j'}P + \cdots  
			\label{eq:Z_operator_expansion}
		\end{equation}
		using this expansion, we obtain 
		\begin{align}
			\langle\psi(t)|Z_{j}Z_{j+d}|\psi(t)\rangle &= \sum_{j_1,j_2} c_{j_1-j}^Z(t)c_{j_2-j-d}^Z(t) \langle \psi_0| Z_{j_1} Z_{j_2}|\psi_0\rangle + c_{j_1-j}^Z(t)c_{j_2-j-d}^{PYP}(t) \langle \psi_0| Z_{j_1} PY_{j_2}P|\psi_0\rangle \\
			&+ c_{j_1-j}^{PYP}(t)c_{j_2-j-d}^{Z}(t) \langle \psi_0| Z_{j_1} PY_{j_2}P|\psi_0\rangle + c_{j_1-j}^{PYP}(t)c_{j_2-j-d}^{PYP}(t) \langle \psi_0| PY_{j_1}P PY_{j_2}P|\psi_0\rangle  
		\end{align}
		It is easy to verify that the first term does not contribute to the connected correlator, since $\langle \psi_0| Z_{j_1} Z_{j_2}|\psi_0\rangle = \langle \psi_0| Z_{j_1} |\psi_0\rangle \langle \psi_0| Z_{j_2}|\psi_0\rangle$ and $|\psi_0\rangle$ is a product state. The second and third terms are also identically zero, and the only term that contributes is the last term, which is only non-zero when $j_1=j_2$. Therefore, keeping only the terms in Eq.~\eqref{eq:Z_operator_expansion}, the connected correlator $C^{ZZ}(d,t)$ is determined by the coefficients $c_{j_1-j}^{PYP}(t)c_{j_1-j-d}^{PYP}(t) \propto \text{tr}[Z_{j}(t)PY_{j_1-j}P]\text{tr}[Z_{j+d}(t)PY_{j_1-j}P]$. As with the autocorrelator $\text{tr}[Z_{0}(t)Z_{j}(0)]$ (Ext.~Dat.~Fig.~\ref{fig:PXP_model_operator_structure_factors}), the real-space correlator $\text{tr}[Z_{0}(t)PY_{j}P(0)]$ is narrowly peaked at $|j| \approx vt$ for some velocity $v$ and decays exponentially as $\exp(-\gamma t)$. This line of reasoning predicts that $C^{ZZ}(d,t) \approx e^{-2\gamma t} \delta_{d,2vt}$, i.e.~that both the velocity and decay constant of the equal-time correlations are doubled relative to the iDSF $S^Z(k,\omega)$, as we verify in the main text.
		
		To further illustrate this connection, in Fig.~\ref{fig:modified_structure_factors} we plot several modified structure factors. Firstly, we plot a \textit{cross structure factor} $S^{Z,PYP}(k,\omega) \equiv \mathcal{F}[\text{tr}(Z_{0}(t)PY_{d}P(0))]$ and verify that it has the same band structure as the iDSFs $S^{Z}(k,\omega)$ and $S^{PYP}(k,\omega)$ (Fig.~\ref{fig:modified_structure_factors}a). Next, we plot the Fourier transform of the connected correlator $\mathcal{F}[C^{ZZ}(d,t)]$ and verify that the same band structure is visible, with twice the frequency of $S^{Z}(k,\omega)$ (Fig.~\ref{fig:modified_structure_factors}b). This can be expressed in the energy eigenbasis as:
		\begin{align}
			&\mathcal{F}[C^{ZZ}(d,t)](k,\omega) = \sum_{i,j,l} \delta(\omega-\mathcal{E}_i+\mathcal{E}_l)~\langle\psi_0|\mathcal{E}_i\rangle\langle \mathcal{E}_l|\psi_0\rangle~ 
			\langle \mathcal{E}_i|Z(-k)|\mathcal{E}_j\rangle \langle \mathcal{E}_j|Z(k)|\mathcal{E}_l\rangle \label{eq:F_CZZ_first_line}\\
			&-\delta_{k}\sum_{i,j,l,m} \delta(\omega-\mathcal{E}_i-\mathcal{E}_l+\mathcal{E}_j+\mathcal{E}_m) \langle\psi_0|\mathcal{E}_i\rangle\langle \mathcal{E}_j|\psi_0\rangle\langle\psi_0|\mathcal{E}_l\rangle\langle \mathcal{E}_m|\psi_0\rangle \langle \mathcal{E}_i|Z(0)|\mathcal{E}_j\rangle \langle \mathcal{E}_l|Z(0)|\mathcal{E}_m\rangle \label{eq:F_CZZ_second_line}
		\end{align}
		where the second line Eq.~\eqref{eq:F_CZZ_second_line}, arising from the disconnected correlator $\langle\psi(t)|Z_j|\psi(t)\rangle\langle\psi(t)|Z_{j+d}|\psi(t)\rangle$, only contributes at $k=0$. Furthermore, since $|\psi_0\rangle$ is translationally invariant, in Eq.~\eqref{eq:F_CZZ_first_line} the eigenstates $|\mathcal{E}_i\rangle, |\mathcal{E}_l\rangle$ have momentum zero, and $|\mathcal{E}_j\rangle$ has momentum $k$, and in Eq.~\eqref{eq:F_CZZ_second_line} all eigenstates $|\mathcal{E}_i\rangle, |\mathcal{E}_j\rangle,|\mathcal{E}_l\rangle,|\mathcal{E}_m\rangle$ have momentum zero.

		Finally, we consider a \textit{state-weighted structure factor}
		\begin{equation}
			S^{Z}_{\psi}(k,\omega) \equiv \sum_{\mathcal{E},\mathcal{E}'} \delta(\omega - \mathcal{E}' + \mathcal{E}) |\langle \mathcal{E}'|O(k)|\mathcal{E}\rangle|^2~ |\langle \mathcal{E}| \psi\rangle |^2
		\end{equation}
		which can be understood as the Fourier transform of the time-averaged autocorrelator $\mathbb{E}_\tau[\langle\psi(\tau)|O_{j+d}(t) O_{j}|\psi(\tau)\rangle]$, for the initial states $|\psi\rangle = |0\rangle^{\otimes N}$ and $|\psi\rangle = |\mathbb{Z}_2\rangle\equiv|1010\cdots \rangle$ [Fig.~\ref{fig:modified_structure_factors}c]. This is the simplest generalization of the iDSF to include effects of the initial state (specifically, its distribution $|\langle \mathcal{E}|\psi\rangle|^2$ over eigenstates) without complications arising from its product-state nature. Furthermore, we can see that it corresponds to the terms in Eq.~\eqref{eq:F_CZZ_first_line} with $i=l$ and a slightly modified condition on $\omega$. While $S^{Z}_{|0\rangle}(k,\omega)$ closely resembles the iDSF $S^{Z}(k,\omega)$, $S^{Z}_{|\mathbb{Z}_2\rangle}(k,\omega)$ has a strongly modified bandstructure, with narrower linewidths and shifted peak frequencies, due to the anomalous overlaps of the $\mathbb{Z}_2$ state with many-body scarred eigenstates~\cite{turner2018weak}. As an illustration, in Fig.~\ref{fig:modified_structure_factors}d we plot cross-sections of these structure factors at $k=\pi$. While $S^{Z}_{|0\rangle}(\pi,\omega)$ is only slightly more narrow than $S^{Z}(\pi,\omega)$, $S^{Z}_{|\mathbb{Z}_2\rangle}(\pi,\omega)$ is markedly different, with a peak frequencies at $\omega_0(\pi)\approx \pm 1.33$, in agreement with the many-body scar oscillation frequency noted in the literature~\cite{turner2018weak}. As a direct consequence of these structure factors, the unequal-time correlators $\langle\psi(\tau)|Z_{j+d}(t) Z_{j}|\psi(\tau)\rangle$ show persistent oscillations in both the $|0\rangle^{\otimes N}$ and $|\mathbb{Z}_2\rangle$ states for all times $\tau$, even after both states are believed to have reached thermal equilibrium.

		In Ext.~Dat.~Fig.~\ref{fig:PXP_model_operator_structure_factors} we also considered a finite-temperature structure factor
		\begin{equation}
			S^O_{\beta}(k,\omega) \equiv \sum_{\mathcal{E},\mathcal{E}'} \delta(\omega-\mathcal{E}'+\mathcal{E}) |\langle \mathcal{E}'|O(k)|\mathcal{E}\rangle|^2 \frac{\exp(-\beta \mathcal{E})}{Z(\beta)},
		\end{equation}
		where $Z(\beta) \equiv \sum_\mathcal{E} \exp(-\beta \mathcal{E})$ is the partition function. This describes the response properties of finite temperature states. In particular, the linear response of $\langle O(t)\rangle_\beta$ of a thermal state $\rho_\beta$ driven by a time-dependent Hamiltonian of the form $H_{PXP}+A \cos(\omega t) O$ can be obtained from the finite-temperature structure factor $S^O_\beta(\omega) - S^O_\beta(-\omega)$~\cite{fradkin2021quantum}. This implies that while the infinite-temperature thermal state (more generally: any particle-hole symmetric state that equally populates eigenstates $|\mathcal{E}\rangle$ and $|-\mathcal{E}\rangle$) will not exhibit such a linear response, a finite-temperature state will.
		
		\subsection{Convolution relationship between iDSFs in the PXP model}
		\label{app:structure_factor_convolutions}
		\noindent In the main text, we discussed the iDSFs of the $PXP$, $PYP$ and $Z$ operators, and found that the iDSF of the $PXP$ operator has a distinct band structure from that of the $PYP$ and $Z$ operators. This is natural from the point of view of the Liovillian graph and the mean field theory, since the $PXP$ operator resides in a different symmetry-number connected component of the Liouvillian graph than the $PYP$ and $Z$, and is therefore nominally independent. However, this neglects the fact that the operators are related. Specifically, they satisfy
		\begin{equation}
			2i PXP(k) = \sum_q [PYP(q), Z(k-q)]~. \label{eq:commutation_for_PXP_k}
		\end{equation}
		As we derive below, this predicts a relationship between the iDSFs $S^{PXP}(k,\omega)$ and $S^{Z}(k,\omega)$, specifically:
		\begin{equation}
			S^{PXP}(k,\omega) \propto \sum_q \left[S^{Z+}(q,\cdot) \ast S^{Z-}(k-q,\cdot)  \right](\omega) + \left[S^{Z-}(q,\cdot) \ast S^{Z+}(k-q,\cdot)  \right](\omega)\label{eq:conv_prediction_for_S_PXP}
		\end{equation}
		where $S^{Z\pm}(k,\omega)$ denote the positive and negative frequency components of the iDSF $S^{Z}(k,\omega) \approx S^{PYP}(k,\omega)$ respectively, and $\ast$ is the convolution symbol. As seen in Ext.~Dat.~Fig.~\ref{fig:PXP_model_operator_structure_factors}, this correctly predicts the band structure of $S^{PXP}(k,\omega)$. This implies that the characteristic velocities of the $PXP$ and $Z/PYP$ band structures are linked, consistent with the fact that there is only one propagating mode in the mutual information and negativity plots of Ext.~Dat.~Fig.~\ref{fig:entanglement_dynamics_ED}.
		
		We derive Eq.~\eqref{eq:conv_prediction_for_S_PXP} with the following approximations. First, using Eq.~\eqref{eq:commutation_for_PXP_k}, we write
		\begin{align}
			S^{PXP}(k,\omega) &\equiv \frac{1}{4}\sum_{i,j} \delta(\omega-\mathcal{E}_j-\mathcal{E}_i)~|\langle \mathcal{E}_i| \sum_q [PYP(q), Z(k-q)]|\mathcal{E}_j\rangle|^2\label{eq:S_PXP_from_commutator}\\
			&\approx \sum_{q} \sum_{i,j} \delta(\omega-\mathcal{E}_j+\mathcal{E}_i) |\langle \mathcal{E}_i|PYP(q)|\mathcal{E}_l\rangle \langle \mathcal{E}_l|Z(k-q)|\mathcal{E}_j\rangle - \langle \mathcal{E}_i|Z(q)|\mathcal{E}_l\rangle \langle \mathcal{E}_l|PYP(k-q)|\mathcal{E}_j\rangle|^2 \label{eq:only_keeping_diagonals}\\
			&\approx \sum_q \int d\omega ' S^{Z+}(q,\omega') S^{Z-}(k-q,\omega-\omega') + S^{Z-}(q,\omega') S^{Z+}(k-q,\omega-\omega')
		\end{align}
		where in going from Eq.~\eqref{eq:S_PXP_from_commutator} to Eq.~\eqref{eq:only_keeping_diagonals} we start with a sum of the form $\sum_{i,j} \sum_{q,q'}\sum_{l,l'} \langle \mathcal{E}_i|PYP(q) |\mathcal{E}_l\rangle \langle \mathcal{E}_l|Z(k-q) |\mathcal{E}_j\rangle \langle \mathcal{E}_j|Z(k-q') |\mathcal{E}_{l'}\rangle \langle \mathcal{E}_{l'}|PYP(q') |\mathcal{E}_i\rangle$ and keep only matching terms where $q=q'$ and $l=l'$, since we expect that for non-matching terms, factors such as $\langle \mathcal{E}_i|PYP(q) |\mathcal{E}_l\rangle\langle \mathcal{E}_{l'}|PYP(q') |\mathcal{E}_i\rangle$ have random sign (in this model, all matrix elements are expected to be real) and average to zero --- the only systematic contributions come from the matching terms. We keep both terms in the commutator since in this model, $Z(k)$ and $PYP(k)$ are related: based on the approximate SU(2) relations (Appendix~\ref{app:Bloch_Wigner}), they can be written in terms of ``raising" and ``lowering" operators $\varepsilon^{\pm}(k) = Z(k)\mp i a PYP(k)$. To arrive at Eq.~\eqref{eq:conv_prediction_for_S_PXP}, note that the commutator only keeps terms of the form $|\langle \mathcal{E}_i|\varepsilon^+(q)|\mathcal{E}_l\rangle \langle \mathcal{E}_l|\varepsilon^-(k-q)|\mathcal{E}_j\rangle|^2$, which in turn can be coarse-grained in terms of (the convolutions of) the iDSFs $S^{Z+}(q,\omega)$ and $S^{Z-}(k-q,\omega)$.

		\subsection{iDSF in integrable and chaotic models}
		\subsubsection{Mixed field Ising model}
		\noindent As our first comparison we consider the mixed field Ising model (MFIM)
		\begin{equation}
			H_\text{MFIM} = \sum_j h_x X_j + h_z Z_j + J X_j X_{j+1}.
			\label{eq:MFIM}
		\end{equation}
		Unless otherwise specified, we use parameters $h_x = 0.8090$, $h_z = 0.9045$ and $J = 1$ as originally specified in Refs.~\cite{kim2013ballistic,kim2014testing} (see Ref.~\cite{rodriguez2024quantifying} for slightly different ``maximally chaotic" parameters). As seen in Ext.~Dat.~Fig.~\ref{fig:MFIM_TFIM_XXZ_structure_factors}, the iDSFs of the local $X,Y$ and $Z$ operators do not exhibit significant structure. For completeness, we also plot the iDSF of the energy density $\varepsilon_j \equiv h_x X_j + h_z Z_j + J Z_j Z_{j+1}$, which is analogous to $PXP_j$ in the PXP model\footnote{This energy density differs slightly from the more-symmetric energy density $\varepsilon^\text{sym}_j \equiv h_x (X_j+X_{j+1})/2 + h_z (Z_j+Z_{j+1})/2 + J Z_j Z_{j+1}$, yet we will argue in App.~\ref{app:hot_band_second_sound} that the former is more well-motivated, since its Fourier transform has fixed operator norm $\Vert h(k) \Vert$ is independent of $k$, and hence the iDSF $\int S(k,\omega) d\omega$ is constant for each $k$.}. This iDSF also does not display significant structure, but see App.~\ref{app:hot_band_second_sound} for slightly different MFIM parameters in which some structure is present.
		
		In Ext.~Dat.~Fig.~\ref{fig:energy_transport}, we use the translationally invariant initial state $|\psi_0\rangle = [\cos(\theta/2)|0\rangle + e^{i\phi}\sin(\theta/2)|0\rangle]^{\otimes N}$ with $(\theta,\phi) = (0.813418\pi,0)$, following Ref.~\cite{maceira2024thermalization}, which is at infinite temperature but has maximum energy variance and hence the largest expected transport of energy correlations.

		\subsubsection{Transverse field Ising model}
		\noindent We next consider the transverse field Ising model (TFIM), an integrable variant of the MFIM in which $h_x=0$:
		\begin{equation}
			H_\text{TFIM} = \sum_j h_z Z_j +  J X_j X_{j+1}.
		\end{equation}
		This model is exactly solvable by a mapping to free fermions~\cite{pfeuty1970one}. In order to keep the strength of the Hamiltonian comparable to those of the MFIM above, we use the values $(h_z,J) = (2,1)$. As seen in Ext.~Dat.~Fig.~\ref{fig:MFIM_TFIM_XXZ_structure_factors}, the iDSF of the $Z$ and energy density operators show considerable structure, while the $X$ and $Y$ iDSFs appear diffuse and structureless. All iDSFs can be understood in terms of the free fermion solution~\cite{sachdev2011quantum}. As an example, we discuss the multiple bands seen in the $Z$ iDSF.
		
		Using the Jordan-Wigner transformation~\cite{sachdev2011quantum} (and ignoring subtleties about periodic boundary conditions which are unimportant for the present discussion), 
		\begin{equation}
			c_i \equiv \left[\prod_{j=1}^{i-1}(-Z_j)\right]\sigma^-_i \Leftrightarrow \sigma^-_i = \left[\prod_{j=1}^{i-1} (1-2 c^\dagger_j c_j^{})\right] c_i~, 
		\end{equation}
		In the Fourier basis, the TFIM Hamiltonian can be written as a non-interacting model: $H = \sum_k \mathcal{E}(k) c^\dagger(k) c(k)$, with single-particle spectrum
		\begin{equation}
			\mathcal{E}(k) = \sqrt{h_z^2+J^2 + 2 h_z J \cos(k)}
		\end{equation}
		The eigenstates of the Hamiltonian are simply labelled by occupation numbers $n_k\in \{0,1\}$ for the fermions at each momentum $k$, and their corresponding eigenvalues are the sums of the single-particle energies $\sum_k n_k \mathcal{E}(k)$.
		
		To understand the $Z$ iDSF, the Jordan-Wigner transformation states that $Z_j = 2c^\dagger_jc^{}_j - 1$. Taking its Fourier transform gives $Z(k) = \frac{2}{\sqrt{N}}\sum_q c^\dagger(q+k)c(q) -\sqrt{N}\delta_k  $. This allows us to determine the peaks of the iDSF, which corresponds to: what are the allowed energy differences $\mathcal{E}_i-\mathcal{E}_j$ for which $\langle \mathcal{E}_i|Z(k)|\mathcal{E}_j\rangle$ is non-zero. These allowed values are simply the differences $\mathcal{E}(k+q) - \mathcal{E}(q)$ for various $q$.
		In the thermodynamic limit, this uniformly fills a region in $(k,\omega)$ space. We note that the XX model $H = \sum J(X_jX_{j+1} + Y_j Y_{j+1})$ can also be solved with the exact same Jordan-Wigner transformation, and as such its $Z$ iDSF (not plotted) is also discrete. In this case, since the non-interacting dispersion $\mathcal{E}(k) = 2J \cos(k)$ is simpler, the differences have analytic form, $\mathcal{E}(k+q)- \mathcal{E}(q) = - 4J \sin(q+k/2)\sin(k/2)$.
		
		\subsubsection{XXZ model}
		\noindent Finally, we consider the XXZ model, which is also integrable with Bethe ansatz solution. The XXZ model has Hamiltonian
		\begin{equation}
			H = \sum_j X_jX_{j+1}+Y_j Y_{j+1}+ \Delta Z_j Z_{j+1},
		\end{equation}
		and is exactly solvable for all $\Delta$. In Ext.~Dat.~Fig.~\ref{fig:MFIM_TFIM_XXZ_structure_factors}, we plot the iDSFs of the $X,Y,Z$ and energy density operators at the Heisenberg model point $\Delta=1$, chosen because it displays superdiffusive spin transport~\cite{bulchandani2021superdiffusion,scheie2021detection, de2020universality}. This can be inferred from the iDSF, as we discuss in App.~\ref{subsec:structure_factor_transport}.
		
		Unlike the TFIM or the XX model, the eigenvalues of the XXZ model cannot simply be expressed as sums of single-particle energies, since the magnon particles have interaction energies in a way specified by the Bethe ansatz solution. As a result, a local operator creates excitations across a range of energies and the iDSF appears diffuse and structureless.
		
		\subsection{iDSF in other constrained models}
		\noindent The Liouvillian graph (Appendix~\ref{app:Liouvillian_graph}) describes how a constrained Hilbert space restricts operator growth which in turn leads to a well-defined operator band structure at infinite temperature. In order to test this hypothesis, we examine the iDSFs of several models which generalize the PXP model in different ways: the PPXPP model~\cite{giudici2019diagnosing}, the 2D PXP model~\cite{bluvstein2021controlling,maskara2021discrete,lin2020quantum,samajdar2020complex} as well as spin-1 and spin-3/2 quantum link models~\cite{zache2022toward}.

		\subsubsection{PPXPP model}
		\noindent We first consider the PPXPP model, a generalization of the PXP model to spin-1/2 chains with a next-nearest neighbor blockade. The Hilbert-space dimension of this model satisfies the recurrence relation:
		\begin{align}
			\text{dim}\mathcal{H}^\text{(OBC)}_N &= \text{dim}\mathcal{H}^\text{(OBC)}_{N-1} + \text{dim}\mathcal{H}^\text{(OBC)}_{N-3},\\
			\text{dim}\mathcal{H}^\text{(PBC)}_N &= \text{dim}\mathcal{H}^\text{(OBC)}_{N-1} + \text{dim}\mathcal{H}^\text{(OBC)}_{N-5},
		\end{align}
		which scales as $\text{dim}\mathcal{H}_N \sim \xi^N$ where $\xi = 1.46557$ is a root of $\xi^3-\xi^2-1 = 0$~\cite{giudici2019diagnosing}. The PPXPP Hamiltonian applies Rabi oscillations subject to a range-2 Rydberg blockade. In Ext.~Dat.~Fig.~\ref{fig:not_scars} we plot the iDSFs for some simple operators in this model. The iDSF exhibits sharp peaks in a quasiparticle band, which is particularly pronounced for the $PPYPP$ operator.
		
		\subsubsection{2D PXP model}
		\noindent We next consider the 2D PXP model, which describes Rydberg dynamics on a spin-1/2 square lattice subject to a nearest-neighbor blockade constraint~\cite{bluvstein2021controlling,maskara2021discrete,lin2020quantum,samajdar2020complex}.  In Ext.~Dat.~Fig.~\ref{fig:not_scars} we plot the iDSF of the $Z$ operator in this model: this structure factor is defined on a two-dimensional momentum space, from which we plot representative slices. Although the system sizes accessible by exact diagonalization are limited, we observe signatures of a well-defined band structure.
		
		\subsubsection{Spin-1 and spin-3/2 quantum link models}
		\noindent Finally, we consider lattice gauge theoretic generalizations of the PXP model. As discussed in the main text, the PXP model can be exactly mapped onto a quantum link model~\cite{surace2020lattice}, a variant of the lattice Schwinger model, in which the unbounded $U(1)$ electric field variable is truncated to a spin-1/2 degree of freedom.
		
		We may consider higher-spin quantum link models~\cite{zache2022toward} as a way to study whether plasma modes persist in the lattice Schwinger model, complementary to the approach taken in Fig.~\ref{fig:Lattice_Schwinger}. The spin-$S$ quantum link model can be written as $PXP$-like models on a constrained spin-$S$ chain: $H_\text{QLM}^{(S)} = \sum_{j} \mathcal{P} S^x_j$, 
		where $S^x_j$ is the spin-$S$ $X$ operator, and $\mathcal{P}$ projects onto the space of allowed nearest-neighbor configurations satisfying the constraints $S^Z_j + S^Z_{j+1}=-1,0$. These constraints reflect the fact that the edges $(j,j+1)$ can either be empty (and hence the electric fields $E_j = (-1)^j S^Z_j$ and $E_{j+1}$ are equal) or occupied by one fermion (hence $E_j$ and $E_{j+1}$ differ by 1), and suffice to describe quantum link models of any spin. For example, the spin-1 link model has allowed nearest-neighbor configurations
		\begin{equation}
			\begin{matrix}
				(-1,0)\\
				(0, -1)
			\end{matrix}~\bigg\}~S^Z_j + S^Z_{j+1}=-1~,~ \begin{matrix}
				(-1,+1)\\
				(~0,~0)\\
				(+1,-1)
			\end{matrix}~\Bigg\}~S^Z_j + S^Z_{j+1}=0,
		\end{equation}
		while the spin-3/2 link model has allowed configurations $(-3/2,+3/2),(-1/2,+1/2),(+1/2,-1/2),(+3/2,-3/2)$ and $(-3/2,+1/2),(-1/2,-1/2),(+1/2,-3/2)$. As seen in Ext.~Dat.~Fig.~\ref{fig:not_scars}, the iDSFs of the $S^Z$ operator feature a band structure, suggesting that plasma modes are a generic property of quantum link models, consistent with our results in the main text.
		
		\subsection{iDSF and ETH}
		\label{subsec:structure_factor_ETH}
		\noindent In this section we briefly discuss the connections between the iDSF and the eigenstate thermalization hypothesis (ETH). The ETH is an ansatz for the statistical properties of matrix elements $\langle \mathcal{E}_i|O|\mathcal{E}_j\rangle$ of local operators $O$~\cite{srednicki1999approach,dalessio2016quantum}. It posits that
		\begin{equation}
			\langle \mathcal{E}_i|O|\mathcal{E}_j\rangle = O(\bar{\mathcal{E}}) \delta_{ij} + \frac{1}{\sqrt{e^{S(\bar{\mathcal{E}})}}} f_O(\bar{\mathcal{E}},\omega)R_{ij},
			\label{eq:ETH}
		\end{equation}
		where $\bar{\mathcal{E}} \equiv (\mathcal{E}_i + \mathcal{E}_j)/2$ is the average energy, $\omega \equiv \mathcal{E}_i-\mathcal{E}_j$ is the energy difference, $O(\bar{\mathcal{E}})$ is a smoothly varying \textit{thermal expectation value}, $S(\bar{\mathcal{E}})$ is the thermodynamic entropy (the logarithm of the density of energy eigenstates) at energy $\bar{\mathcal{E}}$, $f_O(\bar{\mathcal{E}},\omega)$ is another smoothly-varying function (which quickly decays with increasing $\omega$), and $R_{ij}$ are random numbers: random real numbers with zero mean and variance one when $i=j$ and random complex numbers with zero mean and a variance of two when $i\neq j$.
		
		From Eq.~\eqref{eq:struct_fac_one}, we see that the iDSF represents the $\omega$-dependence of the $f_O(\bar{\mathcal{E}},\omega)$ function. Non-trivial peaks have been noted in the literature~\cite{dalessio2016quantum,schonle2021eigenstate}, even in the PXP model~(Fig.~2 of Ref.~\cite{turner2018quantum}) and are fully consistent with the ETH, but have not received detailed attention. In this work, we specifically investigate their \textit{momentum dependence}. In the presence of translational symmetry, we should consider the momentum-resolved operators $O(k)$ (equivalently, consider the momentum difference of the eigenstate pairs $|\mathcal{E}_i\rangle,|\mathcal{E}_j\rangle$). The peaks of the functions $f_{O(k)}(\bar{\mathcal{E}},\omega)$ are momentum-dependent, which leads to the observed transient transport phenomena.
		
		The scaling of the participation ratio of the iDSF in Fig.~\ref{fig:Lattice_Schwinger} is consistent with the ETH:
		even in our system which features a sharp peak in the iDSF, the density of states $\exp(S(\mathcal{E}))$ in Eq.~\eqref{eq:ETH} scales precisely as $O(D)$, up to $\text{poly}(N)$ factors, giving the observed $D^2$ scaling.
		
		\subsection{iDSF and hydrodynamic transport}
		\label{subsec:structure_factor_transport}
		\begin{figure}
			\centering
			\includegraphics[width=\linewidth]{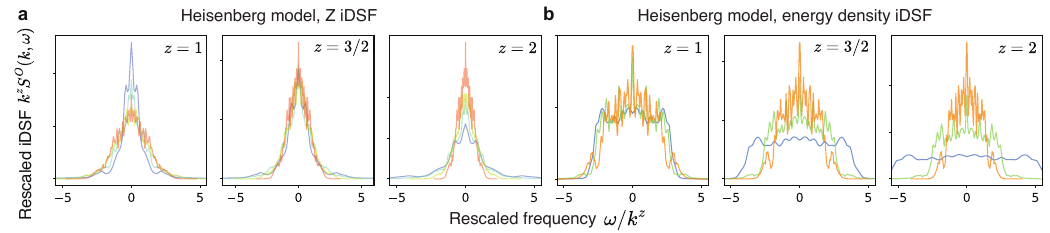}
			\caption{iDSF data collapse to determine hydrodynamic universality class. We demonstrate data collapse under the rescaling $\omega/k^z$ and $k^z S(k,\omega)$, where $z$ is the dynamical exponent. In the Heisenberg model (XXZ model with $\Delta=1$), spin transport is known to be superdiffusive with $z=3/2$, while energy transport is ballistic with $z=1$~\cite{bulchandani2021superdiffusion}. Accordingly, data collapse is only visible for \textbf{a.} $z=3/2$ for the spin iDSF $S^Z(k,\omega)$ and \textbf{b.} $z=1$ for the energy-density iDSF (Ext.~Dat.~Fig.~\ref{fig:MFIM_TFIM_XXZ_structure_factors}).} 
			\label{fig:heisenberg_model_data_collapse}
		\end{figure}
		
		\noindent The iDSF of a conserved quantity can be used to study the universality class of its transport. Specifically, let $O_j$ be the density of the conserved quantity, i.e. $\sum_j O_j$ is conserved. Hydrodynamics predicts that the long-wavelength modes $O(k)$ decay as $k^z$, where $z$ is the dynamical exponent. In turn, this predicts that the iDSF follows the scaling relation~\cite{scheie2021detection}
		\begin{equation}
			\lim_{k\rightarrow 0}k^z S^O(k,\omega) = f(\omega/k^z),
			\label{eq:transport_data_collapse}
		\end{equation}
		where $f$ is a function which could be e.g.~Gaussian~\cite{de2020universality,michailidis2024corrections}. This scaling behavior has been used in Ref.~\cite{scheie2021detection} to establish the presence of superdiffusive transport with $z=3/2$ in a solid-state neutron scattering experiment.
		We can indeed observe this in our numerical iDSF of the Heisenberg model (Ext.~Dat.~Fig.~\ref{fig:MFIM_TFIM_XXZ_structure_factors}): rescaling the data as per Eq.~\eqref{eq:transport_data_collapse} with $z=3/2$ reveals the desired data collapse, while other candidates $z=1$ or $z=2$ do not. In contrast, the iDSF of the energy density displays data collapse with diffusive dynamical exponent $z=2$ instead, as seen in Fig.~\ref{fig:heisenberg_model_data_collapse}. 
		
		This scaling relation, for small $k$, suggests that energy transport in the PXP model is ultimately diffusive (Ext.~Dat.~Fig.~\ref{fig:energy_transport}). Unusual in our model is that the data collapse~\eqref{eq:transport_data_collapse} only appears at very small momenta: this necessitates us to use operator-size truncation (OST)~\cite{yithomas2024comparing} --- an approximate method based on the Liouvillian graph to access small momenta --- to see this scalng.
		
		\subsection{Energy transport in the mixed field Ising model: hot band second sound}
		\label{app:hot_band_second_sound}
		\noindent In Ext.~Dat.~Fig.~\ref{fig:energy_transport}, we described a ballistic-to-diffusive crossover in energy transport. In particular, we observed the iDSF of the energy density $S^{PXP}(k,\omega)$ transition from single-peaked at small $k$ to double-peaked at larger $k$. In the latter regime, the peak positions $\omega_0(k)$ scale roughly linearly with $k$, giving rise to quasi-ballistic transport at short times.  
		
		A similar phenomenon has been noted in Ref.~\cite{yithomas2024comparing}, dubbed \textit{hot band second sound}. They considered the MFIM at parameter value $(h_x,h_z,J) = (0.9045,1.4,1)$~\cite{white2018quantum,rakovsky2022dissipation}, and studied the autocorrelator $\text{tr}[\varepsilon^\text{sym}(k,t) \varepsilon^\text{sym}(-k,0)]$ where $\varepsilon^\text{sym}_j \equiv h_x (X_j+X_{j+1})/2 + h_z (Z_j+Z_{j+1})/2 + J Z_j Z_{j+1}$ is the spatially symmetric energy density, and $\varepsilon(k)$ is its Fourier transform. They observed that this autocorrelator exhibits overdamped decay at small $k$: monotonic relaxing towards 0, but at a critical momentum $k^* \approx 0.5$, transitions to underdamped oscillations, where the autocorrelator oscillates about zero with an exponentially decaying amplitude. They dubbed this phenomenon ``hot band second sound" in reference to earlier work on ``hot band sound"~\cite{bulchandani2022hotbandsound} on underdamped charge transport at infinite temperature.
		
		We reinterpret these results in terms of the iDSF. As we plot in Fig.~\ref{fig:MFIM_hotband_sound}, the over-to-under-damped transition is reflected in the iDSF: $S^{\varepsilon,\text{sym}}(k,\omega)$ is single-peaked at $\omega = 0$ for $k< k^*$, and double-peaked at $\omega = \pm \omega_0$ for $k>k^*$. Unlike the PXP model, hot band second sound in the MFIM only features a small amount of ballistic transport of entanglement (Fig.~\ref{fig:MFIM_hotband_sound}b).
		
		Before going further, we mention a technical point: we argue that the non-symmetric energy density $\varepsilon_j \equiv h_x X_j + h_z Z_j + J Z_j Z_{j+1}$ is more appropriate for the iDSF [as we plot in Ext.~Dat.~Fig.~\ref{fig:MFIM_TFIM_XXZ_structure_factors}]. This is because the iDSF of this object, $S^{\varepsilon}(k,\omega)$ has uniform spectral weight, in the sense that its sum rule [Eq.~\eqref{eq:sum_rule_1}] is independent of $k$: $\int d\omega S(k,\omega) \propto \Vert \varepsilon(k)\Vert_\text{HS}^2 = h_x^2 + h_z^2 + J^2$. In contrast, $\Vert \varepsilon^\text{sym}(k)\Vert_\text{HS}^2 = \cos^2(k/2)(h_x^2 + h_z^2) + J^2$ decreases with $k$. Indeed, while $S^{\varepsilon}(k,\omega)$ appears to vanish near $k=\pi$, $S^{\varepsilon}(k,\omega)$ remains large throughout the Brillouin zone, allowing better visualization of the split peaks in this regime.
		
		The hot band second sound phenomenon is present only in a narrow range of parameter values (Fig.~\ref{fig:MFIM_hotband_sound}c): lowering $h_z$ slightly to $h_z=1.1$ recovers a featureless iDSF similar to the Kim-Huse parameters~\cite{kim2013ballistic,kim2014testing} in Ext.~Dat.~Fig.~\ref{fig:MFIM_TFIM_XXZ_structure_factors}, while increasing it to $h_z=2.2$ shows a multi-band iDSF reminiscent of the integrable TFIM (Ext.~Dat.~Fig.~\ref{fig:MFIM_TFIM_XXZ_structure_factors}).
		
		Ref.~\cite{yithomas2024comparing} provides a simple toy model of the dynamics of the energy operator $\varepsilon(k)$. As discussed in App.~\ref{app:Liouvillian_graph}, the time-evolved operator $\varepsilon(k,t)$ satisfies the Heisenberg equation of motion $(d/dt) \varepsilon(k,t) = i[H,\varepsilon(k,t)]$. We can expand the operator $\varepsilon(k)$ in an orthonormal basis of operators, and the Heisenberg equation gives a coupled first-order system of linear differential equations in an infinitely large (in the thermodynamic limit) basis of operators. The toy model in Ref.~\cite{yithomas2024comparing} truncates this infinite basis into two basis states, spanned by the energy operator $\varepsilon(k)$ and the energy current operator defined as $j_\varepsilon(k)\equiv i[H,\varepsilon(k)]$. Within this approximation, a time-evolved operator $O(t) = c_\varepsilon(t){\varepsilon}(k)+
		c_{j_\varepsilon}(t){j}_\varepsilon$ can be expressed in terms of two coefficients which satisfy the equation:
		\begin{equation}
			\frac{d}{dt}\begin{pmatrix}
				c_\varepsilon\\
				c_{j_\varepsilon}\\
			\end{pmatrix} \approx i\begin{pmatrix}
				0 & v k\\
				v k & i\gamma
			\end{pmatrix}\begin{pmatrix}
				c_\varepsilon\\
				c_{j_\varepsilon}\\
			\end{pmatrix} 
		\end{equation}
		The ${\varepsilon}(k)$ and ${j}_\varepsilon(k)$ operators are coupled with coupling strength that, to leading order, vanishes linearly with $k$. In turn, the current operator $j_\varepsilon(k)$ is connected to a `bath' of large operators. Assuming this behaves as a Markovian bath, we can approximate this with a non-Hermitian dissipation term of strength $\gamma$, to leading order independent of $k$. This $2\times 2$ linear equation has complex eigenvalues $(i\gamma \pm \sqrt{4v^2k^2-\gamma^2})/2$ that correspond to Lorentzian peaks in the iDSF $S^{\varepsilon}(k,\omega)$. For any value of $\gamma$, there is a critical $k^* = \gamma/(2v)$ at which the eigenvalues gain a real component, leading to the over-to-underdamped transition. $\gamma$ depends on the MFIM parameters and is related to the diffusion coefficient in Ref.~\cite{yithomas2024comparing}. When it is too large, $k^*\geq \pi$ and this bifurcation may not be present.  
		
		The above model reproduces the essential features of the ``hot band second sound" phenomenon (and in turn quasiballistic energy transport), yet is completely generic and can be applied to any many-body system. We believe that the PXP model is particularly conducive towards quasiballistic energy transport for the same reason that the $Z$ operator has a non-trivial iDSF: the ``operator growth bottleneck" due to the blockaded Hilbert space and the particle-hole symmetry (Appendix~\ref{app:Liouvillian_graph}) results in a small effective $\gamma$ and a large $v$, hence a small $k^*$. This leads to the natural question of whether any models can exhibit similar quasiballistic energy transport~\cite{bhakuni2025anomalously}.
		
		\begin{figure}
			\centering
			\includegraphics[width=\linewidth]{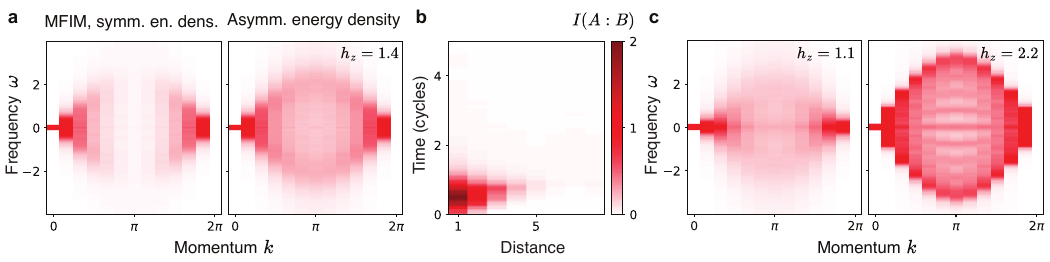}
			\caption{The iDSF of the MFIM in the \textit{hot band second sound} regime (obtained from ED of an $N=20$ periodic chain)~\cite{yithomas2024comparing}. \textbf{a.} We plot the iDSF of the symmetric and asymmetric energy densities, defined in App.~\ref{app:hot_band_second_sound}, at parameter values $(h_x,h_z,J)=(0.9045,1.4,1)$. The asymmetric energy density satisfies the sum rule where $\int d\omega S(\omega,k)$ is independent of $k$, while the symmetric energy density does not. Beyond a critical $k^*\approx 1$, the energy iDSF has two peaks with non-zero frequencies: in real-space and time, this is described as an over-to-under-damped transition in the relaxation of energy modes in Ref.~\cite{yithomas2024comparing}. \textbf{b.} Unlike the PXP model (see Ext.~Dat.~Fig.~\ref{fig:entanglement_dynamics_ED}), even in the hot-band sound regime, the MFIM does not display significant propagation of information. Mutual information plotted for size-3 subsystems of a $N=18$ periodic chain. 
				\textbf{c.} When $h_z$ is modified to $1.1$, we get approach the Kim-Huse MFIM parameters and recover a featureless energy iDSF as seen in Ext.~Dat.~Fig.~\ref{fig:MFIM_TFIM_XXZ_structure_factors}. \textbf{e.} When $h_z$ is doubled to $2.2$, we approach the integrable TFIM limit. 
			}
			\label{fig:MFIM_hotband_sound}
		\end{figure}
		
		\section{Ballistic propagation of entanglement}
		\label{app:ballistic_entanglement}
		\noindent In order to study the ballistic propagation observed in this work without reference to specific quantities, we consider the propagation of information-theoretic quantities such as the quantum mutual information and negativity. We observe ballistic propagation of these quantities as well, indicating that the plasma quasiparticles in this system also carry non-local entanglement. Specifically, we consider small contiguous subsystems $A$ and $B$, and study the reduced states $\rho_A$, $\rho_B$ and $\rho_{AB}$ of the time-evolved state $|\psi(t)\rangle$. The mutual information refers to the difference in von Neumann entropies
		\begin{equation}
			I(A:B) \equiv S(\rho_A)+S(\rho_B) - S(\rho_{AB})
			\label{eq:mutual_info}
		\end{equation}
		A large mutual information $I(A:B)$ indicates strong correlations between the subsystems $A$ and $B$. Intuitively, these correlations can arise when $A$ and $B$ are entangled. This would be reflected in $S_A$ and $S_B$, but not in $S_{AB}$; their difference approximately cancels the entanglement not shared between $A$ and $B$. However, this intuition is not always exact. Subsystems $A$ and $B$ might not be entangled (in the sense that $\rho_{AB}$ can be separable) even if their mutual information $I(A:B)$ is nonzero. To remedy this, we study the negativity $\mathcal{N}(\rho_{AB})$, defined as
		\begin{equation}
			\mathcal{N}(\rho_{AB}) \equiv  \text{log} \Vert \rho_{AB}^\text{PT} \Vert_1,
			\label{eq:negativity}
		\end{equation}
		where $\rho_{AB}^\text{PT}$ is the partial transpose of $\rho_{AB}$ (i.e. only transposing the degrees of freedom associated with either subsystem $A$ or $B$), and $\Vert O \Vert_1$ indicates the sum of the absolute values of its eigenvalues. Negativity is an entanglement witness, a nonzero $\mathcal{N}(\rho_{AB})$ indicating that $\rho_{AB}$ is not separable, and is also an upper bound to the distillable entanglement~\cite{plenio2005logarithmic}. We consider subsystems $A, B$ of sizes $|A|=|B|=$1,2, and 3 respectively. As seen in Ext.~Dat.~Fig.~\ref{fig:entanglement_dynamics_ED}, the negativity and mutual information are nonzero when $A$ and $B$ are separated by a distance $d_{AB}$ that grows linearly with time, indicating non-local entanglement (i.e. $A$ and $B$ have entanglement that is not mediated by the region between them) that is transported ballistically. We associate this with the production of entangled quasiparticle pairs. Due to their dispersion and linewidth, the ballistic propagation is imperfect and the quasiparticle content diffuses locally, supported by the fact that $I(A:B)$ and $\mathcal{N}(\rho_{AB})$ persist for larger distances and longer times for larger subsystem sizes $|A|=|B|$.
		
		\section{Fully packed electric field strings}
		\noindent In Ext.~Dat.~Fig.~\ref{fig:ED_clustering}, we observe certain structure in momentum space: the oscillations in electric field variance persist to long times, and their amplitudes vary over momentum, with peaks near $q\approx 0.18\pi$. Here, we give a toy model that accounts for these peaks. We posit that under PXP evolution, the $|0\rangle^{\otimes N}$ state initially evolves into a superposition of \textit{fully-packed} (FP) bitstring states, i.e.~bitstrings without ``000" substrings and cannot have Rydberg excitations added. Indeed, numerical simulations reveal that the expectation value $\langle \psi(t)|P_{j-1}P_jP_{j+1} |\psi(t)\rangle$ drops from an initial value of 1 to 0.04 within 0.3 Rabi cycles and continues to oscillate after (not plotted), coincident with a maximal Rydberg excitation density of $\langle \psi(t)|(1-P_j)|\psi(t)\rangle$ of 0.4.
		
		Examples of FP bitstrings include $|\mathbb{Z}_2\rangle$ and $|\mathbb{Z}_3\rangle = |100100\dots\rangle$, but there are exponentially many such bitstrings, generated by ``10” and ``100” substrings. We now ask: what is the typical number of ``10” and ``100” substrings in FP bitstrings of length $N$? Answering this question will also allow us to deduce its momentum-space structure.
		
		We calculate the number of FP bitstrings, of total length $N$, composed exclusively of $N_{100}$ ``100" substrings and $N_{10}$ ``10" substrings. This is approximately the binomial coefficient $(N_{10}+N_{100})!/( N_{10}!N_{100}!)$, subject to the constraint $2N_{10}+3N_{100}=N$, where $N_{10}$ is the number of `10' blocks. Using Stirling's approximation, we seek the density of ``100" blocks $x \equiv N_{100}/N$ that maximizes this number: $x=0.18577\dots$, the root of the equation
		$\left(\frac{1}{2}-\frac{3x}{2}\right)^3 =
		\left(\frac{1}{2}-x\right) x^2$. To convert this into the positions of the momenta peaks, we note that each pair of ``00"'s, which we interpret as domain walls separating N{\'e}el domains (bitstrings in Ext.~Dat.~Fig.~\ref{fig:scars}), heuristically adds a field momentum $q = 2\pi/N$: therefore, we expect peaks at $0.18577 \pi$, as seen in Ext.~Dat.~Fig.~\ref{fig:ED_clustering}.
		
		Finally, we remark that this problem bears superficial similarity to the (discrete) R{\'e}nyi parking problem~\cite{gerin2014page}. This predicts a value of $x=e^{-2} = 0.135$ instead. However, the random process in this problem is slightly different from ours: it does not result in a uniform distribution over all fully-packed strings, rather exponentially weights strings by excitation number. Numerical simulations reveal that PXP dynamics is better described by the former uniform distribution.
		
		\section{Semiclassical behavior of electric field and current variables: Approximate $SU(2)$ dynamics, Bloch sphere representations and Wigner distributions}
		\label{app:Bloch_Wigner}
		
		\subsection{Bloch sphere representation for electric field modes}
		\noindent In this work, we discussed plasma oscillations between electric field modes $E(q)=Z(k=q+\pi)$ and current $J(q)=PY\!P(k=q+\pi)$. This dynamics can be equivalently understood in terms of an approximate $SU(2)$ commutation relation. We take the electric field $Z(k)$, the current $PYP(k)$, and the energy $H_\text{PXP}$ to be a triplet of observables analogous to the $SU(2)$ Pauli $X,Y,Z$ operators. In order to obtain real expectation values, we use the Hermitian-symmetrized observables $O_\mathbb{R} (k)\equiv [O(k) + O(-k)]/2$ instead.
		
		This triplet satisfies approximate $SU(2)$ commutation relations~\cite{choi2019emergent,iadecola2019quantum, maskara2021discrete,kerschbaumer2024quantum}, since
		\begin{align}
			[Z_\mathbb{R}(k),PYP_\mathbb{R}(k)] &= -i H_\text{PXP} - iPXP_\mathbb{R}(2k), \label{eq:SU2_1}\\
			[Z_\mathbb{R}(k), H_\text{PXP}] &= 2i PYP_\mathbb{R}(k),\\
			[PYP_\mathbb{R}(k), H_\text{PXP}] &= -2i PZP_\mathbb{R}(k)+2i \cos(k) (P\sigma^+\sigma^-P+\text{h.c.})_\mathbb{R}(k),
		\end{align}
		As a reminder, the $PZP(k)$ operator has high overlap with $Z(k)$ and this set of equations is approximately closed.
		The additional term $PXP_\mathbb{R}(2k)$ in Eq.~\eqref{eq:SU2_1} in fact does not affect the dynamics of $Z_\mathbb{R}(k)$ and $PYP_\mathbb{R}(k)$, since they only depend on the nested commutators $[H_\text{PXP},\cdots [H_\text{PXP},O_\mathbb{R}(k)]]$. However, these terms can affect the rotations of the form $e^{i[\alpha Z_\mathbb{R}(k)+\beta PYP_\mathbb{R}(k)]}$, relevant for the computation of the Wigner function. 
		
		In order to visualize this approximate $SU(2)$ dynamics, we plot the expectation values $\langle \psi(t) | Z_\mathbb{R}(k)| \psi(t) \rangle$, $\langle \psi(t)| PYP_\mathbb{R}(k)| \psi(t)\rangle $, and $\langle \psi(t) | H_\text{PXP} |\psi(t) \rangle$.  As seen in Ext.~Dat.~Fig.~\ref{fig:ED_Wigner}, these expectation values trace out approximately circular orbits which decay over time. While this picture has been known for $k=\pi$, i.e.~the uniform electric field modes, this approach works remarkably well for non-trivial field modes, away from $k=\pi$.
		
		We plot the surface of this Bloch sphere in Ext.~Dat.~Fig.~\ref{fig:ED_Wigner}: this is the boundary of feasible points of expectation values $\langle  H_\text{PXP}  \rangle, \langle PYP_\mathbb{R}(k) \rangle, \langle Z_\mathbb{R}(k)\rangle$.  To find this boundary, we work in spherical coordinates. At each point $(\theta, \phi)$, we solve for the highest energy eigenstate of the Hamiltonian $H(\theta,\phi) \equiv \cos\theta \cos\phi H_\text{PXP} + \cos\theta \sin \phi~PYP_\mathbb{R}(k) + \sin \theta~Z_\mathbb{R}(k)$. The expectation values $(\langle  H_\text{PXP}  \rangle, \langle PYP_\mathbb{R}(k) \rangle, \langle  Z_\mathbb{R}(k)\rangle)$ define the boundary along the direction $(\theta,\phi)$. Taken together, this forms an ellipsoidal boundary. The time-evolved expectation values lie close to this surface (Ext.~Dat.~Fig.~\ref{fig:ED_Wigner}), supporting our picture of the approximate $SU(2)$ dynamics.
		
		\subsection{Many-body Wigner distribution}
		\noindent The Wigner distribution is a common tool in the field of quantum optics to visualize quantum states of light~\cite{scully1997quantum}. Motivated by the oscillator-like dynamics of the electric field modes $Z(k)$ in the PXP model, we adapt the Wigner distribution to the many-body setting in order to visually represent the dynamics of the PXP model.
		
		In quantum optics, the Wigner distribution is given by
		\begin{equation}
			\mathcal{W}(x,p) = \frac{1}{\pi} \int dy \langle x+y |\psi\rangle \langle \psi |x-y \rangle e^{2 i p y},
			\label{eq:wigner_dist_1}
		\end{equation} 
		where $x$ and $p$ are harmonic oscillator variables satisfying the canonical commutation relations $[x,p] = i$ (setting $\hbar=1$ for convenience). We seek to generalize the Wigner distribution to the many-body setting, with $x$ and $p$ replaced by the electric-field and current variables $E_\mathbb{R}(q) = Z_\mathbb{R}(k=q+\pi)$ and $J_\mathbb{R}(q) = PYP_\mathbb{R}(k=q+\pi)$.
		
		Eq.~\eqref{eq:wigner_dist_1} cannot be directly generalized to the many-body setting, because there is a degenerate space of states with the same eigenvalue of $Z_\mathbb{R}(k)$. Furthermore, $Z_\mathbb{R}(k)$ and $PYP_\mathbb{R}(k)$ only satisfy approximate SU(2) commutation relations, preventing even expressions for spin coherent states to be adapted.
		
		Instead, we use the relationship between the Wigner distribution and the (inverse) \textit{Wigner-Weyl transform} to develop the appropriate many-body Wigner distribution~\cite{scully1997quantum}. The Weyl transform is a method to ``quantize" a function $f(x,p)$ into an operator $\Phi[f]$, given by:
		\begin{equation}
			\Phi[f] \equiv \frac{1}{(2\pi)^3} \int d\alpha d\beta\int dx dp~f(x,p) e^{-i[\alpha(X-x) + \beta(P-p)]}
			\label{eq:Weyl_transform} 
		\end{equation}
		The state $|\psi\rangle \langle \psi|$ is the Wigner-Weyl transform $\Phi[\mathcal{W}(x,p)]$, and therefore the Wigner distribution can be understood as the inverse Wigner-Weyl transform of a state. We use Eq.~\eqref{eq:Weyl_transform} to obtain our many-body Wigner distribution:
		\begin{equation}
			W(E,J) \equiv \int d\alpha d\beta~e^{-i(\alpha E + \beta J)}~\text{tr}[\rho e^{i [\alpha Z_\mathbb{R}(k)+\beta PYP_\mathbb{R}(k)]},
			\label{eq:many_body_Wigner}
		\end{equation}
		where we use the variables $(E,J)$ to specify the context of this Wigner distribution. Eq.~\eqref{eq:many_body_Wigner} has many desired properties of a quasiprobability distribution, including:
		\begin{enumerate}
			\item $W(E,J)$ is always real.
			\item $W(E,J)$ is normalized, i.e.~$\int dE dJ~W(E,J) = 1$.
			\item The marginal distribution $w(E)\equiv \int dJ~W(E,J) \sum_z \delta(E - z) \langle z|\rho| z\rangle$, i.e.~is equal to the distribution over the eigenvalues $z$ of $Z_\mathbb{R}(k)$ (and likewise for the integral over $E$). 
		\end{enumerate}
		
		\noindent\textbf{Proof of 1.} From Eq.~\eqref{eq:many_body_Wigner}, we can see that $W(E,J)^* = W(E,J)$, since its complex conjugate amounts to replacing $\alpha \mapsto -\alpha, \beta \mapsto -\beta$, which does not change the value of the integral over all $(\alpha, \beta)$.
		
		\noindent \textbf{Proof of 2.} Integrating over $E$ and $J$, we have:
		\begin{equation}
			\int dE dJ~W(E,J) = \int d\alpha d\beta~ \delta(\alpha)\delta(\beta)~\text{tr}[\rho e^{i [\alpha Z_\mathbb{R}(k)+\beta PYP_\mathbb{R}(k)]} = \text{tr}~\rho = 1
		\end{equation}
		
		\noindent \textbf{Proof of 3.} Integrating over $J$, we have:
		\begin{equation}
			\int dJ~W(E,J) = \int d\alpha d\beta~\delta(\beta)~e^{-i\alpha E}\text{tr}[\rho e^{i [\alpha Z_\mathbb{R}(k)+\beta PYP_\mathbb{R}(k)]} = \sum_z \text{tr}[\rho~\delta(z-E) |z\rangle\langle z|] 
		\end{equation}
		Crucially, the above properties do not rely on any assumptions on $Z_\mathbb{R}(k)$ and $PYP_\mathbb{R}(k)$. Hence, Eq.~\eqref{eq:many_body_Wigner} is a good candidate for a quasiprobability distribution in our case, in which $Z_\mathbb{R}(k)$ and $PYP_\mathbb{R}(k)$ only approximately act as canonical conjugate variables.
		
		\section{Rydberg blockaded Hilbert space: Infinite temperature expectation values and operator basis}
		\noindent The Liouvillian graph suggests that the plasma dynamics we discover is partially enabled by the unique Hilbert space of Rydberg-blockaded systems. While simple to describe, the blockaded Hilbert space hosts many exotic properties and has been studied for its relation to statistical-mechanical models~\cite{fendley2004competing}, anyonic theories~\cite{chandran2018absence}, and most recently has received much interest as the natural setting of quantum systems realized by Rydberg atom arrays~\cite{bernien2017probing,turner2018quantum,nguyen2023quantum,cesa2023uniuversal}. We summarize the properties of the blockaded Hilbert space that are relevant to this work.
		
		In one dimension, the Hilbert space $\mathcal{H}$ is spanned by the set of all bitstrings with no adjacent `1's. With open boundary conditions (OBC), the boundary sites may both be `1', but not with periodic boundary conditions (PBC). As a result, their Hilbert-space dimensions are slightly different: for systems of length $N$, they are given by~\cite{turner2018weak}:
		\begin{align}
			D_{N}^{(OBC)} &= F_{N+2}\\
			D_N^{(PBC)} &=  D_{N-1}^{(\text{OBC})} +
			D_{N-3}^{(\text{OBC})} = F_{N+1}+F_{N-1}\end{align}
		where $F_n = (\varphi^n - (-\varphi)^{-n})/\sqrt{5}$ is the $n$-th Fibonacci number, satisfying $F_n = F_{n-1}+F_{n-2}$ for $n>1$ and $F_0= F_1 = 1$. The expression $F_{n+1}+F_{n-1}$ is also known as the $n$-th Lucas number $L_n$, which also satisfies the recursion relation $L_n = L_{n-1}+L_{n-2}$ for $n>1$ but with $L_0=2,~L_1 = 1$.In both OBC and PBC, the Hilbert-space dimension $D_N$ grows as $\varphi^N$, where $\varphi \approx 1.618$ is the golden ratio.
		
		Among its unusual properties, the blockaded Hilbert space is non-factorizable: if the system is divided into complementary subsystems $A$ and $B$, the joint Hilbert space is not simply the tensor product $\mathcal{H} \neq \mathcal{H}_A \otimes \mathcal{H}_B$. As a consequence, many properties that are straightforward in systems of many spin-1/2 particles require careful treatment. For example, while the set of Pauli operator strings provides a natural basis of operators, there is no such simple choice in blockaded systems. We describe our construction of an othonormal operator basis in Sec.~\ref{eq:orthonormal_basis_blockaded_Hilbert_space}, but we first describe a necessary preliminary step: evaluating the trace of local operators.
		
		\subsection{Operator trace}
		In this work, we evaluate traces of local operators for a variety of purposes: infinite-temperature expectation values of local operators, the Hilbert-Schmidt inner product of two operators, and constructing our mean-field operator equations of motion. Here we describe subtleties of the operator trace in the blockaded Hilbert space.
		
		The most important point to note is that the trace of a local operator $O_j$ depends on the size of the global system, a marked departure from spin-1/2 systems. As a simple illustrative example, consider evaluating the infinite-temperature expectation value of $Z_j$ in a system of $N$ sites with PBC. This operator can be written in terms of projectors $P_j\equiv |0\rangle\langle 0|_j$ and $n_j \equiv |1\rangle\langle 1|_j$ onto the `0' and `1' states at site $j$ respectively:
		\begin{align}
			\langle Z_j \rangle &\equiv \frac{\text{tr}(Z_j)}{D_N^{(PBC)}} = \frac{\text{tr}(P_j) - \text{tr}(n_j)}{D_N^{(PBC)}} = \frac{D_{N-1}^{\text{(OBC)}} -  D_{N-3}^{\text{(OBC)}}}{D_{N-1}^{(OBC)}+D_{N-3}^{\text{(OBC)}}}\\
			&\overset{N\rightarrow\infty}{\longrightarrow} \frac{\varphi^2-1}{\varphi^2+1} =  \frac{1}{\sqrt{5}} = 0.447... \label{eq:Z_expval_TDL}
		\end{align}
		This expectation value, while system-size dependent, quickly saturates to its value in the thermodynamic limit.  A related fact is that even at infinite temperature, there are non-zero two-point (connected) correlations $\langle Z_j Z_{j+d}\rangle_c$ which decay exponentially with distance $d$. These can be evaluated in the thermodynamic limit as
		\begin{align}
			\langle Z_j Z_{j+d}\rangle - \langle Z_j\rangle\langle Z_{j+d}\rangle &=4 \langle P_j P_{j+d}\rangle - 4\langle P_j\rangle\langle P_{j+d}\rangle = 
			4 \frac{D_{d-1}^\text{(OBC)}D_{N-d-1}^\text{(OBC)}}{D_N^\text{(PBC)}} - 4\left(\frac{D_{N-1}^{\text{(OBC)}}}{D_N^\text{(PBC)}}\right)^2\\
			&\overset{N\rightarrow \infty}{\longrightarrow} = \frac{4}{5}(-\varphi^{-2})^d \label{eq:ZZ_spatial_corr}
		\end{align}
		We shall also need its Fourier transform:
		\begin{equation}
			\langle Z(-k)Z(k)\rangle - |\langle Z(k)\rangle|^2 = \frac{4}{\sqrt{5}(3+2\cos k)}
			\label{eq:inf_temp_ZZ_corr}
		\end{equation}
		
		\subsection{Operator inner product}
		
		\noindent In general, the inner product between two real-space operators $O^{(1)}, O^{(2)}$ is nonzero only if the product $O^{(1)}\times O^{(2)}$ consists only of $P$ and $I$ symbols. We then have to compute the trace of this operator string: this is given by the product $\left(\prod_iD_{l_i}^{(OBC)}\right)/D_N^{(PBC)}$, where $l_i$ is the length of each contiguous substring of $I$'s. We summarize the inner products of some operators in Table~\ref{tab:operator_inner_products}.
		\begin{table}[h!]
			\centering
			\begin{tabular}{c}
				\toprule
				$\Vert Z(k) \Vert^2 = \frac{4}{\sqrt{5}(3+2 \cos k)}$\\
				$\Vert PZP(k) \Vert^2 = \langle PZP(-k) PZP(k) \rangle = 2\frac{1+\cos k /\varphi}{1+\varphi^2}$ \\ 
				$\Vert PYP(k) \Vert^2 = \frac{2}{1+\varphi^2}$ \\
				$\Vert (P\sigma^+\sigma^-P+P\sigma^-\sigma^+P)(k)\Vert^2 = \frac{2/\varphi}{1+\varphi^2}$\\
				$\Vert PPYP(k) \Vert^2 = \Vert PYPP(k) \Vert^2 = \frac{2/\varphi}{1+\varphi^2}$\\
				$\langle PYP(k), PYPP(k) \rangle = \langle PYP(k), PPYP(k) \rangle = \frac{2/\varphi}{1+\varphi^2}$\\
				$\langle PPYP(k), PYPP(k) \rangle = \frac{2/\varphi^2}{1+\varphi^2}$\\
				\bottomrule
			\end{tabular}
			\caption{Operator norms and inner products of some small operators in the blockaded Hilbert space}
			\label{tab:operator_inner_products}
		\end{table}

		\noindent There is a complex phase ambiguity in the operator Fourier transforms. For concreteness, in this work we define $PYPP(k) \equiv \sum_j e^{ikj} PY_jPP/\sqrt{N}$, $PPYP(k) \equiv \sum_j e^{ikj} PPY_jP/\sqrt{N}$ and
		$P[\sigma^+\sigma^-+\sigma^-\sigma^+]P(k) \equiv \sum_j e^{ik(j+1/2)} P[\sigma^+_j\sigma^-+\sigma^-_j\sigma^+]P/\sqrt{N}$.
		
		\section{Mean field theory}
		\label{app:mean_field_theory}
		\noindent As discussed in the main text, we observe surprisingly well-defined and long-lived band structures in the iDSFs of the PXP model. 
		In this section, we provide details on our theoretical mean field approximation to estimate properties of these bands. 
		To be specific, we present two different mean field approximations, which make slightly different approximations yet yield similar results.
		These two approximations are united once higher-order effects are included, which is captured in a systematic manner by the Louivillian graph framework developed in the following section.
		
		Our first mean field approximation considers the dynamics of the operator $PZP(k)$. 
		In real space, the $PZ_j P$ operator has an equation of motion
		\begin{equation}
			\frac{d}{dt} PZ_j P = i [H_\text{PXP},PZ_j P] = 2 PY_j P + PPY_{j+1} P + PY_{j-1} P P.
			\label{eq:PZP_EOM}
		\end{equation}
		As presented in the Methods, our mean field method first approximates $PPY_{j+1} P \approx \langle P_{j-1} \rangle' PY_{j+1} P$, where $\langle P_{j-1} \rangle'$ is the infinite-temperature expectation value of $P$ \textit{conditioned} on one of its neighbors being in the zero state.
		A short computation gives $\langle P \rangle'=\langle P_jP_{j+1}\rangle_{\beta=0}/\langle P_j\rangle_{\beta=0} = D_{N-2}^{(\text{OBC})}/D_{N-1}^{(\text{OBC})} = 1/\varphi$; note that this differs from the bare infinite-temperature expectation value, $\langle P \rangle_{\beta=0} = 0.7236$.
		In effect, this approximation treats the $PPYP$ operator identically to the $PYP$ operator, except it is ``dressed" by the factor $\langle P \rangle'= 1/\varphi$.  From the Liouvillian graph perspective (Appendix~\ref{app:Liouvillian_graph}), this is equivalent to ignoring the orthogonal $PYPP^\perp$ component, i.e.~
		\begin{equation}
			PYPP(k) \approx \frac{\langle PYP(k), PYPP(k)\rangle }{\Vert PYP(k)\Vert^2}PYP(k) = \frac{1}{\varphi} PYP(k) = \langle P \rangle' PYP(k),
		\end{equation}
		using the inner products computed in Table~\ref{tab:operator_inner_products}.
		
		Taking the Fourier transform of the equation of motion over the lattice site $j$, we find the following equation of motion in momentum space,
		\begin{equation}
			\frac{d}{dt}  PZP(k) \approx 2(1 + \langle P\rangle' \cos k ) PYP(k)~.
			\label{eq:PZP_EOM}
		\end{equation}
		Meanwhile, the analogous momentum-space equation of motion for the $PYP(k)$ operator is,
		\begin{equation}
			\frac{d}{dt} PYP(k) = i [H_\text{PXP},PYP(k)] = -2PZP(k) + 2\cos(k/2)P[\sigma^+\sigma^-+\sigma^-\sigma^+]P(k). \label{eq:PYP_EOM}
		\end{equation}
		If we make a coarse approximation and set the $P[\sigma^+\sigma^-+\sigma^-\sigma^+]P(k)$ term to zero, then we obtain a closed set of equations for the $PZP(k)$ and $PYP(k)$ operators, with eigenfrequencies $\omega_0(k)=\pm 2\sqrt{1+\cos k/\varphi}$. 
		
		We can include the $P[\sigma^+\sigma^-+\sigma^-\sigma^+]P(k)$ term to enable a more accurate approximation.
		This operator obeys the equation of motion,
		\begin{align}
			\frac{d}{dt}P[\sigma^+_{j}\sigma^-+\sigma^-_{j}\sigma^+]P = & - \left(PP Y_{j+1}P+P Y_{j}PP \right) - iP[\sigma^+\sigma^-_{j-1}\sigma^+-\sigma^-\sigma^+_{j-1}\sigma^-] P \nonumber \\
			& - iP[\sigma^+\sigma^-_{j}\sigma^+-\sigma^-\sigma^+_{j}\sigma^-] P.
		\end{align}
		Approximating the $P\sigma^+\sigma^-\sigma^+P$ term as zero, we obtain
		\begin{equation}
			\frac{d}{dt} P[\sigma^+\sigma^-+\sigma^-\sigma^+]P(k)  \approx - 2\langle P \rangle' \cos \frac{k}{2} PYP(k),
			\label{eq:Oplus_EOM}
		\end{equation}
		which has eigenfrequencies $\omega_0(k) = 0, \pm \sqrt{2(2+\langle P \rangle' + 3 \langle P \rangle' \cos k)}$. We note that the zero-frequency mode is observed in the iDSF of $P(\sigma^+\sigma^- + \text{h.c.})P$. 
		
		Our second mean field approximation consider the dynamics of the $Z(k)$ operator instead of $PZP(k)$.
		The $Z(k)$ operator has an especially simple equation of motion, $\frac{d}{dt}Z(k)=2PYP(k)$. We obtain a closed set of equations with Eq.~\eqref{eq:PYP_EOM} by enforcing a mean-field approximation of $PZP(k)$ by $Z(k)$. This approximation is determined by the operator inner product (App.~\ref{app:Liouvillian_graph}),
		\begin{equation}
			PZP(k) \approx \frac{\langle Z(k),PZP(k)\rangle}{\Vert Z(k)\Vert^2} Z(k) = \frac{\sqrt{5}}{2(1+\varphi^2)} (3+2 \cos k)~Z(k) \equiv \alpha(k) Z(k)
		\end{equation}
		Here, the momentum-dependence arises entirely from the normalization $\Vert Z(k)\Vert = \sqrt{\langle Z(k),Z(k) \rangle}$ (Table~\ref{tab:operator_inner_products}).
		This yields eigenfrequencies $\omega_0(k) = \pm 2 \sqrt{\alpha(k)}$, which are close to our first mean-field approximation's predictions.
		
		Finally, beyond these two approximations, we can perform an analogous mean-field approximation for the energy density operator $PXP(k)$.
		This yields the coupled differential equations,
		\begin{equation}
			\frac{d}{dt} \begin{pmatrix}
				PXP (k)  \\
				iP[\sigma^+ \sigma^- - \sigma^- \sigma^+]P (k)  
			\end{pmatrix} \approx  2 \sin \frac{k}{2} \begin{pmatrix}
				0 & 1 \\
				-\langle P \rangle'  & 0 
			\end{pmatrix}
			\begin{pmatrix}
				PXP (k) \\
				i P[\sigma^+ \sigma^- - \sigma^- \sigma^+]P (k)
			\end{pmatrix},
			\label{eq:mean_field_equations_PXP_op}
		\end{equation}
		which give eigenfrequencies $\omega_0(k) = \pm 2 \sin \frac{k}{2} \sqrt{\langle P \rangle'}$, in agreement with the structure factor of the $PXP$ operator (Fig.~\ref{fig:PXP_model_operator_structure_factors}ef). As discussed in Sec.~\ref{app:structure_factor_convolutions}, the structure factors of the $PXP$, $PYP$ and $Z$ operators are interrelated, a fact that is not obvious from this mean-field analysis. We also note that this mean-field approximation for the energy operator, despite displaying qualitative agreement with numerical observations for some $k$, does not accurately predict the location of the single-to-double-peak crossover (Ext.~Dat.~Fig.~\ref{fig:energy_transport}).

		\section{Liouvillian graph for operator dynamics}
		\label{app:Liouvillian_graph}
		In this section, we describe the construction of the \textit{Liouvillian graph}~\cite{white2023effective,yithomas2024comparing} for the PXP model, which captures the time-evolution of operators under time-independent Hamiltonian dynamics. In the Heisenberg picture, an operator $O(t) \equiv e^{iHt}O e^{-iHt}$ obeys the Heisenberg equation of motion:
		\begin{equation}
			\frac{d}{dt} O(t) = i[H,O(t)] \equiv \mathcal{L} [O] ~,
		\end{equation}
		where we define the commutator with the Hamiltonian as the Liouvillian superoperator $\mathcal{L}$.
		
		The operator equation of motion can be viewed as a single-particle hopping process on a graph in the following way. We begin with an orthonormal basis of operators, $\{O^\wedge_a\}$. Each operator in the orthonormal basis corresponds to a vertex of the graph. The weight of an edge from one vertex $O_a^\wedge$ to another vertex $O_b^\wedge$ is given by the matrix element, 
		\begin{equation} \label{eq:Lab}
			\mathcal{L}_{ba} \equiv \langle O_b^{\wedge \dagger}, \mathcal{L}[O_a^\wedge]\rangle = \langle O_b^{\wedge \dagger}, i [H,O_a^\wedge]\rangle = \frac{1}{D}\text{tr}(O_b^{\wedge\dagger} ~i [H,O_a^\wedge]).
		\end{equation}
		The time-evolution of an initial operator $O(0)$ is given by first decomposing it in the orthonormal basis, and then solving the corresponding hopping problem on the graph, where the hoppings correspond to the edge weights.
		When all operators are Hermitian, the Liouvillian $\mathcal{L}$ is anti-Hermitian and has no self-weights.
		Hence, the Liouvillian is more accurately described as a directed graph.
		
		\subsection{Operator symmetries in the Liouvillian graph: bipartiteness and connected components}
		The presence of symmetries in the Hamiltonian leads to additional structure in the Liouvillian graph. Namely, for the PXP model, the Liouvillian graph is split into two connected components, each of which has a bipartite structure.
		The first component contains the $Z$ and $PYP$ operators, and the second component contains the $PXP$ operators.
		
		This additional structure arises from the presence of time-reversal symmetry (TRS) and particle-hole symmetry (PHS) in the PXP model. As discussed above, TRS refers to the fact that the PXP Hamiltonian is purely real, and hence is invariant under complex conjugation. Meanwhile, PHS refers to the fact that the PXP model is \textit{odd} under the conjugation $\mathcal{C} H_\text{PXP} \mathcal{C} = -  H_\text{PXP}$,
		where $\mathcal{C} = \prod_j Z_j$. This antisymmetry results in an energy spectrum which is \textit{symmetric} about zero: for every eigenstate $|\mathcal{E}\rangle$ with eigenvalue $\mathcal{E}$, the state $\mathcal{C}|\mathcal{E}\rangle$ is also an eigenstate of $H_\text{PXP}$ with eigenvalue $-\mathcal{E}$. 
		
		The symmetry charges of a quantum state $|\psi\rangle$ are defined by the state's eigenvalues under $\mathcal{K}$ and $\mathcal{C}$. A state that is purely real (imaginary) carries TRS charge $+1$ ($-1$), and equivalently for eigenstates of $\mathcal{C}$. In contrast, the symmetry charges of a quantum operator $O$ are defined by how the symmetry charges of a state $|\psi \rangle$ change under the application of $O$, $O|\psi\rangle$. For example, the $Z$ operator carries TRS and PHS charges $(+1,+1)$, the $PXP$ operator carries charges $(+1,-1)$, and the $PYP$ operator carries charges $(-1,-1)$.
		
		Conjugation with the Hamiltonian $i[H_\text{PXP},O]$ flips the signs of both the TRS and PHS symmetry charges carried by an operator $O$. The flip of the TRS charge  is due to the factor of $i$ pre-fixing the conjugation. The flip of the PHS charge is because the Hamiltonian $H_\text{PXP}$ carries PHS charge $-1$. The presence of even one of the symmetries would result in a bipartite Liouvillian graph (for example, the Liouvillian graph of a Hamiltonian with TRS symmetry can only connect operators with opposite TRS charges; one can further show that the entries $\mathcal{L}_{ab}$ are purely real). The presence of \emph{both} of these symmetries results in the Liouvillian graph further splitting into two disconnected components. Each component contains all operators with the same \textit{product} of symmetry charges. Hence, the $Z$ and $PYP$ operators, which have a product of $+1$, belong in the first connected component, while the $PXP$ operator, with a product of $-1$, belongs in the second component, as illustrated in Fig.~\ref{fig:clustering} and Ext.~Dat.~Fig.~\ref{fig:energy_transport}.
		
		\subsection{Mixed Field Ising model}
		As a simple example, we describe the first several hopping elements of the Liouvillian graph of the MFIM~[Eq.~\eqref{eq:MFIM}],
		as illustrated in Ext.~Dat.~Fig.~\ref{fig:MFIM_TFIM_XXZ_structure_factors}. We work directly in the momentum basis.
		The first several hopping elements are given by the equations:
		\begin{align}
			i[H_\text{MFIM},X(k)] &= 2 h_Z~Y(k)~,\\
			i[H_\text{MFIM},Y(k)] &= -2 h_X~Z(k) +2 h_Z~X(k) - 2 J e^{-ik/2}~X\!Z(k)  - 2 J~e^{ik/2} ZX(k)~, \\
			i[H_\text{MFIM},Z(k)] &= 2 h_X~Y(k) + 2 J e^{-ik/2}~XY(k) +2  Je^{ik/2}~Y\!X(k), 
		\end{align}
		where we let $AB(k)\equiv \sum_j e^{ik(j+1/2)} A_j B_{j+1}$ denote the Fourier transform of the product of two Pauli operators, $AB$. Following our discussion above, the Liouvillian graph is bipartite between operators with TRS charge $+1$ and $-1$.
		
		\subsection{Constructing an orthonormal operator basis for the Rydberg blockaded Hilbert space}
		\label{eq:orthonormal_basis_blockaded_Hilbert_space}
		To construct the Liouvillian graph of the PXP model, we first construct an orthonormal operator basis for the blockaded subspace. Such a construction would be straightforward for any local tensor product Hilbert space, however the blockaded subspace does not have  a local tensor product structure.
		
		We begin by enumerating a complete set of operators that act on the blockaded subspace. A suitable  set is comprised of all \textit{contiguous strings} of any length involving the following symbols: $\{P,Z,\sigma^+, \sigma^-\}$. Each string corresponds to an operator and all translations of it. The restriction that the operators act on the blockaded subspace leads to the following rules:
		\begin{enumerate}
			\item The symbol $\sigma^+$ can only be adjacent to $P$ and $\sigma^-$. If $\sigma^+$ was adjacent to another $\sigma^+$, then this creates two adjacent ones and hence has matrix elements outside the blockaded subspace. If $\sigma^+$ was adjacent to a $Z$, then we can assume that the qubit that $Z$ acts on is in the zero state, since otherwise $\sigma^+$ would lead to matrix elements outside the blockaded subspace, in which case $Z$ can be replaced with $P$.
			\item The symbol $\sigma^-$ can only be adjacent to $P$ and $\sigma^+$. This follows by an identical logic to rule 1.
			\item The symbol $Z$ can only be adjacent to $P$. In the bulk of the string, this follows from an identical logic to rules 1 and 2. Moreover, we can assume that $Z$ never occurs at the edge of the string without loss of generality, since we can always replace $Z = \mathbb{I} - 2 P$. 
			\item All strings must start and end with the symbols $P$. This follows from the first three rules.
		\end{enumerate}
		Any operator acting on the blockaded subspace can be written as a linear combination of strings obeying the above rules. For example, the operator $P\sigma^+P I P $ is equal to the sum $2P\sigma^+PPP - P\sigma^+PZP$. Motivated by our plasma theory, we also make a singular exception to rule 1, by replacing the one-body operator $P$ with the one-body operator $Z$. After subtracting off the identity component and normalizing, these are the same operator. Finally we can reject strings composed of entirely $P$ operators, such as $PP$, $PPP$, etc.:
		\begin{enumerate}
			\setcounter{enumi}{4}
			\item No string can be composed of only symbols $P$.
		\end{enumerate}
		This follows because these operators can always be expressed as linear combinations of other strings. For example, $P_jP_{j+1} = I_{j,j+1} - n_j  - n_{j+1} = (Z_j+Z_{j+1})/2$, where we have used the relation $n_j = (I_j - Z_j)/2$, where $I_j$ is the identity operator.
		
		The next step is to orthonormalize this operator basis. To do so, we first address an edge case, since the smallest operator $Z$ is not traceless (one can easily check that all other operator strings allowed by rules 1-5 are traceless).
		As established in Eq.~\eqref{eq:Z_expval_TDL}, the $Z$ operator has trace $\langle Z, I\rangle \equiv \text{tr}(Z)/D \approx 0.447$, which represents its overlap with the identity operator in the blockaded subspace. This trivial component does not evolve under the Hamiltonian dynamics and should be subtracted from the operator to make it orthogonal to the identity. 
		
		We can proceed to orthonormalize the operators. Unlike strings of Pauli operators in a local tensor product space, the operator strings generated above are in general not orthogonal: they can have non-zero inner products in the blockaded subspace, e.g.~$\langle PZ_jP, PZ_jPP\rangle = \text{tr}(PI_jPP)/D \neq 0$. We orthonormalize these operators using the Gram-Schmidt procedure, starting from the smallest operators. Let $\{O_1,\dots O_k, \dots\}$ denote the un-orthonormalized operators in the order that Gram-Schmidt will be applied. For each operator $O_k$, we subtract its components which lie in the linear span of all previous operators $\{O_1,\dots,O_{k-1}\}$, and then normalize the remaining component to obtain the operator $O_k^\wedge$. That is,
		\begin{align}
			O_k' &= O_k - \sum_{j=1}^{k-1} \langle O_j^\wedge,O_k\rangle O_j^\wedge, \label{eq:Gram_Schmidt_1}\\
			O_k^\wedge &= \frac{O_k'}{\Vert O_k'\Vert}, \label{eq:Gram_Schmidt_2}
		\end{align}
		where $\lVert \cdot \rVert \equiv \langle \cdot, \cdot \rangle^{1/2}$.
		This ensures that, in the end, all operators $O_k^\wedge$ are normalized and pairwise orthogonal. Furthermore, this ordering preserves smaller operators in favor of more complicated large operators. For concreteness, the smallest non-trivial operators and their representative orthonormal basis elements are:
		\begin{align}
			Z^\wedge(k) &= \frac{Z(k) - \langle Z(k),I\rangle I}{\Vert Z(k) - \langle Z(k),I\rangle I\Vert }, \\
			PXP^\wedge(k) & = \frac{PXP(k)}{\Vert PXP(k) \Vert}~, \\
			PYP^\wedge(k) & = \frac{PYP(k)}{\Vert PYP(k) \Vert}~, \\
			PZP^\wedge(k) &= \frac{PZP(k) - \langle Z^\wedge(k), PZP(k)\rangle Z^\wedge(k)}{\Vert PZP(k) - \langle Z^\wedge(k), PZP(k)\rangle Z^\wedge(k)\Vert},
		\end{align}
		where the $PXP$ and $PYP$ operators are time-reversal symmetric and anti-symmetric superpositions of $P\sigma^\pm P$.
		
		The normalizations of the operators in momentum space can be computed from the inner products $\langle O_{a}, O_{b} \rangle$ of their constituent local operators in real-space. These inner products are non-zero only when the supports of $O_a$ and $O_b$ overlap. The sole exception is when both $O_{a}$ and $O_{b}$ are $Z$ operators, in which case the connected correlator $\text{tr}(Z_i Z_j)/D - (\text{tr}(Z_i)/D)(\text{tr}(Z_j)/D)$ is non-zero for all $i,j$ [Eq.~\eqref{eq:ZZ_spatial_corr}].
		
		Finally, we construct the Liouvillian graph on the orthonormal basis as follows.
		First, we construct the Liouvillian in the un-orthonormalized operator string basis, by computing the matrix elements
		\begin{equation}
			[\mathcal{L}^\text{(str)}_{(k)}]_{ba} \equiv \big\langle O_{b}(k),~ [H,O_{a}(k)] \big\rangle~.
		\end{equation}
		Using the above Gram-Schmidt procedure, we compute the orthonomalized operators $O^{\wedge}_a(k)$, then form the (non-unitary) basis-change matrix $R$, with entries
		\begin{equation}
			[R_{(k)}]_{ba} \equiv [O^{\wedge}_a(k)]_b
		\end{equation}
		i.e.~each column of $R(k)$ is the basis element $O^{\wedge}_a(k)$ in the operator string basis. The Liouvillian in the Gram-Schmidt basis is then simply
		\begin{equation}
			\mathcal{L}^\text{(orth)}(k) \equiv R^\dagger(k) \mathcal{L}^\text{(str)}(k) R(k)~.
		\end{equation}
		While $\mathcal{L}^\text{(str)}$ is in general not a Hermitian matrix, since the original operator strings are not orthonormal, its similarity transform $\mathcal{L}^\text{(orth)}$ is Hermitian.
		As a concrete example, the simplest non-zero edge weight in the Liouvillian graph is given by
		\begin{equation}
			\langle PYP^\wedge (k), \mathcal{L}^\text{(orth)}[Z^\wedge(k)] \rangle = 2 \frac{\Vert PYP (k) \Vert}{\Vert Z(k) - \langle Z(k), I\rangle I\Vert} = \sqrt{\frac{6+4 \cos k}{\varphi}}~.
		\end{equation}
		The other edge weights can be numerically computed systematically following the procedure above.
		
		We conclude with a short remark on improving the numerical efficiency of this procedure.
		Namely, we do not have to perform the Gram-Schmidt orthonormalization Eq.~(\ref{eq:Gram_Schmidt_1},~\ref{eq:Gram_Schmidt_2}) over \textit{all} operator strings at once together. Rather, we only need to perform the Gram-Schmidt procedure on smaller \textit{equivalence classes} of strings that can have non-zero overlap with one another.  For example, the operators $Z$, $PZP$ and $PZPP$ belong to the same equivalence class, and thus must be orthonormalized together, but they have zero overlap with operators outside of this class, such as the $PXP$ or $PYP$ operator, and so can be orthonormalized independently from these operators.
		More generally, for a given operator string, one can find its equivalence class by:
		\begin{enumerate}[i]
			\item Replacing the $P$ and $Z$ characters with the identity $I$.
			\item Removing $I$ characters at the start and end of the string.
		\end{enumerate}
		As a further example, the operators $PXP$, $PXPP$ and $PXPZP$ belong to the same equivalence class. While the first and third operators have zero overlap: $\langle PXP,PXPZP\rangle = 0$, the other pairs $(PXP,PXPP)$ and $(PXPP,PXPZP)$ have non-zero overlap and hence all three operators belong to the same equivalence class. Indeed, applying steps i and ii above reduces all three operators to the string $X$. Orthonormalizing only within each equivalence class enables substantial speed-ups in our numerical computations. 
		
	\end{widetext}
	
\end{document}